\documentclass[longauth]{aa}

\usepackage[varg]{txfonts}
\usepackage{graphicx}
\usepackage{lscape}
\usepackage{url}
\usepackage{natbib}
\usepackage{multirow}
\usepackage{color}

\bibpunct{(}{)}{;}{a}{}{,}

\begin{document}



\title{Stellar populations of galaxies in the ALHAMBRA survey up to $z \sim 1$}
\subtitle{I. \textit{MUFFIT}: A Multi-Filter Fitting code for stellar population diagnostics}

%
\author{L.~A.~D\'iaz-Garc\'ia\inst{\ref{a1}} \thanks{\email{diaz@cefca.es}} \and
A.~J.~Cenarro\inst{\ref{a1}} \and
C.~L\'opez-Sanjuan\inst{\ref{a1}} \and
I.~Ferreras\inst{\ref{a1.5}} \and
J.~Varela\inst{\ref{a1}} \and
K.~Viironen\inst{\ref{a1}} \and
D.~Crist\'obal-Hornillos\inst{\ref{a1}} \and
M.~Moles\inst{\ref{a1},\ref{a2}} \and
A.~Mar\'in-Franch \inst{\ref{a1}} \and
P.~Arnalte-Mur\inst{\ref{a3}} \and
B.~Ascaso\inst{\ref{b3},\ref{a2}} \and
M.~Cervi\~no\inst{\ref{a2},\ref{a10},\ref{a11}} \and
R.~M.~Gonz\'alez~Delgado\inst{\ref{a2}} \and
I.~M\'arquez\inst{\ref{a2}} \and
J.~Masegosa\inst{\ref{a2}} \and
A.~Molino\inst{\ref{b2},\ref{a2}} \and
M.~Povi\'c\inst{\ref{a2}} \and
E.~Alfaro\inst{\ref{a2}} \and
T.~Aparicio-Villegas\inst{\ref{b1},\ref{a2}} \and
N.~Ben\'itez\inst{\ref{a2}} \and
T.~Broadhurst\inst{\ref{a6},\ref{a7}} \and
J.~Cabrera-Ca\~no\inst{\ref{a8}} \and
F.~J.~Castander\inst{\ref{a9}} \and
J.~Cepa\inst{\ref{a10},\ref{a11}} \and
A.~Fern\'andez-Soto\inst{\ref{a12},\ref{a5}} \and  
C.~Husillos\inst{\ref{a2}} \and
L.~Infante\inst{\ref{a13}} \and  
J.~A.~L.~Aguerri\inst{\ref{a10},\ref{a11}} \and
V.~J.~Mart\'inez\inst{\ref{a3},\ref{a4},\ref{a5}} \and
A.~del~Olmo\inst{\ref{a2}} \and
J.~Perea\inst{\ref{a2}} \and
F.~Prada\inst{\ref{a2},\ref{a14},\ref{a15}} \and 
J.~M.~Quintana\inst{\ref{a2}}
}
%
\institute{Centro de Estudios de F\'isica del Cosmos de Arag\'on (CEFCA), Plaza San Juan 1, Floor 2, E--44001 Teruel, Spain\label{a1}\\  \email{diaz@cefca.es} \and
Mullard Space Science Laboratory, University College London, Holmbury St Mary, Dorking, Surrey RH5 6NT, United Kingdom\label{a1.5} \and
IAA-CSIC, Glorieta de la Astronom\'ia s/n, 18008 Granada, Spain\label{a2} \and
Observatori Astron\`omic, Universitat de Val\`encia, C/ Catedr\`atic Jos\'e Beltr\'an 2, E-46980, Paterna, Spain\label{a3} \and
Departament d'Astronomia i Astrof\'isica, Universitat de Val\`encia, E-46100, Burjassot, Spain\label{a4} \and
Unidad Asociada Observatorio Astron\'omico (IFCA-UV), E-46980, Paterna, Spain\label{a5} \and
Department of Theoretical Physics, University of the Basque Country UPV/EHU, 48080 Bilbao, Spain\label{a6} \and
IKERBASQUE, Basque Foundation for Science, Bilbao, Spain\label{a7} \and
Departamento de F\'isica At\'omica, Molecular y Nuclear, Facultad de F\'isica, Universidad de Sevilla, 41012 Sevilla, Spain\label{a8} \and
Institut de Ci\`encies de l'Espai (IEEC-CSIC), Facultat de Ci\`encies, Campus UAB, 08193 Bellaterra, Spain\label{a9} \and
Instituto de Astrof\'isica de Canarias, V\'ia L\'actea s/n, 38200 La Laguna, Tenerife, Spain\label{a10} \and
Departamento de Astrof\'isica, Facultad de F\'isica, Universidad de La Laguna, 38206 La Laguna, Spain\label{a11} \and
Instituto de F\'isica de Cantabria (CSIC-UC), E-39005 Santander, Spain\label{a12} \and
Departamento de Astronom\'ia, Pontificia Universidad Cat\'olica. 782-0436 Santiago, Chile\label{a13} \and
Instituto de F\'{\i}sica Te\'orica, (UAM/CSIC), Universidad Aut\'onoma de Madrid, Cantoblanco, E-28049 Madrid, Spain \label{a14} \and
Campus of International Excellence UAM+CSIC, Cantoblanco, E-28049 Madrid, Spain \label{a15} \and
Observat\'orio Nacional-MCT, Rua Jos\'e Cristino, 77. CEP 20921-400, Rio de Janeiro-RJ, Brazil\label{b1} \and
Instituto de Astronom{\'{\i}}a, Geof{\'{\i}}sica e Ci\'encias Atmosf\'ericas, Universidade de S{\~{a}}o Paulo, S{\~{a}}o Paulo, Brazil\label{b2} \and
GEPI, Observatoire de Paris, CNRS, Universit\'e Paris Diderot, 61, Avenue de l’Observatoire 75014, Paris France\label{b3} 
}

\date{Received ? / Accepted ?}

%

\abstract{}     
         {To present \textit{MUFFIT}, a new generic code optimized to retrieve the main stellar population parameters of galaxies in photometric multi-filter surveys, and check its reliability and feasibility with real galaxy data from the ALHAMBRA survey.}
         {Making use of an error-weighted $\chi^2$-test, we compare the multi-filter fluxes of galaxies with the synthetic photometry of mixtures of two single stellar populations at different redshifts and extinctions, to provide the most likely range of stellar population parameters (mainly ages and metallicities), extinctions, redshifts, and stellar masses. To improve the diagnostic reliability, \textit{MUFFIT} identifies and removes from the analysis those bands that are significantly affected by emission lines. The final parameters and their uncertainties are derived by a Monte Carlo method, using the individual photometric uncertainties in each band. Finally, we confront the accuracies, degeneracies, and reliability of \textit{MUFFIT} using both simulated and real galaxies from ALHAMBRA, comparing with results from the literature.} 
         {\textit{MUFFIT} is demonstrated to be a precise and reliable code to derive stellar population parameters of galaxies in ALHAMBRA. Using as input the results from photometric-redshift codes, \textit{MUFFIT} improves the photometric-redshift accuracy by $\sim 10$--$20\%$. \textit{MUFFIT} also detects nebular emissions in galaxies, providing physical information about their strengths. 
The stellar masses derived from \textit{MUFFIT} show an excellent agreement with the COSMOS and SDSS values. In addition, the retrieved age-metallicity locus for a sample of $z \le 0.22$ early-type galaxies in ALHAMBRA at different stellar mass bins are in very good agreement with the ones from SDSS spectroscopic diagnostics. Moreover, a one-to-one comparison between the redshifts, ages, metallicities, and stellar masses derived spectroscopically for SDSS and by \textit{MUFFIT} for ALHAMBRA reveals good qualitative agreements in all the parameters, hence reinforcing the strengths of multi-filter galaxy data and optimized analysis techniques, like \textit{MUFFIT}, to conduct reliable stellar population studies.}     
         {}  

%

%
\titlerunning{\textit{MUFFIT}: A MUlti-Filter FITting code for stellar population diagnostics}
%
\authorrunning{L.~A.~D\'iaz-Garc\'ia et al.}

\maketitle

%


\section{Introduction}\label{sec:introduction}

Studying the stellar content of galaxies is crucial to understand their star formation histories (SFH), what in turn provides us with valuable information on the possible evolutive paths since their formation at high redshift down to the present time. Despite the large efforts and advances achieved in this topic during the last decades, it still remains as one of the most challenging and promising ways to understand galaxy evolution.

Early attempts to study the stellar content of early-type galaxies were based on colours, from wide and narrow band photometry \citep{Baum1959,Tifft1963,Wood1966,Mcclure1968,Faber1973}, and on empirical synthesis of the populations using as reference basis the observed colours of nearby early-types. These early methods can be considered as the pioneers of the current photo-spectral fitting techniques, which are the main topic of the present paper. The above methods were gradually displaced by techniques based in more specific features \citep{Faber1973,Pritchet1977} that were defined in narrow spectral ranges. 

The arrival of absorption line-strength indices to study the stellar content of galaxies \citep{Burstein1984,Faber1985} brought a significant breakthrough in the field. In this front, it is worth noting the Lick system of indices \citep{Gorgas1993,Worthey1994b}, which for the last decades has been the standard for most spectroscopic studies in stellar populations in the optical \citep[e.~g.][]{Trager1998,Jorgensen1999,Kuntschner2001,Thomas2005,Bernardi2006,Sanchezblazquez2006a,Gorgas2007}. The combination of a certain number of absorption lines mainly sensitive to age, e.~g. the Balmer lines, or to the metallicity, as traced by certain elements such as \ion{Fe}, \ion{Mg}, \ion{Ti}, \ion{C}, \ion{Ca}, \ion{Na}, etc., were proven to be an efficient way to break, at least to some extent, the well known degeneracy between these two parameters \citep{Worthey1994a}. The way to measure these features is delicately chosen to be very sensitive to a parameter of interest, focusing its study in small spectral ranges. By construction, line-strength indices are quite insensitive to the influence of extinction, and by fine-tuning their definition or combining the sensitivities of different indices, some of them may end up being almost independent from other parameters, e.~g. metallicity \citep{Vazdekis1999, Cervantes2009} and $\alpha$-element overabundances \citep{Thomas2003}.

In the last fifteen years, the development of stellar libraries in spectral ranges other than the optical has driven the definition of new indices that allowed to extend this kind of studies to other regions with unexplored sensitivities \citep{Cenarro2002,Marmol-Queralto2008}. In addition, the index system of reference in the optical spectral range has been revisited and improved \citep[see e.~g.][]{Vazdekis2010} thanks to the availability of much better stellar libraries at much better spectral resolution. 

It was with the arrival of improved stellar libraries, such as CaT \citep{Cenarro2001a, Cenarro2001b}, ELODIE \citep{Prugniel2001}, STELIB \citep{LeBorgne2003}, INDO-US \citep{Valdes2004}, \citet{Martins2005}, and MILES \citep{Sanchezblazquez2006b, Cenarro2007}, and the consequent evolutionary stellar population synthesis models \citep[e.~g.][]{Bruzual2003, Vazdekis2003, GonzalezDelgado2005,Maraston2009, Vazdekis2010, Conroy2012, Vazdekis2012}, that fitting techniques over the full spectral energy distribution (SED) of galaxies appeared as an alternative to line-strength indices. SED-fitting can also be used to derive several physical properties of galaxies \citep{Mathis2006,Koleva2008,Coelho2009,Walcher2011,Liu2013}. In fact, there is a growing number of public codes specifically devoted to carrying out SED-fitting with different procedures, e.~g. \textit{hyperz} \citep{Bolzonella2000}, \textit{Le PHARE} \citep{Arnouts2002,Ilbert2006}, \textit{STARLIGHT} \citep{Cid2005}, \textit{STECKMAP} \citep{Ocvirk2006}, \textit{VESPA} \citep{Tojeiro2007}, \textit{ULySS} \citep{Koleva2009}, \textit{FAST} \citep{Kriek2009}, \textit{SEDfit} \citep{Sawicki2012}.

Nowadays, there is an increasing number of present and future multi-filter surveys, e.~g. COMBO-17 \citep{Wolf2003}, MUSYC \citep{Gawiser2006}, COSMOS \citep{Scoville2007}, ALHAMBRA \citep{Moles2008}, CLASH \citep{Postman2012}, SHARDS \citep{PerezGonzalez2013}, J-PAS \citep{Benitez2014}, and J-PLUS \citetext{Cenarro et al.~2015, in prep.}, each of them with a vast volume of high-quality multi-filter data. These kinds of surveys pursue diverse goals with a common feature: sampling the SEDs of galaxies using top-hat and/or broad-band filters that mainly cover the optical range. Owing to this configuration, the retrieved SEDs are half-way between classical photometry and spectroscopy, being in practice like a low-resolution spectrum whose resolution depends on the filter system (e.~g. $R \sim 20$ for ALHAMBRA; $R \sim 50$ for J-PAS). Although multi-filter observing techniques suffer from the lack of high spectral resolution, their advantages over standard spectroscopy are worth to be noted: (i) The galaxy samples of multi-filter surveys do not suffer from selection criteria other than the photometric depth in the detection band of the survey, as all the objects in the field of view are observed. For a fixed observational time and similar telescopes, this leads to much larger galaxy samples than in multi-object spectroscopy, where achieving multiplexities larger than $\sim1\,000$ is a challenge at present; (ii) Unlike standard spectroscopy, the SED of galaxies observed in multi-filter surveys does not suffer from the typical uncertainties in the flux calibration that lead to systematic colour terms, as the photometric calibration of each individual band is independent from the rest. This advantage is crucial, as it is the overall continuum of the stellar population what in most cases dominates the diagnostic with SED-fitting techniques; (iii) With similar telescopes, the depth of multi-filter surveys is usually much larger than in that of spectroscopic survey, as direct imaging is much more efficient that spectroscopy; and (iv) Multi-filter surveys provide spatially resolved photo-spectra, similar to an IFU technique, allowing us to perform 2D stellar population studies in galaxies whose apparent sizes are not dominated by the point spread function (PSF) of the system.

It is therefore clear that multi-filter surveys open a profitable way to advance in our understanding of galaxy evolution by providing complete and homogeneous sets of galaxy SEDs down to a certain magnitude depth. Although there are several SED-fitting codes available, to cope with the calibration particularities of multi-filter surveys (see e.g. Molino et al. 2014), and given the vast amount of high-quality photometric data already available in the literature, and still to come in the next years, in this paper we present \textit{MUFFIT} (\textit{MUlti-Filter FITting for stellar population diagnostics}), a  code specifically designed for analysing the stellar content of galaxies with available multi-filter data. This paper is mainly aimed to describe the code and its functionalities, set the accuracy and typical uncertainties in the retrieved stellar population parameters, and demonstrate its reliability confronting with already existing stellar population results in the literature. The development of \textit{MUFFIT} has been performed within the framework of the ALHAMBRA survey (see Sect.~\ref{sec:data}), so, even though the code is generic and can be easily employed for any kind of photometric system, many sections in this paper are particularized for the ALHAMBRA dataset. This allows us to show the code performances on real galaxy data, which is ultimately the best sanity check for any stellar population code. Despite in this paper we use galaxy data from ALHAMBRA, it is not our intention to exploit scientifically the data set here. In the next papers of this series, we will provide and exploit the stellar population parameters retrieved for the whole galaxy sample in the ALHAMBRA survey. 

This paper is organized as follows. Section~\ref{sec:data} presents a quick overview of the ALHAMBRA survey, i.~e. the photometric dataset employed to develop the present work. In Sect.~\ref{sec:bigcode}, we summarize the main technical aspects carried out by our code, \textit{MUFFIT}, as well as the processes to obtain photometric colour predictions from models of single stellar populations (SSP) and the Milky Way (MW) extinction corrections. We show the accuracy and reliability of the stellar population parameters retrieved with our code, together with the uncertainties and degeneracies expected for ALHAMBRA data in Sect.~\ref{sec:feasibility}. Section~\ref{sec:testing} presents a comparison study of the results retrieved from ALHAMBRA galaxy data using \textit{MUFFIT} with previous studies, including spectroscopic ones, and data from the literature, hence testing the reliability of our results. Finally, we provide the summary and conclusions of this research in Sect.~\ref{sec:conclusions}.

Throughout this paper we assume a $\Lambda$CDM cosmology with $H_0 = 71 $~km s$^{-1}$, $\Omega_\mathrm{M}=0.27$, and $\Omega_\mathrm{\Lambda}=0.73$.

\section{The ALHAMBRA survey}\label{sec:data}
The stellar population code that we present in this paper is generically designed for all types of multi-filter surveys. However, we make use of the data in the ALHAMBRA survey\footnote{\url{http://www.alhambrasurvey.com}} to prove and test the reliability of our techniques, as in fact this code will be employed to analyse the stellar population properties of ALHAMBRA galaxies in forthcoming papers \citetext{D\'iaz-Garc\'ia et al.~2015; in prep.}. Therefore, throughout this work,we mainly present results, from both simulations and real observations, that are based either on the ALHAMBRA data or on its technical setup. In the following paragraphs we present a short summary of the ALHAMBRA survey.

The ALHAMBRA survey provides a photometric dataset of $20$ contiguous, medium-band (FWHM $\sim300$~\AA), top-hat filters, that cover the complete optical range $\lambda\lambda\ 3\,500$--$9\,700$~\AA\ \citep[see][for further details]{Aparicio2010} over $8$ non-contiguous regions of the northern hemisphere, amounting a total area of $4$~deg$^2$ of the sky \citep[including areas in common with other cosmological surveys such as COSMOS, see ][for further overlapping areas]{Molino2014}. All filters in the optical range have very steep side transmission slopes, close to zero overlap in wavelength, a flat top, and transmissions $80$--$95$\% \citep{Moles2008}. The magnitude limit is $m_\mathrm{AB}\sim24$ ($5$--sigma, measured on $3\arcsec$) for the $14$ filters ranging from $3\,500$ to $7\,700$~\AA, decreasing smoothly in the six reddest filters reaching down to $m_\mathrm{AB} \sim 21.5$ in the reddest one \citep{Molino2014}, which is centred at $9\,550$~\AA. The optical coverage is supplemented with the standard NIR \textit{J}, \textit{H}, and $K_\mathrm{s}$ filters which have a $50$\% detection efficiency depth (point-like sources, AB magnitudes) of \textit{J} $\sim 22.4$, \textit{H} $\sim 21.3$, and $K_\mathrm{s} \sim 20.0$, analysed in \citet{Cristobal2009}. The ALHAMBRA filter set\footnote{\url{http://svo2.cab.inta-csic.es/theory/fps3/}} is designed to optimize the accuracy of photometric redshifts \citep[photo-\textit{z},][]{Benitez2009}, but due to their characteristics, it also provides low-resolution photo-spectra composed of $23$ bands, corresponding to a resolving power $R\sim20$ in the optical. All the observations were done under a quality criterion, seeing  $< 1.6\arcsec$ and airmass $< 1.8$, using the $3.5$~m telescope in the Calar Alto Observatory\footnote{\url{www.caha.es}} (CAHA) with two cameras, the imager LAICA in the optical range and Omega--2000 for the NIR filters. At present, all the ALHAMBRA fields have not been imaged yet, being the effective area for this work $2.8$~deg$^2$ with a total on-target exposure time of $\sim700$~h ($\sim608$~h were dedicated for the optical bands, and $\sim92$~h for the NIR ones), although the rest of fields will be completed reaching the expected total area of $4$~deg$^2$.

The ALHAMBRA Gold catalogue \footnote{\url{http://cosmo.iaa.es/content/alhambra-gold-catalog}} \citep{Molino2014}, hereafter Gold catalogue, is the reference catalogue for this work. As explained in \citet{Molino2014}, synthetic $F814W$ images were created, as a linear combination of individual filters, to be used for both detection and completeness purposes, emulating the $F814W$ band of the Advanced Camera for Surveys (ACS) in the Hubble Space Telescope (\textit{HST}). Therefore, the Gold catalogue provides $23+1$ photometric AB magnitudes \citep{Oke1983} and errors for $\sim 95\,000$ bright galaxies ($17 < m_{F814W} < 23$), being complete up to $m_{F814W}=23$. Throughout this work, the synthetic $F814W$ photometry is removed from the analysis. Due to the existence of a PSF variability among different filters, the photometry is corrected of PSF and aperture effects. In addition, and for the specific ALHAMBRA case, we add quadratically an extra uncertainty of $\sim0.025$~(AB magnitudes) in each photometric measurement to account for potential calibration issues or uncertainties.

For further details of the ALHAMBRA survey, we refer the readers to \citet{Moles2008} and \citet{Molino2014}.

\section{The code}\label{sec:bigcode}
Although there are many public codes devoted to carrying out SED fitting in many different ways, e.~g. \textit{hyperz}, \textit{STARLIGHT}, \textit{ULySS}, \textit{VESPA}, \textit{Le PHARE}, \textit{FAST}, or \textit{SEDfit}; we are performing our own analysis techniques to retrieve stellar population parameters from photometric SEDs, specifically designed for analysing the stellar content of galaxies from the ALHAMBRA survey, but being generic and easily adaptable to any multi-photometric survey. Secondarily, there is an increasing number of large-scale multi-filter surveys; e.~g. ALHAMBRA, J-PLUS and J-PAS, SHARDS, CLASH, MUSYC, COSMOS, or COMBO-17; offering a huge amount of photometric data that we can exploit to study the evolution of galaxies, opening a new path to explore the stellar population of galaxies, overall at intermediate and high redshifts. Although these photometric data are like low-resolution spectra, these techniques present remarkable advantages respect to spectroscopy: they can go deeper with a better flux calibration (the calibration of each filter is independent of the rest of them), we can study the stellar content in each resolution element (similar to IFU techniques) with one exposure, and we can work with larger galaxy samples; hence, it would be a pity not taking advantage of these studies and not exploiting all the opportunities that they offer.

The collection of analysis techniques, routines, and other tools that we are performing, are collected under the code name \textit{MUFFIT} (\textit{MUlti-Filter FITting in photometric surveys}), which is written in \texttt{Python} language, and it is mainly focused on retrieving the stellar populations of galaxies whose SEDs are dominated by their stellar content.

This section is subdivided in two extended sections. On the one hand, we show in Sect.~\ref{sec:ingredients} the main ingredients or inputs required to develop the analysis. These preliminary elements are basically composed by the SSP models, the photometric system, and the selection of a dust extinction law to treat properly the impact of dust on the model SEDs. On the other hand, the performing of the code is described in Sect.~\ref{sec:analysis}, being emphasized the description of some specific tasks.

\begin{figure*}
\centering
\includegraphics[width=17cm]{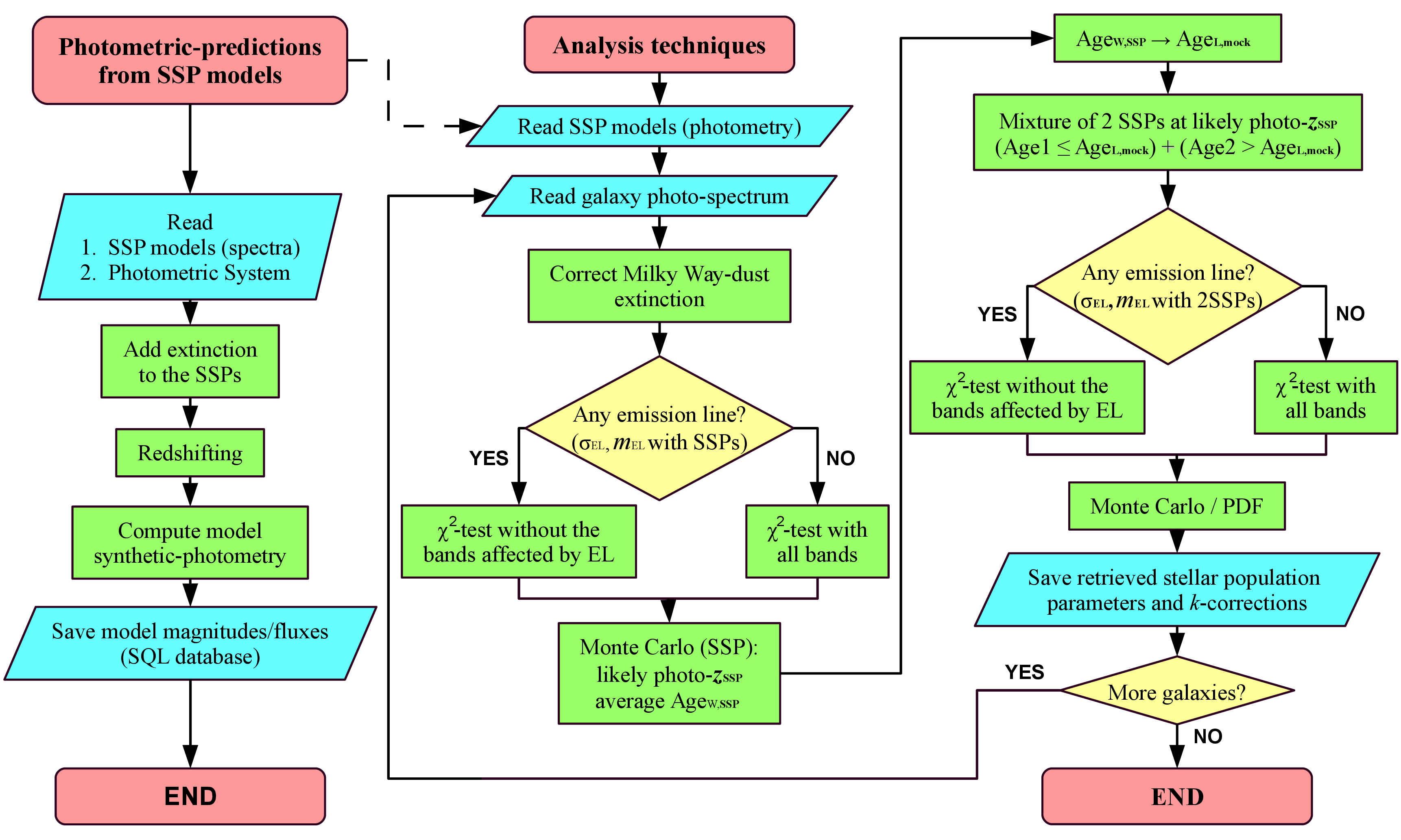}
\caption{Flowcharts of the photometric model predictions (left) and the analysis techniques (right). A more detailed explanation of each step can be found in Sects.~\ref{sec:synphot}, \ref{sec:ext}, and \ref{sec:analysis}. The dashed row marks where both processes are related. Flowchart symbols represent standard tasks: ovals, start/end of a process (red); arrows, the direction of logic flow in the process; parallelograms, input/ouput operation (cyan); diamonds, a decision or branch to be made (yellow); and rectangles, a processing step (green).}
\label{fig:techniques}
\end{figure*}

In Fig.~\ref{fig:techniques}, we outline the main structure of the code by a brief flowchart that summarizes the main features of the followed processes to set constrains in the stellar populations. We caution that the purposes of the flowchart is to help the reader follow the development of Sects.~\ref{sec:ingredients} and ~\ref{sec:analysis}, and support the schematic comprehension of the large number of stages.

The reader who may be primarily interested either in the reliability of the code or in the comparison with results retrieved from the literature, may skip this section to continue with the self-contained Sects.~\ref{sec:feasibility} and~\ref{sec:testing}.


\subsection{Main ingredients of the stellar population code}\label{sec:ingredients}

In this section we describe the main input ingredients and preparatory tasks that are considered for the development of the stellar population analysis code that is presented in this paper. In particular, our code requires an input set of reference SSP models (Sect.~\ref{sec:sspmodels}), the photometric system of the data to be analysed (Sect.~\ref{sec:synphot}), and a set of recipes to take into account the intrinsic and Milky Way extinction (Sect.~\ref{sec:ext}). The redshifts of the target galaxy data can be managed as an input ingredient or an output of the code, as it is explained in Sect.~\ref{sec:redsh}. The flowchart on the left side of Fig.~\ref{fig:techniques} particularly illustrates the main ingredients and preliminary work carried out by the code before starting with the analysis of the data.


\subsubsection{The SSP models}\label{sec:sspmodels}

The code has been designed to use SSP models as input templates for the comparative analysis of the stellar populations of galaxies. Currently the code is ready to account for \citet[][hereafter BC03]{Bruzual2003}\footnote{\url{http://bruzual.org/}} and MIUSCAT\footnote{\url{http://miles.iac.es/}} \citep{Vazdekis2012,Ricciardelli2012} SSP models, although any other SSP spectral dataset can be easily implemented. 

BC03 is perfectly suited for SED fitting given the large spectral coverage of the models, from $91$~\AA\ to $160\ \mu\mathrm{m}$, allowing us to cope with most kind of multi-filter galaxy data, irrespective of the redshift. For the present work, we assume ages up to $14$~Gyr and metallicities [Fe/H]$= -1.65$, $-0.64$, $-0.33$, $0.09$, and $0.55$, Padova 1994 tracks \citep[for further details and references, see][]{Bruzual2003}, and a \citet{Chabrier2003} initial mass function (IMF).

MIUSCAT provides a sample of SEDs with a spectral range $\lambda\lambda\ 3\,465$--$9\,469$~\AA\ and a resolution of $\mathrm{FWHM} \sim 2.5$~\AA, almost constant with wavelength \citep{FalconBarroso2011}. Despite the great colour calibration of these models, its spectral range is not enough for galaxies at intermediate redshift and further, missing the observed ALHAMBRA colours in the UV range. For this purpose, we extend the lower end of MIUSCAT models up to $1\,860$~\AA\ \citetext{A.~Vazdekis 2015, priv.~comm.}, using the Next Generation Spectral Library \citep[NGSL,][]{Heap2007}. In addition, we complement these models with their photometric predictions for \emph{J}, \emph{H}, and \emph{K}, which are adapted to predict the ALHAMBRA NIR bands. Throughout this work, we use the models up to $14.13$~Gyr with metallicities [Fe/H]$= -1.31$, $-0.71$, $-0.40$, $0.0$, and $0.22$. We assume a Kroupa Universal-like initial mass function \citep{Kroupa2001}, despite of its universality being a current matter of debate \citep[see, e.~g.][]{Ferreras2013}. In future works, we will shed light on the systematic variation of the IMF for the more massive galaxies in the ALHAMBRA database.

By construction, the code can also use not only any other set of SSP models, but also any other kind of reference template spectra, e.g. spectra of real galaxies, as long as their main stellar population parameters (age, metallicity, IMF, extinction, and over-abundances) are assigned by the user. Throughout this paper we do not present this possibility, but concentrate on the performance of the code on the basis of the two SSP model sets mentioned above.


\subsubsection{Photometric system and synthetic photometry}\label{sec:synphot}

\begin{figure*}
\centering
\includegraphics[width=17cm]{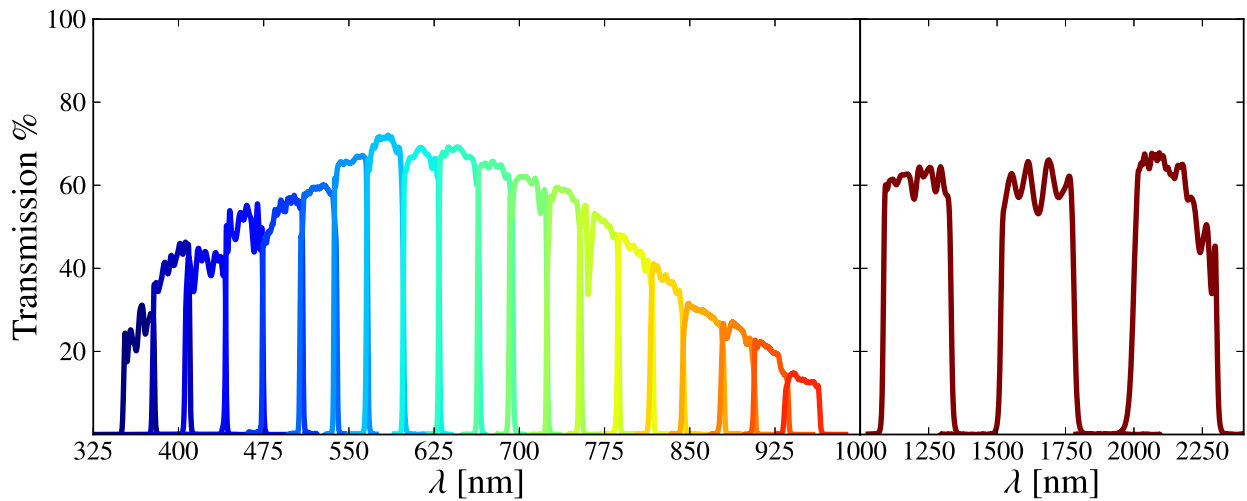}
\caption{Response curves of the ALHAMBRA filter set for the CCD 1 in the optical range (LAICA camera; one colour from blue to red per band) together with the ALHAMBRA \textit{J}, \textit{H}, and $K_\mathrm{s}$ filters (Omega--2000 camera; dark red) to make the model synthetic photometry.}
\label{fig:filters_alh}
\end{figure*}

For a proper comparison between input SSP models and galaxy data, it is essential to build a reliable estimation of the synthetic magnitudes (or integrated fluxes) of the SSP template models in the same photometric system of the galaxies that need to be analysed. This is computed by convolving the SSP model or galaxy reference spectra with the response functions of the photometric system. In addition to taking the empirical filter transmission curves into account, to obtain a reliable photometric prediction it is advisable to account for specific characteristics of the observing conditions and the instrumental setup employed for the photometric observations of the galaxies to be analysed. Among others, e.g. the transmittance of the optical system and/or the sky absorption spectrum where the observations were taken. It is remarkable the wavelength dependence of the quantum efficiency of charge-coupled devices (CCDs), being typically less sensitive in the bluer and redder ends. If not accounted for properly, this effect modifies the effective wavelength of such filter bandpasses creating a fictitious colour term in the synthetic photometry of the reference models. In Fig.~\ref{fig:filters_alh}, the response functions of the ALHAMBRA photometric system is presented. It consists of $20$ optical bands (left-hand side) and the ALHAMBRA \textit{J}, \textit{H}, and $K_\mathrm{s}$ NIR-bands (right-hand side). In this figure, all the effects explained above are already embedded.

We compute the synthetic photometry following the methodology described in \citet{Pickles2010}, which is based in the \textit{HST} \texttt{synphot}\footnote{\url{http://www.stsci.edu/institute/software_hardware/stsdas/synphot}} package and in \citet{Bessel2005}. 

As current detectors are photon-counting detectors, the number of photons detected across a pass-band \textit{X} is 
\begin{equation}
N_{X}^\mathrm{ph} = \int \frac{\lambda}{\mathrm{hc}} F_{\lambda} R_X(\lambda) \: {\mathrm{d}}\lambda \ , 
\label{eq:nph}
\end{equation}
where $F_{\lambda}$ is the spectrum to convolve and $R_X(\lambda)$ is the response function of the filter \textit{X} (also called sensitivity function in some previous work). Normalizing Eq.~\ref{eq:nph}, we get the weighted mean photon flux density,
\begin{equation}
\overline{F^{\mathrm{ph}}_{X}} = \frac{\int \lambda F_{\lambda} R_X(\lambda) \: {\rm d}\lambda}{\int \lambda R_X(\lambda) \: {\rm d}\lambda}.
\label{eq:meanflux}
\end{equation}

Some catalogues provide photometry in AB magnitudes, defined as 
\begin{equation}
m_{\rm AB} = -2.5 \log_{10}{f_{\nu}} - 48.6 \ ,
\label{eq:mab}
\end{equation}
where $f_{\nu}$ is the flux in ergs cm$^{-2}$ Hz$^{-1}$ s$^{-1}$. To transform the weighted mean photon flux density into AB magnitudes, we compute the magnitude of the flux in the STMAG system \citep[system for calibrating \textit{HST} stars,][]{Stone1996}, which can be easily transformed to the AB magnitude system (Eq.~\ref{eq:mab_mst}). This intermediate step is necessary due to the fact that the weighted mean photon flux density is established per unit wavelength, whereas the AB magnitude system is given per unit frequency. The magnitude across the bandpass \textit{X} in the STMAG system, $m_{\mathrm{ST},X}$, and in the AB system, $m_{\mathrm{AB},X}$, are
\begin{equation}
m_{{\rm ST},X} = -2.5 \log_{10}{\overline{F^\mathrm{ph}_{X}}} - 21.1 \ ,
\label{eq:mst}
\end{equation}
\begin{equation}
m_{\mathrm{AB},X} = m_{\mathrm{ST},X} - 5 \log_{10}{\lambda_\mathrm{pivot}} + 18.692 \ ,
\label{eq:mab_mst}
\end{equation}
where $\lambda_\mathrm{pivot}$ is the source-independent pivot wavelength, which is defined as
\begin{equation}
\lambda_\mathrm{pivot} = \sqrt{\frac{\int \lambda R_X(\lambda) \: \mathrm{d}\lambda}{\int R_X(\lambda) \: \mathrm{d}\lambda / \lambda }} \ .
\label{eq:pivot}
\end{equation}

\begin{figure*}
\centering
\includegraphics[width=17cm]{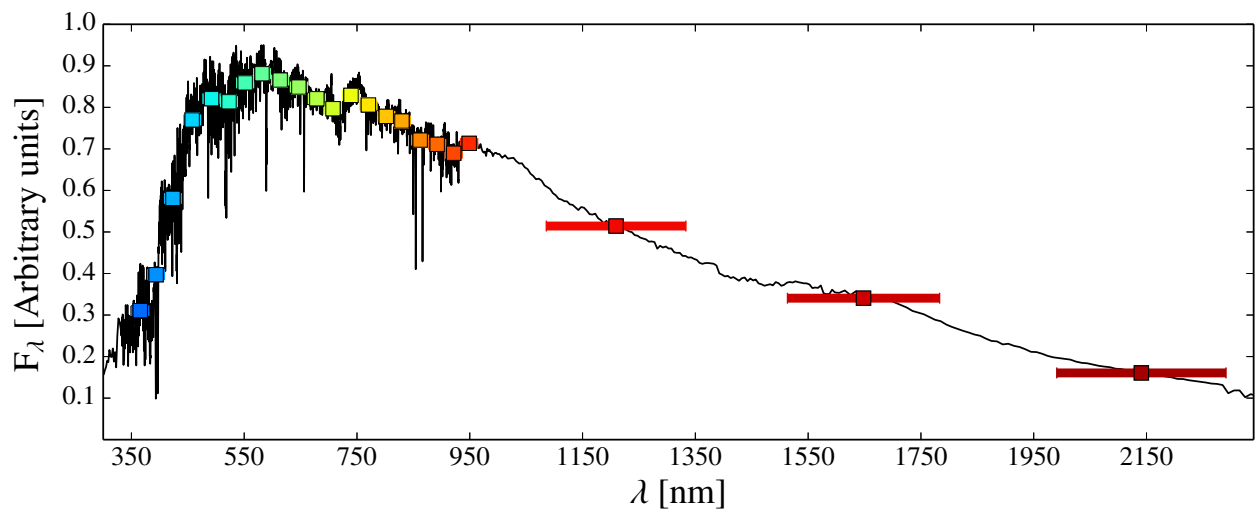}
\caption{Synthetic photometry of a SSP using the ALHAMBRA photometric system. The black line is the SSP flux, the colour squares are the expected pass-bands and the horizontal bars represent the FWHM of each filter.}
\label{fig:synphot}
\end{figure*}

To illustrate this, Fig.~\ref{fig:synphot} shows an example of the synthetic photometry determined for a SSP from BC03 with the ALHAMBRA filter set. The black line is the flux of a SSP at rest-frame with solar metallicity, intermediate-age ($5$~Gyr), Chabrier IMF, and without intrinsic extinction. The colour squares correspond to the spectrum synthetic photometry following the process explained above, centred at their effective wavelengths ($\lambda_\mathrm{eff} = \int \lambda R_X(\lambda) \: \mathrm{d}\lambda / \int R_X(\lambda) \: \mathrm{d}\lambda$), and the horizontal bars represent the FWHM of each filter. This example is also useful to show that the main, broader spectral features are easily distinguishable after convolving, emphasizing the power of the ALHAMBRA photo-spectra as halfway between classical photometry and spectroscopy.

For the ALHAMBRA specific case, and because of the configuration of LAICA, we compute four photometric databases for the optical bands, one per CCD, as there exists discrepancies among CCD sensitivities and each CCD has its own set of filters. For the NIR-filters \textit{J}, \textit{H}, and $K_\mathrm{s}$, we repeat this process taking the Omega--2000 configuration (only one detector plate). In both optical and NIR synthetic photometry, we take into account the filter transmission curves, the quantum efficiency of every CCD/camera, the sky absorption spectrum at CAHA, and the reflectivity of the $3.5$~m-telescope primary mirror with the transmittance of the optical system.

Due to both the large number of input model parameters (ages, metallicities, extinctions, IMF slopes and redshifts) and the intermediate-high spectral resolution of current SSP models, in general it is more efficient to build up our set of convolved models once at the beginning, rather than recomputing the model synthetic photometry every time the code is run. After computing the synthetic photometry of all models, the photometric predictions (fluxes and magnitudes) along with the main characteristics of each model are stored in a structured query language (SQL) database.

A straightforward flowchart of the process to estimate photometric predictions is shown on the left hand-side of Fig.~\ref{fig:techniques}.

\subsubsection{Dust Extinction}\label{sec:ext}

Stellar population diagnostic techniques based on SED fitting over a large spectral coverage, as in this case, require the reddening by extinction to be thoroughly taken into account to avoid potential misinterpretations of the integrated colours of the population, e.~g. older ages or higher metallicities, as well as to derive reliable stellar masses.

Many authors have tried to parametrize the shape of the dust extinction curve \citep[e.g.][]{Prevot1984,Massa1987,Mathis1990,Cardelli1989,Calzetti2000,Odonnell1994,Fitzpatrick1999}, overall on the bluer parts where the dust reddening is more complex. The dust extinction curve is well reproduced using the parameter $R_V \equiv A_V/E(B-V)$ \citep{Cardelli1989}, which varies between 2.2 and 5.8 depending on the environmental characteristics of the diffuse inter stellar medium (ISM). Although the values of $R_V$ may change depending of the line-of-sight, throughout this work we assume that the value of this parameter is $R_V = 3.1$, which is the mean value in the diffuse ISM of the MW \citep{Cardelli1989,Schlafly2011}. Among the available extinction laws in our code \citep{Prevot1984, Cardelli1989, Fitzpatrick1999, Calzetti2000}, throughout this work we choose the Fitzpatrick reddening law \citep{Fitzpatrick1999} as it reproduces well the extinction observed for MW stars with a preferred mean value around $R_V = 3.1$ \citep{Schlafly2011}. 

For extragalactic objects, there are two main sources of extinction to account for. On the one hand, the dust intrinsic to the observed galaxy, which is redshifted with the galaxy system. On the other hand, the observed SED is reddened by the foreground MW dust in the observer's reference system. It is important to note that this local MW extinction cannot be corrected together with the intrinsic galaxy reddening as the emitted flux is red-shifted before being scattered by the dust in our galaxy. As we present below, we separately face both extinction effects.

Following a given extinction law, the intrinsic extinction is applied to the SSP template models before they are red-shifted and convolved with the photometric system. Throughout this work the values of $A_V$ range from $0.0$ to $3.1$ (in bins of $0.1$ in the range $0.0$--$1.0$, and in bins of $0.3$ from $1.0$--$3.1$). The intrinsic extinction can be added as 
\begin{equation}
F_\lambda = F_{\rm \lambda, 0} \times 10^{-0.4 A_\lambda} \ ,
\label{eq:add_ext}
\end{equation}
where $F_{\rm \lambda, 0}$ is the SSP-model/template flux at rest-frame, $F_\lambda$ after adding extinction, and $A_\lambda$ is determined by the extinction law, which can be chosen by the user. Since it is not clear how $R_V$ varies within a host galaxy and among different types of galaxies, we keep constant the value to $R_V = 3.1$, i.~e. the mean value in the MW. This helps to avoid degeneracies and to reduce the number of free parameters, which is already very high and time consuming. In spite of the different reddening laws have intrinsic differences \citep[see][]{Fitzpatrick1999}, we do not assume errors in the SSP template models owing to such uncertainties.

We use the dust maps of \citet{Schlegel1998}, hereafter SFD, in order to deal with the MW reddening in the line-of-sight of each galaxy in our sample. The SFD dust maps provide $E(B-V)$ values in different positions of the sky by estimating the dust column density. These estimations were calibrated using galaxies and assuming a standard reddening law, to infer the existence of galactic dust between the observer and the sources beyond the MW limits. As the spatial resolution of SFD is low, FWHM $\sim 6.1 \arcmin$ and pixel size $(2.372 \arcmin)^2 $, \textit{MUFFIT} makes a bilinear interpolation in the $E(B-V)$ grid for every position of the target galaxies.

\textit{MUFFIT} applies a foreground extinction correction for each individual galaxy photo-spectrum using an extinction law for a value of $E(B-V)$ and $R_V$. The most simple way to de-redden the photo-spectrum of a given galaxy is to compute the extinction in the effective wavelengths of the different filters and then correct the source photometry using the equation

\begin{equation}
F_{\lambda,{\rm c}} = F_{\rm \lambda, red} \times 10^{0.4 A_\lambda} \ ,
\label{eq:ext}
\end{equation}
where $F_{\lambda,{\rm c}}$ is the flux corrected of MW extinction for a given wavelength, $F_{\rm \lambda, red}$ is the observed flux (reddened), and $A_\lambda$ is the extinction factor given for a extinction law. Since the transmission curves of the filters are not completely flat and the shape of the continuum is source dependent, this approximation may be inappropriate for those filters that exhibit a gradient in their transmission curves (e.~g. the lower and upper end of the ALHAMBRA optical bands, see Fig.~\ref{fig:filters_alh}), especially in the spectral ranges where the observed spectrum is not flat. This effect would be interpreted as a shift in the filter effective wavelength \citep{Fitzpatrick1999}, and finally, as a colour term in the spectral regions with strong gradients in flux, e.~g. the $4\,000$~\AA-break. To get a more reliable correction in this sense, the code carries out the de-reddening process of the data in three steps:

\begin{itemize}

\item First, we pick up a set of models from BC03 ($29$ ages, from $0.1$ to $10$~Gyr, four metallicities, $[{\rm Fe/H}] = -0.64$, $-0.33$, $0.09$, and $0.56$, and a Chabrier IMF) to be redshifted (redshift bin $0.01$) and convolved with the survey photometric system. Before redshifting and computing the synthetic photometry, we add the intrinsic extinction ($A_V$ from $0.0$ to $1.0$, in bins of $0.2$ ) to the rest-frame BC03 models. Then, we carry out an error-weighted $\chi^2$ test to find the best fitting between the above models and the observed galaxy photometry. The aim of this step is not deriving physical parameters from the best fitting, but setting constraints on the shape of the continuum.

\item Secondly, we re-normalize the BC03 spectroscopic model associated to the best-fitting photo-spectrum. The synthetic photometry of this re-normalized model has to exactly reproduce all the observed photometric bands.

\item Finally, we apply Eq.~\ref{eq:ext} on the re-normalized model derived in the previous step, in order to obtain a de-reddened spectrum that we convolve with its related filter response-curve. We use the \citet{Fitzpatrick1999} extinction laws to calculate $A_\lambda$, the value $E(B-V)$ provided by SFD and $R_V=3.1$, to de-redden all the galaxies of our sample.

\end{itemize}

In particular, the \citet{Fitzpatrick1999} extinction law was built from the superposition of the extinction curves derived for a set of stars. Consequently, this extinction law contains intrinsic uncertainties, although we would accurately know the values of $R_V$ and $E(B-V)$. We account for the particular uncertainties of this law adding an error to the de-reddened photometry of MW dust, $F_{\lambda,{\rm c}}$, following the methodology explained in \citet{Fitzpatrick1999} and assuming $\sigma_{R_V} = 0$.

Cosmological fields, often the targets of multi-filter photometric surveys, use to be regions of the sky with low extinction values. In the particular case of ALHAMBRA, our main galaxy sample has MW extinction values of $E(B-V)$ down to $0.04$ ($A_V < 0.12$ for $R_V = 3.1$) in all the cases. The colour term due to the MW dust in the ALHAMBRA survey may reach a maximum of $\Delta m_{\rm AB} \sim 0.15$, and the stellar masses may be underestimated by $3$\% ($8$\%) if we use the \textit{K}$_s$ (\textit{R}) filter to estimate it. Although the stellar mass is not primarily affected by MW extinction in these fields, the colour term might change the retrieved stellar populations and consequently the derived stellar mass (see Sect.~\ref{sec:mass}). In ALHAMBRA there are no galaxies at low Galactic latitudes, $\mid b\mid \ < 5$, where the MW temperature structures are not duly resolved in the SFD maps \citep{Schlegel1998}.

\subsubsection{Redshifts}\label{sec:redsh}

The code is generically prepared to provide, together with the mass and the stellar population parameters of the galaxy, an estimation of the photo-\textit{z}. It is worth noting however that this code is not intended to be a photo-\textit{z} code. Due to the large number of potential model parameters that the code plays with, when the redshift is set as a completely free parameter in the fitting process there exists a slight degeneracy with other parameters (like extinction; see Sect.~\ref{sec:photoz_uncer}) that tends to overestimate the derived photo-\textit{z}. To overcome this effect, the code is also prepared to accept as initial constraint a list of redshift values for each target galaxy: either a list of nominal redshift values, hence the code only performs the fitting process at exactly these redshifts, or a complete probability distribution functions (PDF) of redshifts, and then the code only accounts for the model redshifts within the PDF interval. Because of the good results we obtain, throughout this work we use as input redshift constraints the photo-\textit{z} PDFs provided by the ALHAMBRA Gold catalogue using \textit{BPZ2.0} \citep{Molino2014}. It is noteworthy that the combination of our code with the ALHAMBRA photo-\textit{z} constraints further improves the quality of the input photo-\textit{z} alone (see Sect.~\ref{sec:photoz}).


\subsection{The core of the \textit{MUFFIT} analysis techniques}\label{sec:analysis}
This section is devoted to the main technical features and processes carried out by our code in order to constrain the stellar population parameters of galaxies in multi-filter data samples. We first describe in Sect.~\ref{sec:sspmixing} the way in which the $\chi^2$ minimization is performed, with the addition of a mixture of SSPs giving a remarkable improvement, specifically computed for each galaxy, in order to set more precise constrains in the stellar populations. In Sect.~\ref{sec:elines} we explain in detail the process to detect those bands that may be affected by strong emission lines, helping to understand the overall fitting process. Section~\ref{sec:mass} details how the stellar masses are calculated from the fittings. In addition, a Monte Carlo approach is performed to set constraints on the confidence intervals of the parameters provided by the code, detailed in Sect.~\ref{sec:mc}. Finally, we describe how we manage the \textit{k}-corrections as result of the fittings in Sect.~\ref{sec:kcorr}. The content of this section is outlined on the right-hand side of the flowchart (see Fig.~\ref{fig:techniques}).


\subsubsection{The $\chi^2$ minimization and mixture of SSPs}\label{sec:sspmixing}
Our stellar population analysis technique is based on error-weighted $\chi^2$--tests between the multi-filter galaxy data and the template SSP models of choice. Since SSP models are usually normalized to a initial stellar mass and both the galaxy distance and its luminosity are uncertain in a general case (in fact these are parameters generally derived from the fit), it is required to add a normalization term, $\varepsilon$, in the classical $\chi^2$ equation. This term minimizes the $\chi^2$ value for every model--galaxy pair, being the result only colour dependent. Our normalization way takes into account all observed bands and associated errors, being more robust for multi-filter surveys as, at most, they only contain a few dozens of filters (e.~g. $23$ in ALHAMBRA). This way, all the meaningful filters contribute to determine the best solution of the fitting (without giving up one of the best bands in order to normalise), and there is no risk that the normalization band is affected by emission lines or cosmetic defects.

Due to the fact that the number of reliable bands in each source may be different from one object to the other (for some objects, some filters may be rejected due to observational, cosmetic or calibration issues), in general we divide every $\chi^2$ by the number of available, safe filters in each case. Depending on whether we are working with bandpass fluxes or magnitudes, the $\chi^2$ definition can be expressed as

\begin{equation}
\chi^2_\mathrm{m} = \frac{1}{N_{\rm p}}\sum\limits_{X=1}^{N_{\rm p}}{\left[\frac{O^\mathrm{m}_X-\left(\varepsilon_\mathrm{m} + m_X\right)}{\sigma^\mathrm{m}_X}\right]^2} \ , {\rm and}
\label{eq:chi2_m}
\end{equation}
\begin{equation}
\chi^2_\mathrm{f} = \frac{1}{N_{\rm p}}\sum\limits_{X=1}^{N_{\rm p}}{\left[\frac{O^\mathrm{f}_X-\varepsilon_\mathrm{f} f_X}{\sigma^\mathrm{f}_X}\right]^2} \ ,
\label{eq:chi2_f}
\end{equation}
where $N_{\rm p}$ is the number of available filters in an observed galaxy, $O^\mathrm{m,f}_{X}$ is the observed $X$-filter (magnitude or flux), $\sigma^\mathrm{m,f}_X$ its error, $m_X$ ($f_X$) is the {\it X}-filter model prediction (single SSP or SSP mixture, more details below) and $\varepsilon_\mathrm{m}$ ($\varepsilon_\mathrm{f}$) is the normalization term. For our purposes, $\varepsilon_\mathrm{m}$ and $\varepsilon_\mathrm{f}$ are written as
\begin{equation}
\varepsilon_\mathrm{m} = \left(\sum\limits_{X=1}^{N_{\rm p}}{\frac{O^\mathrm{m}_X-m_X}{{\sigma^\mathrm{m}_X}^2}}\right) \times \left(\sum\limits_{X=1}^{N_{\rm p}}{\frac{1}{{\sigma^\mathrm{m}_X}^2}}\right)^{-1} \ , {\rm and}
\label{eq:offset}
\end{equation}
\begin{equation}
\varepsilon_\mathrm{f} = \left(\sum\limits_{X=1}^{N_{\rm p}}{\frac{O^\mathrm{f}_X f_X}{{\sigma^\mathrm{f}_X}^2}}\right) \times \left(\sum\limits_{X=1}^{N_{\rm p}}{\frac{f_X^2}{{\sigma^\mathrm{f}_X}^2}}\right)^{-1} \ ,
\label{eq:norm}
\end{equation}
which correspond respectively to minimizing Eqs.~\ref{eq:chi2_m} and \ref{eq:chi2_f} for each galaxy, i.~e., $\partial \chi^2_\mathrm{m,f} / \partial \varepsilon_\mathrm{m,f} = 0$. As we will show later (Sect.~\ref{sec:mass}), by finding the best stellar population solutions for each galaxy, we can estimate its stellar mass from the $\varepsilon$ values. 

Note that Eq.~\ref{eq:chi2_m} (the equation used throughout this work) is assuming that the distribution of errors is Gaussian, when in general the distribution of magnitudes is not Gaussian since these are logarithmic measurements of flux. From certain signal-to-noise ratios, S/N$ \ga 5$ (or uncertainties $\sigma^m_X \la 0.22$), the magnitude uncertainties are quasi-normally distributed, being this approach valid. 
Consequently, we encourage potential \textit{MUFFIT} users to take fluxes instead of magnitudes when several galaxy bands are compromised by very low signal-to-noise ratios, S/N $\la 4$--$5$. It must be also taken into account that a certain minimum signal-to-noise ratio is required to determine reliable stellar population parameters without being dominated by degeneracies, as it will be shown later on in this paper.

Once we have defined how to compute the fitting goodness, the next step is to compare our set of models to retrieve the most likely stellar population parameters. We carry out this process in two different steps. 

\begin{itemize}

\item First, we run the $\chi^2$-test described above with the set of SSP models selected by the user (base models), making a first  determination of the bands that may be affected by strong emission lines. 
In short, for every redshift step of the SSP models, the code looks for a flux excess in the galaxy SED with respect to the SSP model SEDs, for all those bands that could be affected by emission lines at the given redshift. A more extensive explanation on our technique of detection of emissions lines in multi-filter galaxy data is presented in Sect.~\ref{sec:elines}. When this is the case, those bands potentially affected by emission lines are removed from the fitting process, and the $\chi^2$-test is repeated again without the affected bands. In addition, rather than taking the parameters of the best SSP fitting, we carry out a Monte Carlo simulation using the proper signal-to-noise ratios in each filter (further details in Sect.~\ref{sec:mc}). From the set of parameters retrieved during the Monte Carlo approach, we map the parameter space of compatible solutions (overall age, metallicity, extinction, redshift, stellar mass and IMF), although at this stage we only focus on the retrieved distributions of age and redshift to carry out the next step: the mixture of two SSPs and its sub-sequent SED-fitting process.

\begin{figure}
\resizebox{\hsize}{!}{\includegraphics{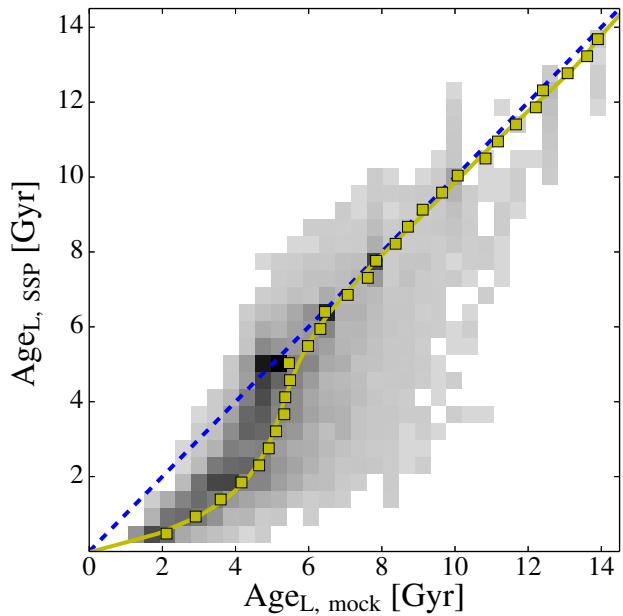}}
\caption{Empirical relation between the luminosity-weighted ages of mock galaxies made of random mixtures of two SSPs, Age$_\mathrm{L,\ mock}$, and the best age determination for such mock galaxies derived from a single SSP fitting, Age$_\mathrm{SSP}$. The yellow curve is the Age$_\mathrm{L,\ mock}$ median for a given value of Age$_\mathrm{SSP}$, and represents the typical offset in age that one may expect when interpreting the SED of a mixture of two SSPs by fitting a unique SSP.}
\label{fig:ssp_age}
\end{figure}

\item Secondly, according to the age and redshift distributions derived from the initial SSP analysis, we make a new database of models consisting of a mixture of two individual base SSP models. The mixture is computed only for the best redshift solutions determined in the previous step. For each redshift value, the two-model mixture is constrained to combine two SSPs, respectively younger and older than a certain age threshold, age$_\textrm{T}$, that is related with the most likely age, age$_\textrm{SSP}$, inferred from the Monte Carlo analysis performed in the previous step. This is a reasonable assumption given that the stellar content of galaxies are usually the result of complex SFHs with multiple stellar populations \citep{Ferreras2000,Kaviraj2007,Lonoce2014}, and the age solutions derived from comparisons with single SSPs can be considered, in first order, luminosity-weighted means of the ages of the individual, true populations. To determine the age$_\textrm{T}$ value that allows us to define the limit between \textit{younger} and \textit{older} SSP mixtures for each galaxy, we have studied the empirical relation between the luminosity-weighted ages of mock galaxies made of random mixtures of two SSPs, age$_\textrm{L,\ mock}$, and the best age determination for such mock galaxies derived from a single SSP fitting, age$_\textrm{SSP}$. In Fig.~\ref{fig:ssp_age} we present the result of this study. As expected, we observe that age$_\textrm{SSP}$ underestimates the \textit{real} age, in particular for age$_\textrm{L,\ mock} \la 6$~Gyr. The yellow curve in Fig.~\ref{fig:ssp_age} represents age$_\textrm{T}$ as a function of age$_\textrm{SSP}$. Once the age threshold is established, we generate all the possible SSP combinations (younger and older than age$_\textrm{T}$), including as a new degree of freedom the stellar mass weight of each component. For a general case with $n$ components per mixture, each magnitude in the band $X$ of the new mixture model is expressed as
\begin{equation}
m_{X,\mathrm{mix}} = -2.5 \log_{10} \left( \sum\limits_{i=1}^{n}{\alpha_i 10^{-0.4\ m_{X}^{i}}} \right) \ ,
\label{eq:mix_m}
\end{equation}
\begin{equation}
f_{X,\mathrm{mix}} = \sum\limits_{i=1}^{n}{\alpha_i f_{X}^{i}} \ ,
\label{eq:mix_f}
\end{equation}
where $m_{X}^{i}$ ($f_{X}^{i}$) is the magnitude (flux) in the band \textit{X} for the \textit{i}-th SSP model and $\alpha_i$ is the relative flux-contribution of the SSP model in the \textit{i}-th component, with $\sum\limits_{i=1}^{n}{\alpha_i}=1$ and $0 \leq \alpha_i \leq 1$. Note that in our case, we are mixing two SSPs and  consequently $n=2$.

After mixing the SSP models as explained above, the code searches again the best fitting solution, repeating the detection of emission lines with the mixture of models as explained in Sect.~\ref{sec:elines}. As in the first step using a single SSP, we do not only provide the best solution but map the compatible stellar-population parameters by a Monte Carlo approach, treating properly the errors in each band. This provides an extra advantage when carrying out a statistical treatment of the results. We devote Sect.~\ref{sec:mc} to explain in detail how we explore the compatible space of derived parameters for each galaxy.

\end{itemize}

Notice that, with this method and two SSPs, one database of mixed SSPs is particularly created for each galaxy, being more adequate and realistic than a single SSP fitting. As shown above, for a non-parametric SFH this represents a substantial improvement with respect to using one SSP only, which is not able to reproduce the colour of an underlying main \textit{red} population plus less massive and later events of star formation. The mixture of two populations is a reasonable compromise, to improve significantly the reliability in the determination of the stellar population parameters of multi-filter galaxy data \citep{Ferreras2000,Kaviraj2007,Lonoce2014}. 
In fact, it has been demonstrated \citep[e.~g.][]{Rogers2010} that the mixture of $2$ SSPs turns out to be the most reliable approach to describe the stellar populations of young early-type galaxies, as well as a very reasonable approach for older galaxies, in this latter case only slightly surpassed by the use of chemically enriched exponential models. So, the two SSP model fitting approach may be considered in general, as a reasonable method for analysing the stellar populations of most kind of galaxies in a consistent way. Moreover, given that \textit{MUFFIT} does not impose constraints on the metallicities of the $2$ SSP mixture, this can provide hints not only for age evolution but also for a metallicity build-up. That being said, future versions of \textit{MUFFIT} will also account for the use of different sets of SSP or $\tau$-models for the best choice of the user.


\subsubsection{Emission lines}\label{sec:elines}
Nebular emission lines appear frequently in the SEDs of galaxies, even if these are  dominated by the light contribution of their stellar content. In particular, dealing with multi-filter galaxy data, filters affected by emission lines may present a substantial excess in flux with respect to any combination of SSP models, as the later typically do not account for the nebular emission physics. To guarantee the accuracy and reliability of the stellar population parameters derived during the fitting process, it is crucial to detect and remove those bands that can be significantly affected by emission. Not only due to the fact that they are not comparable to SSP models, but also since filters contaminated by strong emission lines tend to exhibit much large luminosities, hence lower photonic errors, than the rest of bands, and dominate our error-weighted SED-fitting techniques (see Eqs.~\ref{eq:chi2_m} and~\ref{eq:chi2_f}). On the other hand, it is worth reminding that the presence of strong emission lines may also provide a fruitful information, since they contribute to the restriction of the feasible redshift intervals of the galaxy. The redshift constraints due to nebular emissions are additionally considered during the analysis.

\begin{figure}
\resizebox{\hsize}{!}{\includegraphics{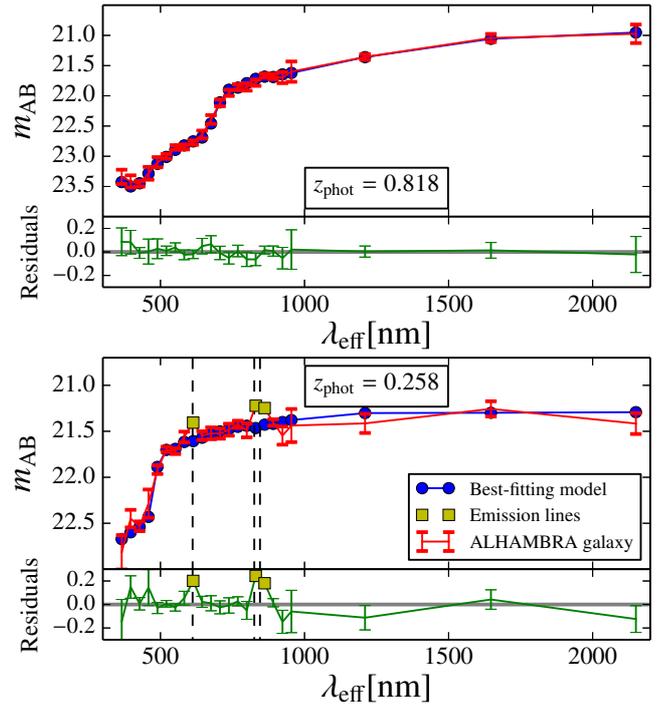}}
\caption{Spectral fitting examples for galaxies from the ALHAMBRA survey using \textit{MUFFIT} and the MIUSCAT SSP models. The galaxy photo-spectra and their errors are given in red, whereas the best-fitting models are given in blue. Top panel corresponds to a quiescent galaxy for which no emission lines are detected. Bottom panel illustrates the case for a star-forming galaxy for which \textit{MUFFIT} detects three bands affected by emission lines, in yellow. The dashed black lines indicate the wavelengths for \ion{H}{$\beta$}, \ion{H}{$\alpha$}, and [\ion{S}{II}]. Photometric redshifts are given at the labels inside.}
\label{fig:eline}
\end{figure}

The emission line detection process of our code is dependent on the specific photometric system of the galaxy sample, as it only accounts for those emission lines that are typically strong enough to affect the photometry of the given filter set. It is obvious that the broader the spectral filter width, the larger the equivalent width (hereafter EW) of the line that may be potentially detected at a fixed signal-to-noise. In this sense, the code is initially fed with a list of target emission lines that depends particularly on the filter set, customizable by the user, with emission lines such as [\ion{O}{II}]\ $\lambda\lambda 3726,3729$, [\ion{O}{III}]\ $\lambda\lambda 4959,5007$, \ion{H}{} Balmer's series, [\ion{S}{II}]\ $\lambda\lambda 6717,6731$, etc. Due to the design of \textit{MUFFIT}, we can also provide a list of typical AGN emission lines to reduce their effects in the fittings, but broad AGN lines may affect $2$ or more ALHAMBRA filters, hence the AGN emission line detection criteria in \textit{MUFFIT} might fail or be inaccurate. It is very important to note that an excessive list of emission lines, mostly when they are spread in the large wavelength ranges, will eventually derive wrong results, as some bands may be removed in excess. It is therefore advisable to restrict this list to the lines that can present a measurable excess in flux, which mainly depends on both the filter widths and the line intensity. For this reason, some bands can be forced by the user to remain in the SED-fitting analysis irrespective of whether they can be potentially affected by emission lines. For instance, in ALHAMBRA we do not expect to be sensitive to any flux excess in the NIR bands due to the presence of emission lines, so they are never removed during the fitting process even if the code detects a flux excess in any of them.

Once we specify in the code the emission line list, the emission line detection process is carried out in two steps. First, taking into account the model redshift, we fit our models (single SSP or SSP mixture) without all the bands that could be potentially affected by the specified emission lines, and explore the residuals of the best fitting. If the residuals of any of the potentially affected bands present an excess in flux/magnitude larger than a limit value provided by the code user, $\Delta m_\mathrm{EL}$, and these residuals are deviated beyond a band-error factor, $\sigma_\mathrm{EL}$, the bands are considered to be affected by emission lines and are removed from the fitting process. Both constraints $\Delta m_\mathrm{EL}$ and $\sigma_\mathrm{EL}$ are required, the latter to assure that the excess in flux is not due to photometric uncertainties, and the first one to avoid removing those bands with tiny observational errors that present little discrepancies with respect to the models. Finally, we repeat the fitting without the bands identified in the previous step, getting a new set of reliable $\chi^2$ values, clean from nebular contributions.

For the ALHAMBRA case, we use $\Delta m_\mathrm{EL} = 0.1$, as for lower contributions the affected bands do not affect significantly the SED-fitting results retrieved with \textit{MUFFIT}. In addition, we set $\sigma_\mathrm{EL} = 2.5$ as a reasonable statistical threshold to detect emission lines over the noise. Figure~\ref{fig:eline} illustrates two SED-fitting examples of two galaxies from ALHAMBRA. The top panel of Fig.~\ref{fig:eline} illustrates a fitting of a quiescent galaxy in ALHAMBRA without strong nebular emissions, whereas the bottom panel shows a galaxy for which \textit{MUFFIT} detects that some bands may be affected by emission lines (yellow squares). The red curves represent the observed photo-spectra, while the blue curve is the best-fitting model after the detection of emission lines process. The yellow squares are the bands where the influence of an emission line is ticked, in this particular case \ion{H}{$\beta$}, \ion{H}{$\alpha$}+[\ion{N}{II}], and [\ion{S}{II}]. The dashed black lines point out the wavelengths for \ion{H}{$\alpha$}, \ion{H}{$\beta$}, and [\ion{S}{II}] at the galaxy photo-\textit{z}. For this case, the detection of the emission lines particularly contribute to strongly constrain the redshift range. Despite the ALHAMBRA resolution (FWHM $\sim300$~\AA), we note that strong emission lines can modify the fitting results. In some cases, even \ion{H}{$\beta$} shows significant contributions in the ALHAMBRA data set.

Since we are providing both those bands that may be affected by strong emission lines and the residuals from the SED-fittings, we can easily estimate the flux excesses in order to posteriorly transform them to equivalent widths. The advantage of this method is that our best SED-fittings are mixtures of SSPs that already include the corresponding stellar absorptions, hence the residuals can be directly related to the absolute nebular emission. The main limitation, in general, comes from the low resolution of the data, as in many cases some filters can be affected by more than one emission line, like e.g. \ion{H}{$\alpha$} and [\ion{N}{II}]. Still, as it will be presented in Section~\ref{sec:ew}, this technique opens new paths for future work on emission-line galaxies with multi-filter data.


\subsubsection{Stellar masses}\label{sec:mass}

As we explain in Section~\ref{sec:sspmixing}, both the normalization term $\varepsilon$ introduced in the $\chi^2$ minimization equation and the intrinsic luminosities of the two best fitted SSPs are directly related with the total stellar mass of the galaxy. SSP models are usually normalized to an initial stellar mass of $1\mathrm{M_\sun}$, but this decreases with time accounting for the evolution of the most massive  stars. This effect is properly taken into account for determining the final galaxy mass by applying a correction term to each SSP, $\kappa_{\rm SSP}$, which is usually provided by the models.

Taking into account the above considerations, the total stellar mass, M$_{\star,{\rm T}}$, of a mixture of \textit{n} SSPs (for this work $n=2$) can be expressed as
\begin{align}
M_{\star,{\rm T}} = \sum\limits_{i=1}^{n}{M_{\star,\ i}} & = 10 ^ {-0.4\ \varepsilon_\mathrm{m}} \times 4\pi d_{\rm  L}^2 \times \sum\limits_{i=1}^{n}{\kappa_{i,\ {\rm SSP}}\ \alpha_i} \nonumber \ , \\ 
                                                         & = \varepsilon_\mathrm{f} \times 4\pi d_{\rm  L}^2 \times \sum\limits_{i=1}^{n}{\kappa_{i,\ {\rm SSP}}\ \alpha_i} \ ,
\label{eq:mass}
\end{align}
where ${M_{\star,\ i}}$ is the stellar mass of each population in the mixture, $\varepsilon_\mathrm{m}$ is the normalization term defined in Eq.~\ref{eq:norm}, $d_{\rm  L}$ is the luminosity distance in cm units \citep[see ][]{Hogg2000}, $\kappa_{i,\ {\rm SSP}}$ is the relative stellar mass correction for the \textit{i}-th component in the SSP mixture, and $\alpha_i$ is the relative flux-contribution of the SSP model in the \textit{i}-th component (see Eqs.~\ref{eq:mix_m} and ~\ref{eq:mix_f}). Throughout this work, the derived stellar masses are quoted including stellar remnants through $\kappa_{i,\ {\rm SSP}}$, but for a more general case this parameter may not include remnants.


\subsubsection{Stellar population parameters of the SSP mixture}\label{sec:stellar_par}
The stellar-population parameters of the mixture of SSPs are estimated from the parameters of each component in the mixture. This can be done in different ways, which mainly depend of the weights assigned to the parameters of the different components. The most common definitions, provided by \textit{MUFFIT} and employed in this paper, are luminosity-weighted and mass-weighted. The latter provides a more realistic information since it accounts for the total mass of stars in each population, hence assigning larger weights to the more abundant or dominant stellar populations. However, these populations may have very different luminosities. In this sense, luminosity-weighted parameters are more representative of the populations that dominate the observed spectrum, since the galaxy SEDs are predominantly leaded by the brighter populations, even if they are not dominant in relative mass. 

Throughout this work, the luminosity-weighted and mass-weighted stellar population parameters of a mixture of \textit{n} SSPs (for this work $n=2$), $p_\mathrm{L}$ and $p_\mathrm{M}$ respectively, are defined from the stellar population parameters of each \textit{i}-th component ($p_i$; age, metallicity, extinction, IMF slope, or $[\alpha/\mathrm{Fe}]$) as
\begin{equation}
p_\mathrm{L} = \frac{\sum\limits_{i=1}^{n}{\alpha_i \times L_i \times p_i}} {\sum\limits_{i=1}^{n}{\alpha_i \times L_i}} \ ,
\label{eq:pL}
\end{equation}
\begin{equation}
p_\mathrm{M} = \sum\limits_{i=1}^{n}\frac{M_{\star,\ i}}{M_{\star,{\rm T}}} p_i = \frac{\sum\limits_{i=1}^{n}{\alpha_i \times \kappa_{i,\ {\rm SSP}} \times p_i}}{\sum\limits_{i=1}^{n}{\alpha_i \times \kappa_{i,\ {\rm SSP}}}} \ ,
\label{eq:pM}
\end{equation}
where $\alpha_i$ is the relative flux-contribution of the SSP model in the \textit{i}-th component, $\kappa_{i,\ {\rm SSP}}$ is the relative stellar mass correction for the \textit{i}-th component in the SSP mixture, and $L_i$ is the luminosity of the SSP model in the observed spectral range. Note that both definitions agree when the parameter value is the same in each component. 


\subsubsection{Determining the space of best solutions}\label{sec:mc}

Because of the well known degeneracies among stellar population parameters, it is essential to perform a reliable analysis of the possible solutions (as mixtures of two SSPs) for each galaxy according to the uncertainties of the data. For this reason, rather than providing only the best fitting solution for each galaxy (it is well known that the most likely parameters are not always the best-fitting model parameters), our code accounts for the photometric errors of the multi-filter galaxy data to provide a set of the best fitting solutions, hence providing a set of probable values of redshifts, stellar masses, extinctions, and stellar population parameters (ages, metallicities and IMFs) for each object. These values can be ultimately averaged according to their weights and frequencies to derive the average final parameters assigned to each galaxy and their errors. In this section we explain the processes and applied criterion to carry out this analysis.

The determination of the best solutions space is based on a Monte Carlo method that, using the proper signal-to-noise ratio of each filter, seeks to obtain which parameter values are compatible within the photometric errors of the data. Since photometric uncertainties usually follow a normal distribution (or Gaussian), we assume an independent Gaussian distribution in each filter, centred in the band flux/magnitude, and with a standard deviation equal to its photometric error. It is worth noting that each filter is observed and calibrated independently from the remaining ones, so the errors of different filters are not expected to be correlated. 

For each galaxy, on the basis of the above Gaussian error distributions for its multi-filter data, \textit{MUFFIT} generates Monte Carlo simulations (the number of realizations is defined by the user), ending up with a set of multi-filter data realizations for the same galaxy, all of them compatible within the errors. Ideally, the next step would be to run the $\chi^2$ test individually for each realization of the galaxy using the complete set of models, but this is extremely time-consuming as the code plays with million of models (for the present research: $21$ ages, $5$ metallicities, $18$ extinctions, $1$ IMFs, $300$ redshifts, and solar $[\alpha/\mathrm{Fe}]$) for each fitting. Instead, to speed up this computational process, for each galaxy we perform a preliminary selection of SSP and mixture models that can play an important role in the fitting given the specific SED and errors of the galaxy. This pre-selection of models is carried out as it follows:

\begin{itemize}

\item i) After having run our code for a certain galaxy SED, and having obtained the $\chi^2$ values for all the possible mixture of two SSP models ($\chi^2_\mathrm{mix}$), we take the $\chi^2_\mathrm{mix}$ value of the best fitting model (hereafter BFM), $\chi^2_\mathrm{BFM}$, i.~e. the mixture of two SSPs with the greatest probability to be the solution, which corresponds to the lowest $\chi^2_\mathrm{mix}$ value.

\item ii) Since the parameter space of best solutions depends not only on the filter photometric uncertainties but also on the shape of the SED, the next step is to determine, for each individual galaxy SED, the range of plausible $\chi^2$ values that are expected according to the set of photometric uncertainties. To do this, \textit{MUFFIT} performs $10\,000$ Monte Carlo realizations of the BFM bands according to the Gaussian error distributions of the real galaxy multi-filter data. The corresponding $10\,000$ $\chi^2$ values between these realizations and the BFM, namely $\chi^2_\mathrm{M}$, represent the range of $\chi^2$ values that one would expect just due to the photometric uncertainties of the real galaxy data. Note that this range can be very different among different galaxies. In \textit{MUFFIT}, the limiting plausible-value, $\chi^2_\mathrm{phot}$, is set to the value that encloses the $68.27$\% (a Gaussian $1\sigma$) of the cumulative distribution function of the $\chi^2_\mathrm{M}$ values. 

\item iii) Finally, the sub-sample of possible solutions for a given galaxy SED is constituted by those ones that fulfil the criterion $\chi^2_\mathrm{mix} \leq \chi^2_\mathrm{BFM}+\chi^2_\mathrm{phot}$. This sub-sample is consequently restricted to those models whose colours are statistically compatible within the galaxy photon-errors.

\end{itemize}

This way, the set of compatible best solutions for each galaxy is determined by generating $N_\mathrm{m}$ Monte Carlo realizations of the galaxy SED data (throughout this work $N_\mathrm{m}=100$) according to their errors, and then running our $\chi^2$ minimization test for each galaxy realization using as input the sub-sample of preselected models. In each realization, we get a new BFM whose parameters are ultimately weighted ($\omega_j$) with its $\chi^2$ value to provide the most likely stellar population parameters together with their errors. Formally,
\begin{equation}
\omega_j = \frac{{1/\chi^2_{\mathrm{MC},j}}} {\sum\limits_{j=1}^{N_\mathrm{m}} {1/\chi^2_{\mathrm{MC},j}}},
\label{eq:wMC}
\end{equation}
\begin{equation}
<p> = \sum\limits_{j=1}^{N_\mathrm{m}} {\omega_j p_j}
\label{eq:mc}
\end{equation}
\begin{equation}
\sigma_{<p>} = \sqrt{\sum\limits_{j=1}^{N_\mathrm{m}}{\omega_j \left(<p> -\ p_j\right)}^2},
\label{eq:mc_err}
\end{equation}
where $<p>$ and $\sigma_{<p>}$ are, respectively, the average stellar population parameters (age, metallicity, extinction, redshift, stellar mass, IMF, and $[\mathrm{\alpha/Fe}]$, in a general case) and their errors, and $p_j$ are the stellar population parameters associated to each BFM in the Monte Carlo realization with a $\chi^2$ value equal to $\chi^2_\mathrm{MC,j}$.

In addition, the essential stellar parameters of each BFM obtained in the $N_\mathrm{m}$ Monte Carlo iterations are also provided.

Finally, we remark that the uncertainties of the parameters retrieved in this stage comprise not only the main parameters of the models, like ages, metallicities and IMFs, but also the extinction, the redshift (if it is the case, within the interval provided by an external photo-\textit{z} code) and the stellar mass.

\subsubsection{K-corrected luminosities}\label{sec:kcorr}

Once we have computed the best fitting models, we end up with a combination of SSP models that reproduce the colours of the galaxy photometric SED. Hence, the luminosity of the galaxy, its absolute magnitudes at any band, and the mass-luminosity relation is estimated from exactly the same combination of SSP models taken at rest-frame. Note that, independently of the physical parameters linked to the best combination of models, the \textit{k}-correction is model-independent since it properly reproduces the colours of a galaxy SED at a given cosmological distance, as long as the redshift is well constrained. If we compute the magnitudes for the different bands following this method, the main parameter that determines the \textit{k}-correction goodness of fitting is the photo-\textit{z} accuracy. Since the set of SSP models does not contain emission lines templates, and our code removes them automatically during the fitting process, the provided \textit{k}-corrections/luminosities only contains rest-frame predictions about the stellar continuum, not about the nebular content.

To determine rest-frame magnitudes with the corresponding errors, for each galaxy we take all the best-fitting models recovered in the Monte Carlo approach (see Sect.~\ref{sec:mc}), average them and provide the average rest-frame magnitudes/colours and their standard deviations, hence  considering the uncertainties in the photometry thanks to the Monte Carlo approach. It is noteworthy that, at low redshifts, the uncertainties of rest-frame magnitudes may be very high, as the apparent magnitudes depend of the luminosity distance ($\propto d_{\rm L}^{-2}$), which diverges at $z=0$. This suggests that, to have better \textit{k}-corrections in the most local Universe, more accurate photo-\textit{z} are necessary.  Despite this, the colour terms among different filters are not so affected by this effect, as the major impact is in the source luminosity and not in the rest-frame colours. To minimize this effect, we provide a second \textit{k}-correction, for which we study the variability of the colours respect to an anchor band. In short, once we have all the rest-frame models recovered during the Monte Carlo method, the anchor band is the one that presents the lowest variability at rest-frame. In ALHAMBRA, this anchor band is usually a band in the red optical part (higher signal-to-noise ratios). This approach turns out to be very useful, e.g. in order to make reliable colour-magnitude diagrams (CMD) at low redshift.


\section{Intrinsic uncertainties and degeneracies with ALHAMBRA galaxy data}\label{sec:feasibility}

After having presented the main technical aspects of our SED fitting code in Section~\ref{sec:bigcode}, and before presenting a comparison study between our stellar population results and similar previous data from the literature (see Sect.~\ref{sec:testing}), the goal of this section is to study the accuracy and reliability of the stellar population parameters retrieved with our code. Since this strongly depends on the photometric system of the data under study, it is important to note that, along this section, all the tests and predictions about uncertainties, degeneracies, etc. are particularly performed for the ALHAMBRA filter system.

Since the code presented in this paper is particularly suited for the study of the stellar populations of galaxies whose SEDs are dominated by their stellar content, we begin with Sect.~\ref{sec:rs} to build the CMD of the ALHAMBRA galaxy data, allowing us to make a proper selection of our target of galaxies and to compare our results with those published in the literature (Section~\ref{sec:testing}). In Sect.~\ref{sec:uncertainties} we check how the intrinsic uncertainties in the photometry of the ALHAMBRA filter system affect the typical errors of the derived parameters, using a set of mock galaxies with well known input parameters. Furthermore, the impact that the uncertainties of the input ALHAMBRA photo-\textit{z}, Sect.~\ref{sec:photoz}, have on the derived stellar population parameters is analysed in Sect.~\ref{sec:photoz_uncer}. Finally, we quantify the expected degeneracies among the derived galactic parameters of typical red-sequence galaxies, for the ALHAMBRA photometric system and different signal-to-noise ratios.


\subsection{Selection criteria of ALHAMBRA red sequence galaxies}\label{sec:rs}

It is well known that the CMD of galaxies exhibits a bimodal distribution with two main populations, usually referred as the "red sequence" (hereafter RS) and the "blue cloud" \citep{Bell2004, Baldry2004, Faber2007, Fritz2014}. A great fraction of RS galaxies is mainly composed of early-types \citep{Strateva2001, Cassata2007}, but since the RS definition is clearly based on the observed galaxy colours, there is also a fraction of star forming dusty-galaxies that may lie on the RS \citep{Williams2009}. To break the degeneracies between quenched galaxies and dusty star-forming galaxies, there exist colour-colour diagnostics using NIR bands \citep{Williams2009,Arnouts2013}, and even methods to split the CMD into three populations ("red", "blue", and "green") by fitting to a set of SED type classes \citep{Fritz2014}. For the aims of this work, we just follow the \textit{classical} method of the CMD \citep{Bell2004, Faber2007, Fritz2014}. A more detailed study of the contamination of star-forming, reddened galaxies in the RS will be given in \citetext{D\'iaz-Garc\'ia et al. 2015, in prep.}.

To build the sample of RS galaxies, we firstly choose all the galaxies from the Gold catalogue\footnote{\url{http://cosmo.iaa.es/content/alhambra-gold-catalog}} with a statistical STAR/GALAXY discriminator parameter lower or equal to $0.5$ (\texttt{Stellar\_flag} $\le 0.5$), and imaged with 70\% photometric weight on the detection image (\texttt{PercW} $\ge 0.7$), to avoid photometric errors in the galaxies close to the image edges. Secondly, we apply our analysis techniques over the full sample of ALHAMBRA galaxies, using the set of MIUSCAT SSP models and the photo-\textit{z} predictions included in the Gold catalogue, to automatically get their \textit{k}-corrections (see Sect.~\ref{sec:kcorr}). From the \textit{k}-corrections and the stellar masses, we can easily estimate their absolute magnitudes, that together with the rest-frame colours compose the CMD. We note that our CMD does not change significantly if we use another set of models, e.~g. BC03, instead of MIUSCAT. In fact, this method is roughly model-independent as we are reproducing the luminosity and colours of the galaxy through the best mixture of two SSP models, irrespective of their parameters, hence the key here is to have a well-constrained photo-\textit{z} (see Sect.~\ref{sec:photoz_uncer}). 

\begin{figure*}
\resizebox{\hsize}{!}{\includegraphics{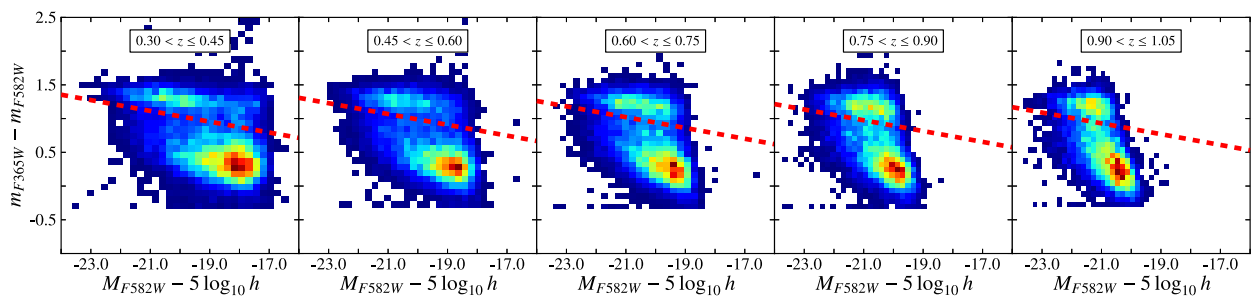}}
\caption{CMD of the ALHAMBRA galaxies at different redshift bins. The filters \textit{F365W} and \textit{F582W}, both in Vega magnitudes, are used as a proxy for the Johnson filters \textit{U} and \textit{V}. Redder and  bluer colours indicate, respectively, regions of the CMD with higher and lower galaxy densities. The dashed red line is the colour-magnitude relation that splits the RS from the blue-cloud for the mean redshift of the bin, as expressed in Eq.~\ref{eq:rs}.}
\label{fig:cmd}
\end{figure*}

The RS and the blue cloud appear clearly separated when the CMD is constructed using the Johnson-like filters \textit{U} and \textit{V} \citep{Johnson1953}. In our case, for simplicity we select the ALHAMBRA filters \textit{F365W} and \textit{F582W}, as these are the ones whose effective wavelengths are most similar to \textit{U} and \textit{V}, respectively. The CMD of the ALHAMBRA galaxies based on the \textit{F365W} and \textit{F582W} filters is presented in Fig.~\ref{fig:cmd}, where redder and bluer colours indicate higher and lower galaxy densities respectively. Following the equation provided in \citet{Bell2004}, which is compatible with the relation obtained in \citet{Fritz2014}, we define the RS as those galaxies redder than the following colour-magnitude relation:
\begin{equation}
m_{F365W}-m_{F582W} = 1.15 - 0.3 z - 0.08 (M_{F582W} - 5 \log h + 20), 
\label{eq:rs}
\end{equation}
where $m$ and $M$ indicate apparent and absolute magnitudes in the Vega system. By simply visual inspection, it is clear that Eq.~\ref{eq:rs}, illustrated in Fig~\ref{fig:cmd} as a red dashed line, splits properly the RS from the blue cloud, which already constitutes a first order check about the goodness of the SED-fitting.

\subsection{Photon-noise uncertainties}\label{sec:uncertainties}
To analyse the intrinsic uncertainties in the derived stellar population parameters of the galaxies due to the photon-noise errors of the ALHAMBRA photometry, we create mock galaxies consisting of a mixture of two random SSPs, in which we add photon noise according to the sensitivity of the ALHAMBRA filters and to the SED of the mock galaxies. Note that, by construction, this test is rather representative of the performance of RS galaxies. After adding noise, we run our code in order to derive the stellar population parameters of these mock galaxies, treating them as observed galaxies, but for which we know the real values of their parameters. The comparison between the input and the output parameters, as a function of the signal-to-noise ratio of the filters, allows us to conclude on the main topic of this section. 

We take the \emph{extended} version of the MIUSCAT models (see Sect.~\ref{sec:sspmodels}) with the Kroupa's IMF to develop these simulations. After adding different extinction values (Fitzpatrick's law, $A_{V}$ values ranging from $0.0$ to $0.8$ in steps of $0.1$) at different redshifts (from $0$ to $1$ and a step of $0.01$), we randomly mix two SSPs with a series of constraints: 

\begin{itemize}
\item The weight of the younger population shall not be larger than $30$\% in mass, and its age shall not be larger than $4$~Gyr. The mass limit is required to avoid too low luminosity-weighted ages, unlikely in RS galaxies \citep{Kaviraj2007}, and to guarantee the presence of old galaxies at all redshifts.
\item The age of the random SSPs cannot be much older than the age of the Universe at that redshift. Since the ages of SSP models are a discrete set of values, we state the limit to the first model that surpasses the age of the universe at each redshift.
\item The extinction of both SSPs is the same. Although this may not be necessarily the case in general, it is a reasonable assumption as we are studying integrated stellar populations, which translates to an average intrinsic extinction that affects the projected incoming light from different populations. 
\end{itemize}

To properly sample the galaxy mass range, we assign random stellar masses in the range $9.5 \le \log_{10} (M_\star/{\rm M}_\sun) \le 12.5$. We repeat this process $2\,000$ times per interval of redshift, from $0$ to $1$ in bins of $0.2$, getting $10\,000$ mock galaxies. As we explain below, we study the impact of the signal-to-noise ratio for three cases ($S/N = 10$, $20$, and $50$). In each case, we construct a new random sample of mock galaxies, being the total number of simulations $30\,000$.

\begin{figure*}
\resizebox{\hsize}{!}{\includegraphics{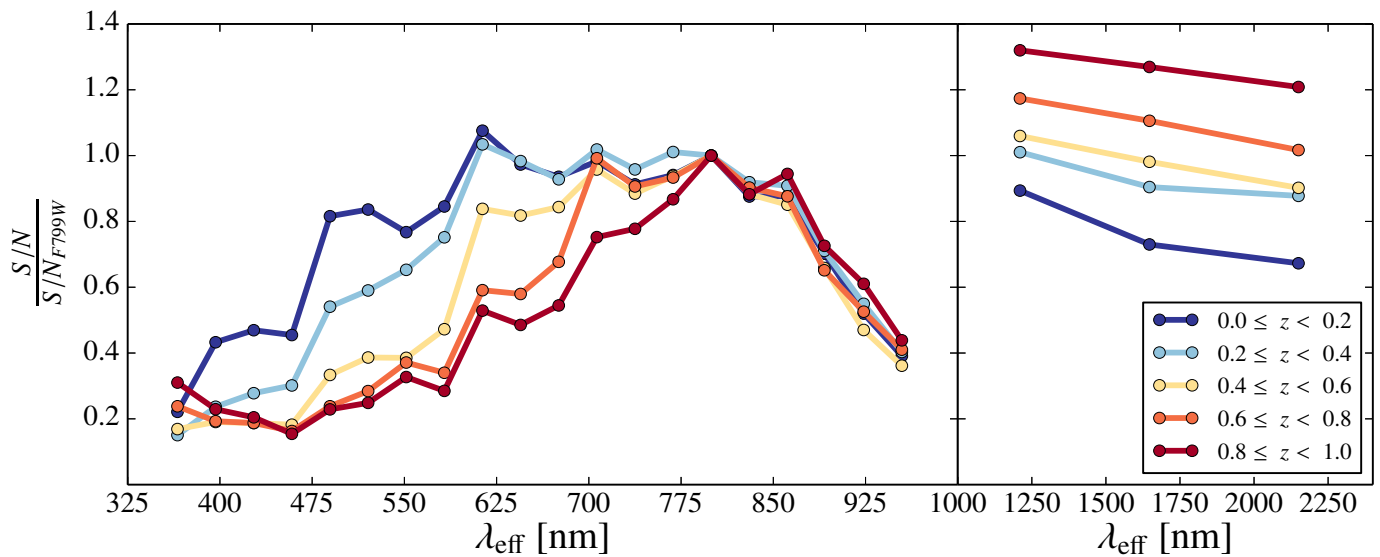}}
\caption{Typical signal-to-noise ratios per filter for real ALHAMBRA red sequence galaxies at different redshift bins. The signal-to-noise ratios are normalized to that of the $F799W$ filter. The observed data points account for the effective throughput of the telescope plus camera system, and for the average SEDs of red sequence galaxies.}
\label{fig:sn_curve}
\end{figure*}

After having built the set of toy mock galaxies, it is important to model accurately the way in which the galaxies are seen by the ALHAMBRA photometric system. It is in this point where the ALHAMBRA configuration plays an important role. The ALHAMBRA characteristics (see Fig.~\ref{fig:filters_alh} and Sect.~\ref{sec:synphot}) are such that the reddest bands are not so deep as the rest of the LAICA filters. On the other hand, the SED of typical RS galaxies, even the youngest ones, exhibit a clear flux drop in the blue region, with a prominent $4\,000$~\AA-break in the middle. Therefore, we cannot assume either that all the filters present the same signal-to-noise ratio or that the signal-to-noise ratio among filters does not depend on the redshift. To carry out a realistic simulation, we take all the galaxies from the ALHAMBRA Gold catalogue for which the best-fitting corresponds to a RS galaxy (see Sect.~\ref{sec:rs}), and compute, for every galaxy, the signal-to-noise ratios in each filter relative to the $F799W$ filter, which is on average the band with the maximum signal-to-noise ratio at any redshift. By repeating this process in different redshift bins, we determine how the signal-to-noise ratio changes along the SED as a function of the signal-to-noise in the anchor band $F799W$. These curves are shown in Fig.~\ref{fig:sn_curve}, and they account for the effective throughput of the telescope plus camera system, and for the average SEDs of RS galaxies. Note that the signal-to-noise ratios of the reddest filters up to $z=1$ are strongly affected by technical features of the survey (mainly the depth in these bands), whereas the bluest filters are also affected by the SED shape. From the curves in this figure, it is easy to see how the $4\,000$~\AA-break moves from blue to red wavelengths when the redshift increases. Interestingly, at larger redshifts, the signal-to-noise ratio of the bluest filters starts to grow indicating larger fluxes in these bands, probably due to the presence of young populations in the galaxy \citep{Ferreras2000}, which are easily observable at larger redshifts. Regarding the NIR filters, we check that typical RS galaxies become redder, on average, when they are observed at larger redshifts.

To study the impact of different signal-to-noise ratios on the derived stellar population parameters, we add noise to the mock galaxies, taking in each case the suitable signal-to-noise ratio curve depending of its redshift. We build three samples of $10\,000$ mock galaxies, and in each sample we force that the mean signal-to-noise ratio per mock photo-spectrum is $S/N = 10$, $20$, and $50$ respectively; that is, for a galaxy with $S/N = 20$ at redshift $0.5$, the mean signal-to-noise for the $23$ filters is $20$, but in the bluest filter $S/N_{F365W} \sim 6$ and in the anchor band (maximum) $S/N_{F799W} \sim 30$. The values of $S/N = 10$, $20$, and $50$ respectively correspond to median apparent magnitudes for the detection band of $m_{F814W} \sim 22.6$, $21.4$, and $19.8$ (ALHAMBRA RS galaxies and AB magnitudes). For the anchor band $F799W$, these values are almost identical. In ALHAMBRA, typical errors in the zero points due to calibration issues are $\sim 0.025$ (AB magnitudes), that correspond to a signal-to-noise ratio of $\sim50$. Furthermore, most ($\ga 80$\%) of our ALHAMBRA RS subsample has a mean signal-to-noise ratio larger than $10$, whereby these values ($S/N = 10$, $20$, and $50$) are suitable for our simulations.

Although for the mock galaxies we take models with $0.0 \le A_V \le 0.8$ and $0.0 \le z \le 1.0$, for the mock analysis we use SSP models with redshifts up to $1.2$ and extinctions up to $1.0$ to avoid border effects in the parameter estimation. Concerning the age estimation, we use the same constraint as in the mocks, i.~e. depending on the redshift, the oldest ages are not allowed.

\begin{figure*}
\resizebox{\hsize}{!}{\includegraphics{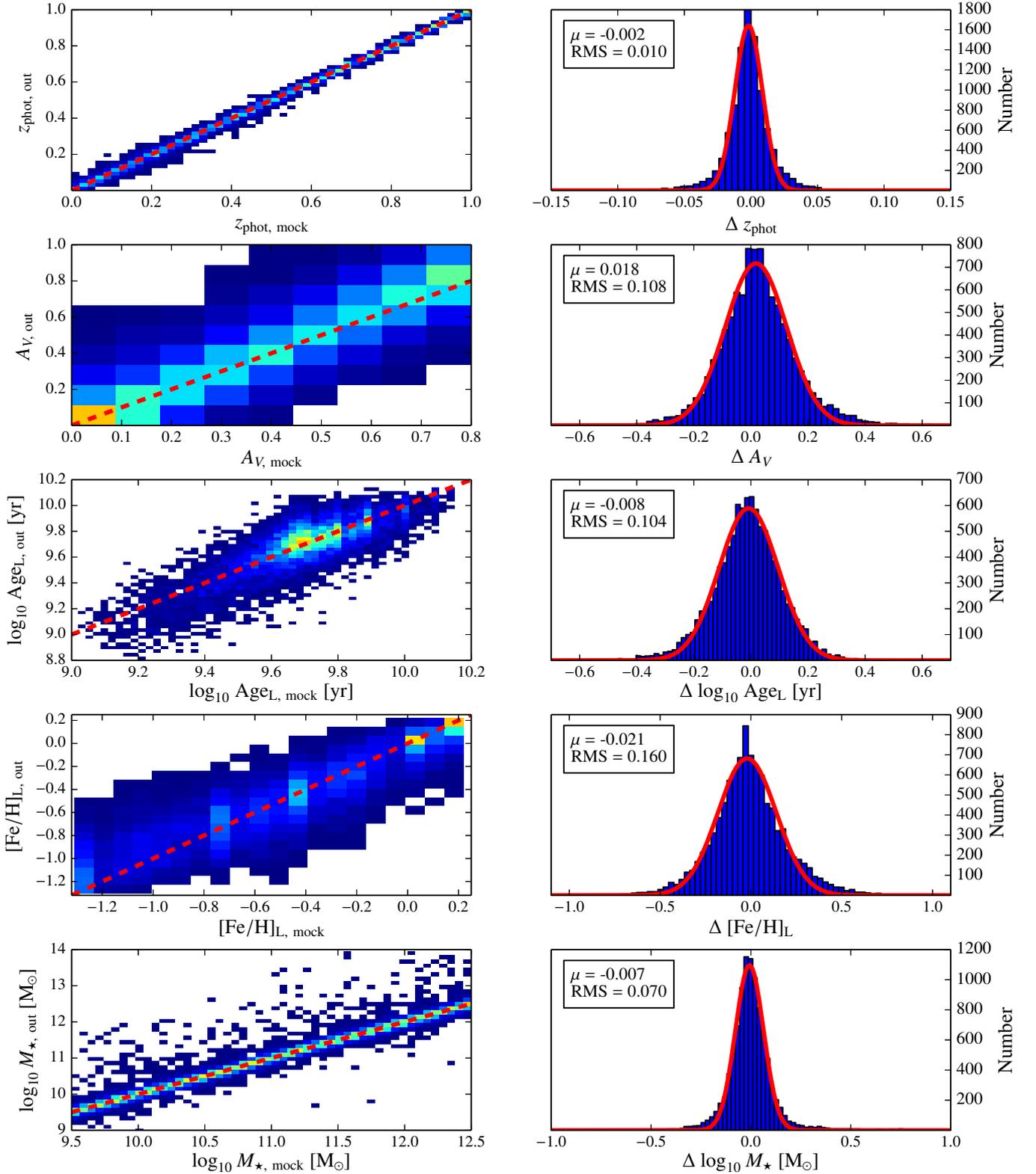}}
\caption{Comparison between the input parameters of mock ALHAMBRA galaxies, consisting of random mixtures two SSP models, and the output parameters retrieved with \textit{MUFFIT}. The mock galaxies (RS-like galaxies) have average signal-to-noise ratios per filter of 20, according to the typical signal-to-noise ratio distribution presented in Fig.~\ref{fig:sn_curve}. Left panels show, from top to bottom, the one-to-one comparisons in redshift, extinction, luminosity-weighted age, luminosity-weighted metallicity, and stellar mass. Redder and  bluer colours indicate, respectively, regions with higher and lower solution densities. The red dashed line indicates the one-to-one relationship. Right panels present the distributions of the differences between the input and output values in each case, fitted to a Gaussian function (in red) whose mean and RMS are indicated within the box.
}
\label{fig:simulation20}
\end{figure*}

Figure~\ref{fig:simulation20} illustrates the comparison between the input parameters of the mock galaxies and the output parameters retrieved with \textit{MUFFIT}, for the case $S/N=20$ and all redshifts. Left panels present one-to-one comparisons for the input and output photometric redshifts, extinctions, luminosity-weighted ages and metallicities, and stellar masses. Right panels illustrate the distributions of the differences between the input and output values in each case, fitted to a Gaussian function (in red) whose mean and RMS are therein indicated. In addition, Table~\ref{tab:simulations} provides the typical mean differences and their RMS for different redshift bins and $S/N = 10$, $20$, and $50$.

As expected, overall there is a very good agreement between the input and output derived parameters. This is not surprising as we are analysing mock galaxies made of mixtures of two SSP models, with the same SSP models as input of our code. In this sense, this test must be considered as a lower limit to the parameters uncertainties that we can expect for the forthcoming analysis of ALHAMBRA galaxies, just due to the photon-noise photometric errors. As a matter of fact, the total errors in the derived parameters are expected to be larger, due to potential differences between the spectro-photometric systems of the ALHAMBRA data and the models, independently of the SSP models of choice. Note also that real galaxies may be affected by ISM emissions or AGNs, which modify their SEDs with respect to a classical mixture of SSPs. 

Looking at the stellar mass plot in Fig.~\ref{fig:simulation20}, there seems to be a slight overestimation of the stellar mass. These cases correspond to galaxies with $z \la 0.02$, for which little variations in the redshift cause big changes in the luminosity distance, and therefore, in the retrieved stellar mass (see Eq.~\ref{eq:mass}). This result suggests that in the very local Universe, more accurate redshifts are required to provide reliable stellar masses using the analysis techniques explained above. Fortunately, the very few local galaxies in the ALHAMBRA survey have a very high signal-to-noise ratio as well, whereby this overestimation is negligible in our case. 

Another case that worth to be explained is the one of mock galaxies with low extinctions and low metallicities. According to Fig.~\ref{fig:simulation20}, we are getting on average larger values. However this is an artifact of the simulations since there are not lower values in our set of SSP models. The important result in these plots is that we are still retrieving the right trend in the parameters, despite the border effects in the parameter space.

\begin{table*}
\caption{Typical uncertainties in the determination of redshifts, extinctions, luminosity-weighted ages, luminosity-weighted metallicities, and stellar masses, expected from running our code on ALHAMBRA red sequence galaxies at different redshift bins and for different $S/N$ ($10$, $20$, and $50$). The random errors in the parameters are given as the mean and RMS of the best Gaussian function that reproduces the distribution of the differences between the input and output parameter values, as illustrated in the right panels of Fig.~\ref{fig:simulation20}.}
\label{tab:simulations}
\centering
\begin{tabular}{lrrrrr}
\hline
\hline
Parameters & $ 0.0 \le z \le 0.2 $ & $ 0.2 \le z \le 0.4 $ & $ 0.4 \le z \le 0.6 $ & $ 0.6 \le z \le 0.8 $ & $ 0.8 \le z \le 1.0 $ \\
\hline
\multirow{2}{*}{$S/N = 10$} & & & & & \\
& & & & & \\
\hline
$z_{\rm phot}$                     & $  0.01 \pm 0.03 $  & $  0.00 \pm 0.04 $  & $  0.00 \pm 0.03 $  & $  0.00 \pm 0.02 $  & $  0.00 \pm 0.02 $ \\
$A_{V}$                            & $  0.10 \pm 0.22 $  & $  0.08 \pm 0.20 $  & $  0.04 \pm 0.17 $  & $  0.04 \pm 0.15 $  & $  0.02 \pm 0.14 $ \\
$\log_{10}$ Age$_{\rm L}$\ [yr]    & $ -0.01 \pm 0.19 $  & $ -0.03 \pm 0.17 $  & $ -0.03 \pm 0.14 $  & $  0.00 \pm 0.11 $  & $  0.03 \pm 0.10 $ \\
$[\mathrm{Fe/H}]_{\rm L}$          & $ -0.09 \pm 0.29 $  & $  0.01 \pm 0.30 $  & $  0.03 \pm 0.26 $  & $  0.00 \pm 0.23 $  & $  0.03 \pm 0.22 $ \\
$\log_{10}M_\star\ [{\rm M_\sun}]$ & $  0.10 \pm 0.28 $  & $  0.02 \pm 0.12 $  & $  0.01 \pm 0.10 $  & $  0.01 \pm 0.08 $  & $ -0.01 \pm 0.06 $ \\
\hline
\multirow{2}{*}{$\rm S/N = 20$}& & & & & \\
& & & & & \\
\hline
$z_{\rm phot}$                     & $  0.00 \pm 0.01 $  & $  0.00 \pm 0.01 $  & $  0.00 \pm 0.01 $  & $  0.00 \pm 0.01 $  & $  0.00 \pm 0.01 $ \\
$A_{V}$                            & $  0.06 \pm 0.15 $  & $  0.04 \pm 0.13 $  & $  0.01 \pm 0.10 $  & $  0.02 \pm 0.09 $  & $  0.00 \pm 0.08 $ \\
$\log_{10}$ Age$_{\rm L}$\ [yr]    & $ -0.03 \pm 0.14 $  & $ -0.03 \pm 0.13 $  & $ -0.01 \pm 0.11 $  & $  0.00 \pm 0.09 $  & $  0.01 \pm 0.07 $ \\
$[\mathrm{Fe/H}]_{\rm L}$          & $ -0.06 \pm 0.20 $  & $ -0.03 \pm 0.19 $  & $ -0.01 \pm 0.16 $  & $ -0.01 \pm 0.14 $  & $  0.00 \pm 0.12 $ \\
$\log_{10}M_\star\ [{\rm M_\sun}]$ & $  0.02 \pm 0.15 $  & $  0.01 \pm 0.09 $  & $  0.00 \pm 0.07 $  & $  0.00 \pm 0.06 $  & $ -0.01 \pm 0.05 $ \\
\hline
\multirow{2}{*}{$\rm S/N = 50$}& & & & & \\
& & & & & \\
\hline
$z_{\rm phot}$                     & $  0.00 \pm 0.00 $  & $  0.00 \pm 0.00 $  & $  0.00 \pm 0.00 $  & $  0.00 \pm 0.00 $  & $  0.00 \pm 0.00 $ \\
$A_{V}$                            & $  0.01 \pm 0.08 $  & $  0.00 \pm 0.06 $  & $  0.00 \pm 0.05 $  & $  0.00 \pm 0.03 $  & $  0.00 \pm 0.02 $ \\
$\log_{10}$ Age$_{\rm L}$\ [yr]    & $ -0.02 \pm 0.11 $  & $ -0.02 \pm 0.09 $  & $  0.00 \pm 0.08 $  & $  0.00 \pm 0.06 $  & $  0.00 \pm 0.05 $ \\
$[\mathrm{Fe/H}]_{\rm L}$          & $ -0.03 \pm 0.08 $  & $ -0.02 \pm 0.08 $  & $ -0.02 \pm 0.07 $  & $ -0.02 \pm 0.06 $  & $ -0.01 \pm 0.05 $ \\
$\log_{10}M_\star\ [{\rm M_\sun}]$ & $  0.00 \pm 0.08 $  & $  0.01 \pm 0.06 $  & $  0.00 \pm 0.05 $  & $  0.00 \pm 0.04 $  & $  0.00 \pm 0.03 $ \\
\hline
\end{tabular}
\end{table*}

The results in Table~\ref{tab:simulations} are divided in different redshift bins as old ages are not allowed at high redshifts. It is noteworthy that all the parameters are better determined at high-redshifts than at low-redshifts at the same mean signal-to-noise ratio. First, because at higher redshifts the galaxy SEDs are sampled with an equivalent higher spectral resolution at rest frame, whereby both redshift and age, that are sensitive to the $4\,000$~\AA-break, are better established (and consequently, the rest of parameters as well). Also, at high redshift the range of possible ages is shorter, and in turn they are younger, with lower degeneracies that their older counterparts. Finally, the bluest part of the $z \ga 0.5$ SEDs have larger signal-to-noise ratios. These filters act as anchoring bands to constrain blue-sensitive parameters (like extinction or metallicity). This growth in the signal may be due to an underlying young and less massive population in the galaxy \citep{Ferreras2000}, that is not strong enough to contribute in the optical range, but that dominates the flux in the NUV rest-frame regime (being visible at $z \ga 0.5$ in ALHAMBRA), reinforcing the necessity of using two components in the fittings.

To conclude, these simulations are key to give us an idea of the typical issues that may appear in this kind of studies and the uncertainties that we expect just due to photon-noise photometric uncertainties. These results show that one can robustly explore the stellar populations of galaxies in the ALHAMBRA dataset by use of the \textit{MUFFIT} code presented here.

\subsection{Photometric redshifts in the ALHAMBRA survey}\label{sec:photoz}

Although the main aim of our code is not to determine redshifts, it is very important to check whether, for a general case in which the galaxies do not have any redshift information, the code is self-sufficient to estimate photo-\textit{z} properly, at least to some extent. Otherwise, the derived galaxy parameters may be wrongly estimated.

To do this, we run our code on the sub-sample of RS galaxies with spectroscopic redshifts in ALHAMBRA, from \citet{Molino2014}, to set the accuracy of the ALHAMBRA photo-\textit{z}. This sub-sample is built by the publicly available data of the spectroscopic surveys that overlap with ALHAMBRA (zCOSMOS, \citealt{Lilly2009}; GROTH, \citealt{Davis2007}; and GOODS-N, \citealt{Cooper2011}), amounting to $\sim900$ RS galaxies up to magnitude $m_{F814W}<22.5$. For the purpose of this test, we make use of the photo-\textit{z} predictions provided by \textit{BPZ2.0} in the Gold catalogue. In addition, we also take the photo-\textit{z} constraints provided by \textit{EAZY} \citep{Brammer2008}, with the default configurations and templates, to asses whether \textit{MUFFIT} also works similarly when the input photometric redshifts come from an external source.

To provide a numerical value about the quality and accuracy of the photo-\textit{z}, we simultaneously use definitions, which can be more or less useful depending of our purposes, for both accuracy and catastrophic outliers. \citet{Brammer2008} proposed the normalized median absolute deviation, $\sigma_\mathrm{NMAD}$, as a measurement of the photo-\textit{z} uncertainties, since it estimates the deviation of the photo-\textit{z} distribution without being affected by catastrophic errors. It is defined as
\begin{equation}
\sigma_\mathrm{NMAD} = 1.48 \times median \left( \frac{|\Delta z -median(\Delta z)|}{1+z_\mathrm{spec}} \right),
\label{eq:nmad}
\end{equation}
where $\Delta z = z_\mathrm{phot} - z_\mathrm{spec}$. Furthermore, we provide the RMS of the distribution $\Delta z/(1+z_\mathrm{spec})$ that, in the following, we denote as $\sigma_z/(1+z_\mathrm{spec})$. Additionally, we use two definitions for the rate of photo-\textit{z} catastrophic outliers, as in \citet{Molino2014}, formally expressed as
\begin{equation}
\eta_1 = \frac{\Delta z}{1+z_\mathrm{spec}} > 0.2, {\rm and}
\label{eq:eta1}
\end{equation}
\begin{equation}
\eta_2 = \frac{\Delta z}{1+z_\mathrm{spec}} > 5 \times \sigma_\mathrm{NMAD}.
\label{eq:eta2}
\end{equation}

\begin{table}
\caption{Quality of the photo-\textit{z} retrieved for a subsample of RS galaxies in ALHAMBRA, when different method are applied: our code alone, \textit{BPZ2.0}, \textit{EAZY}, and our code using \textit{BPZ2.0} and \textit{EAZY} as input values of redshift. Details on the definition of the quality values are given in Eqs.~\ref{eq:nmad}, ~\ref{eq:eta1}, and \ref{eq:eta2}.}
\label{tab:photoz}
\centering
\begin{tabular}{lcccc}
\hline
\hline
& $\sigma_\mathrm{NMAD}$ & $\sigma_z/(1+z_\mathrm{s})$ & $\eta_1$ & $\eta_2$  \\
\hline
\textit{MUFFIT} & $0.0157$ & $0.0105$ & $1.6$\% & $5.8$\% \\
\textit{BPZ2.0} & $0.0104$ & $0.0076$ & $0.9$\% & $7.7$\% \\
\textit{EAZY}   & $0.0102$ & $0.0083$ & $0.8$\% & $4.0$\% \\
\hline
\textit{MUFFIT + BPZ2.0} & $0.0087$ & $0.0070$ & $0.9$\% & $6.3$\% \\
\textit{MUFFIT + EAZY}   & $0.0092$ & $0.0071$ & $0.7$\% & $5.5$\% \\
\hline
\end{tabular}
\end{table}

\begin{figure*}
\resizebox{\hsize}{!}{\includegraphics{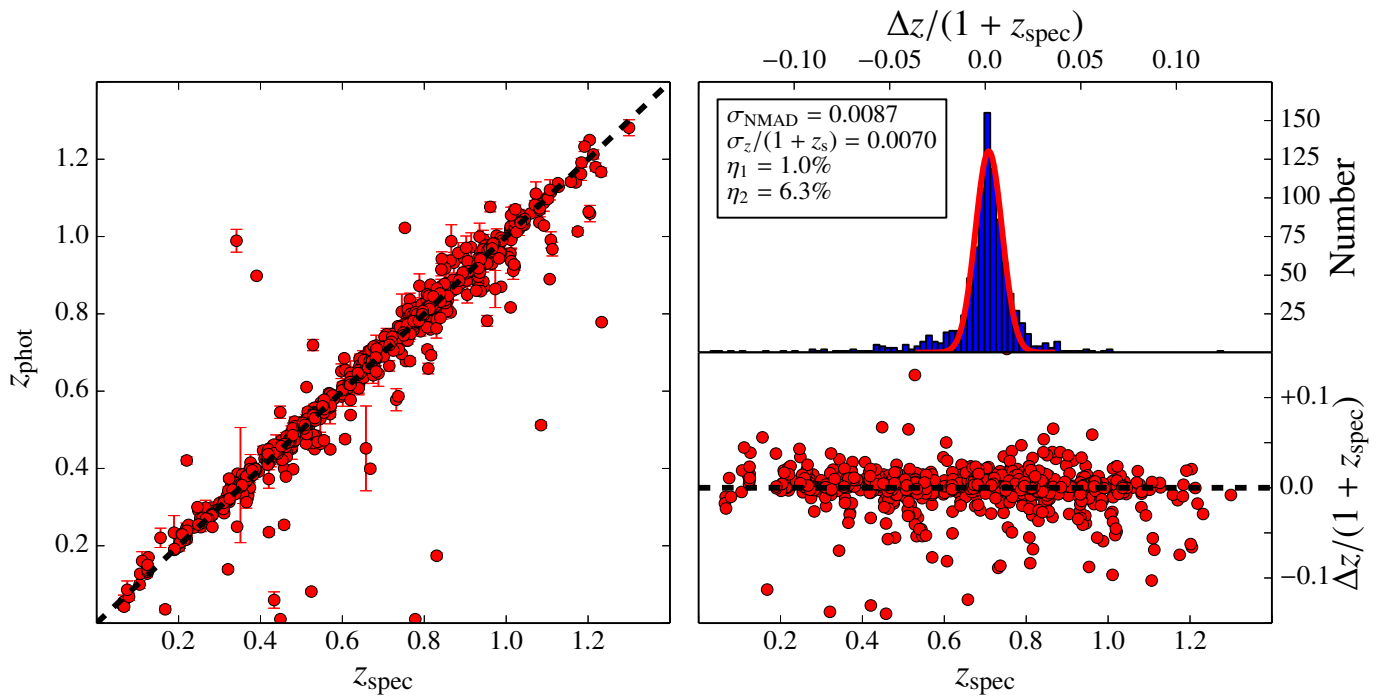}}
\caption{Comparison of the photo-\textit{z} retrieved with our code using the redshift PDFs of \textit{BPZ2.0} as input values. The data correspond to a sub-sample of RS galaxies from ALHAMBRA with spectroscopic redshifts in the literature. Left panel illustrates the one-to-one redshift comparison for every galaxy, with the dashed line being the one-to-one relation. The bottom-right panel presents the differences between the photo-\textit{z} and their spectroscopic counterparts for each galaxy, normalized by (1+$z_\mathrm{spec}$). The top-right panel shows the obtained distribution $\Delta z/(1+z_\mathrm{spec})$, indicating the accuracy parameters and the rate of outliers (Eqs.~\ref{eq:nmad}, ~\ref{eq:eta1}, and \ref{eq:eta2}) at the inner box.}
\label{fig:zspec_zphot}
\end{figure*}

On the basis of the above equations, Table~\ref{tab:photoz} presents the quality of the photo-\textit{z} determined for the subsample of RS galaxies under different methods. First, we analyse the reliability and accuracy of the photo-\textit{z} derived from our own code alone, that is, not using any photo-\textit{z} value as input to constrain the solution. This case can be directly compared with the values directly derived from the \textit{BPZ2.0} and \textit{EAZY} codes, showing, as expected, that photo-\textit{z} codes make a better job to determine redshifts from scratch. In addition, we analyse in the same way the quality of the photo-\textit{z} derived from our code when the redshift PDFs of \textit{BPZ2.0} are used as input parameters. \textit{MUFFIT} explores the plausible stellar population parameters, managing the photo-$z$ as another free parameter inside the redshift range of choice for the user. For the present work, we use the $1-\sigma$ PDF range provided in the Gold catalogue, hence being all the photo-$z$ weights equal to 1 inside the provided range ($1-\sigma$) and $0$ beyond this range. According to Eqs.~\ref{eq:nmad}, ~\ref{eq:eta1}, and \ref{eq:eta2}, we obtain $\sigma_\mathrm{NMAD} \sim 0.0087$, yielding a rate of catastrophic outliers $\eta_1 = 0.97$\% and $\eta_2 = 6.26$\%. For this specific case, the resultant photo-\textit{z} are compared with their spectroscopic counterparts in Fig.~\ref{fig:zspec_zphot}. Similarly, using the photo-\textit{z} constrains from \textit{EAZY} as an input to our code (also $1-\sigma$, for consistency in the comparison), we find $\sigma_\mathrm{NMAD}=0.0092$, $\eta_1 = 0.76$\%, and $\eta_2 = 5.51$\%. From these results we conclude that, at least for RS galaxies, our stellar population code improves the redshift accuracy of classical photo-\textit{z} codes, when these are used as input for our technique. This can be explained as our method plays with mixtures of much larger numbers (million) of SSP models, allowing to perform flexible SED fittings, hence providing a fine-tuned, second order correction to the redshift values of photo-\textit{z} codes. In addition, the number of catastrophic outliers, $\eta_1$ and $\eta_2$, is also marginally decreased on average. Regarding the shifts between photo-\textit{z} and spectroscopic values, these are statistically insignificant ($\la 0.002$), as we see in Sect.~\ref{sec:photoz_uncer}, on our stellar population results. As we show in Sect.~\ref{sec:photoz_uncer}, devoted to setting constraints on the uncertainties of the stellar population parameters due to the photo-\textit{z} uncertainties, it is more important for our stellar population aims to minimize the number of outliers, rather than decreasing a little $\sigma_\mathrm{NMAD}$.

\subsection{Impact of the photometric-redshift uncertainties}\label{sec:photoz_uncer}

One of the most critical parameters to determine reliable stellar populations and masses for a galaxy is the redshift. 
If the redshift is unknown or uncertain, it is obvious that the rest of derived parameters will not be reliable either. In this section, we aim to quantify the impact of the typical redshift uncertainties on the reliability of the retrieved parameters, and the maximum redshift errors allowed to reach our goals. 

To answer these questions, we focus again on the ALHAMBRA data. To determine the direct impact of redshift uncertainties over the rest of derived parameters, we compare the results obtained from our code for the spectroscopic sub-sample of RS galaxies (see Sect.~\ref{sec:photoz}) using the redshift PDFs of \textit{BPZ2.0} as an input, with the results that we obtain with our code for the same galaxy sample assuming exactly their spectroscopic redshifts. It is worth noting that, in contrast to the previous simulations (e.~g. Sect.~\ref{sec:uncertainties}), we are using real galaxies to estimate these uncertainties.

\begin{figure*}
\resizebox{\hsize}{!}{\includegraphics{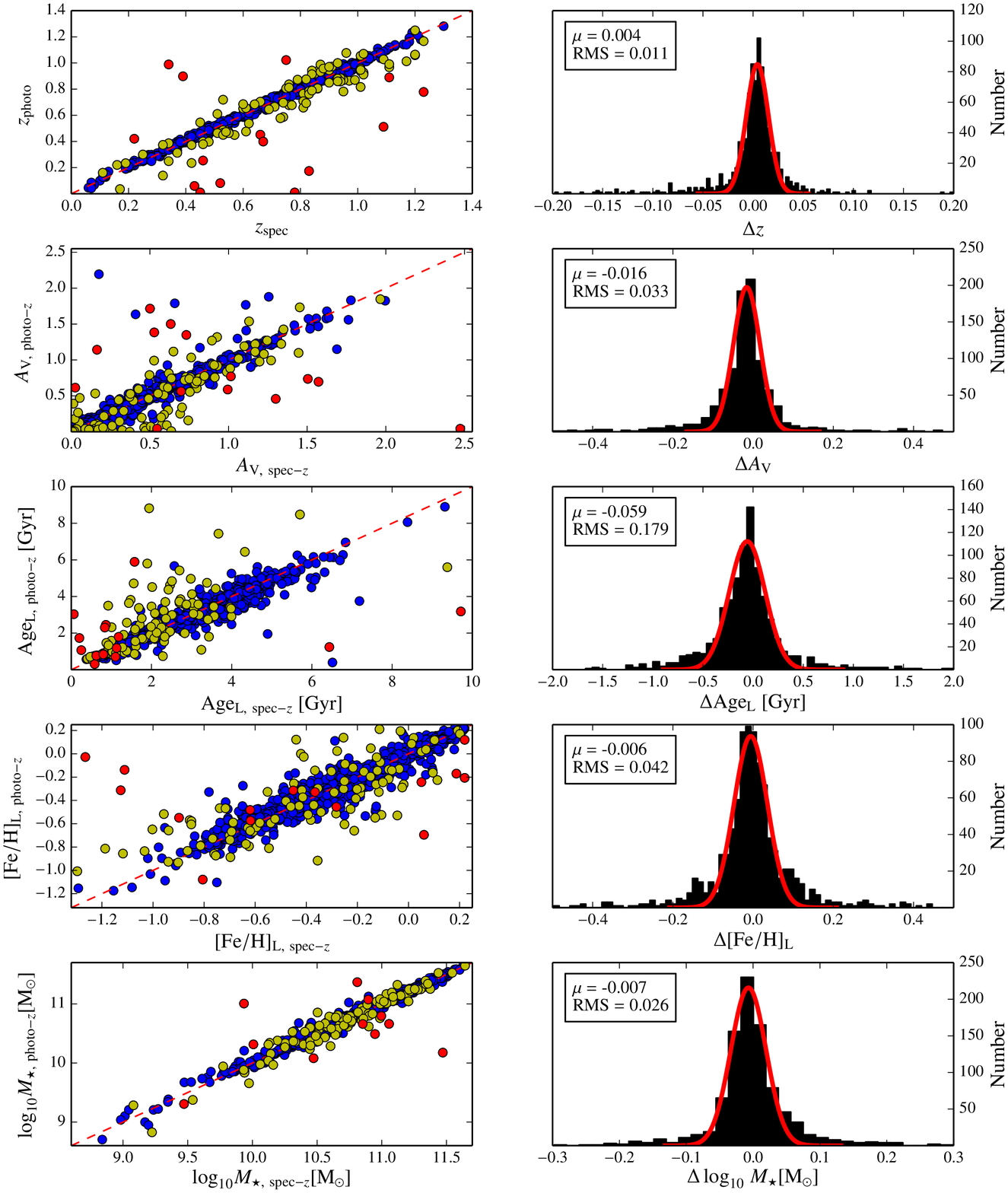}}
\caption{Impact of a redshift uncertainty $\sim 1\%$ on the stellar population parameters. On the left, we present the comparative one-to-one of the obtained parameters without any constrain in the photo-\textit{z} (Y--axis) versus the results forcing the redshift to its spectroscopic value (X--axis). The dashed red line represent the one-to-one relationship. Red, yellow, and blue dots are respectively galaxies for which $|z_{\rm spec}-z_{\rm phot}|  > 0.2$, $0.044 \le |z_{\rm phot}-z_{\rm spec}| < 0.2$, and $|z_{\rm phot}-z_{\rm spec}| < 0.044$. On the right, we have the histograms of the differences between the obtained results with and without the spectroscopic constraint. The solid red line is the best-fitting of the distribution to a Gaussian function, in the boxes there are showed both Gaussian mean and RMS. From top to bottom, we show redshift, extinction, age, metallicity, and stellar mass.}
\label{fig:zphot_uncer}
\end{figure*}

Figure~\ref{fig:zphot_uncer} summarizes the results of this test. Left panels present the one-to-one comparison for the redshift, extinction, luminosity-weighted age, luminosity-weighted metallicity, and stellar mass. To facilitate the visual interpretation of the comparison, red dots indicate those galaxies for which there exist large discrepancies between the photometric and spectroscopic redshifts, hence symbolizing the catastrophic outliers ($\eta_1$), as defined in Sect.~\ref{sec:photoz}, whereby yellow dots are the outliers related to $\eta_2$. This is clearly seen in the top-left panel of Fig.~\ref{fig:zphot_uncer}, where the redshift values are compared. In all cases, to guide the eye, the red dashed line indicates the one-to-one relationship. The right panels show the histograms of the differences between the obtained results with and without the spectroscopic redshift constraint. The solid red line represents the best fitting to a Gaussian function, with its mean and RMS indicated in the panels. From the data in Fig.~\ref{fig:zphot_uncer} and the RMS of the differences given in the right panels, we observe that, overall, the stellar population parameters present very minor changes due to typical redshift uncertainties. On average, such uncertainties account for $\sigma_{A_V}^{\ z} = 0.03$, $\sigma_\mathrm{Age}^{\ z} = 0.18$~Gyr ($0.03$~dex), $\sigma_\mathrm{[Fe/H]}^{\ z} = 0.04$~dex, and $\sigma_{M_\star}^{\ z} = 0.03$~dex, all negligible when we compare them with the uncertainties introduced by the typical photon-noise (see Sect.~\ref{sec:uncertainties}). 
As expected, catastrophic outliers (red and yellow dots) exhibit a larger spread in most parameters. 

To conclude, the typical uncertainties on the redshifts present a negligible impact on the main stellar population parameters, except if the galaxy is a catastrophic outlier. The uncertainties on the ALHAMBRA photometry and the model systematics are more crucial uncertainties in the present work.


\subsection{Degeneracies}\label{sec:degeneracies}

In addition to estimating the uncertainties, it is crucial to know which kind of degeneracies may alter our results, to avoid unveiling a finding that really is a degeneracy aftermath. To interpret the output properly, we must keep these degeneracies under control, reducing their impact as much as possible. 

Unlike stellar population diagnostic techniques based on local absorption features, e.~g. classical line strength indices, which also present the well known age-metallicity degeneracy \citep{Worthey1994a}, our multi-filter stellar population code is colour dependent, as it tries to reproduce the galaxy SED by mixing SSPs over a wide wavelength range. Therefore, considering only the age-metallicity degeneracy may not be enough, as we must evaluate any parameter that can modify the colour in a wide wavelength-range, i.e. we also have to include the intrinsic extinction as another degenerated parameter in our analysis. As we mention above, in this paper we assume an universal IMF. Otherwise, this parameter should be considered for the degeneracies as well, as bottom-heavy IMFs exhibit redder colors than top-heavy ones. Moreover, since the degeneracies among parameters strongly depend on the number and width of the filters, the total spectral coverage, etc., it is worth noting that the results presented in this section only apply to the use of our code on ALHAMBRA data.  

To address the degeneracy issues in the most realistic way, we take as targets all the ALHAMBRA RS galaxies (see Sect.~\ref{sec:rs}) with mean signal-to-noise ratios of $\sim20$ and in certain ranges of age and metallicity. These results come up after having run \textit{MUFFIT} with the ALHAMBRA galaxies using the MIUSCAT SSP models. The degeneracy estimation is done by taking all the stellar population values recovered during the Monte Carlo approaches at different stellar population bins (detailed below), and then stacking each retrieved distribution to build a whole distribution per bin, getting distributions among pairs of parameters (age, metallicity, and extinction). We characterise each distribution by setting confidence ellipses (2D confidence intervals) that enclose the results provided during the Monte Carlo process. These ellipses are obtained by the covariance matrix of each distribution, and they allow to parametrise the degeneracies through two parameters: by the ellipticity, denoted as $e$, and by $\theta$, the angle  between the X--axis and the ellipse semi-major axis. The angle $\theta$ is determined by the eigenvectors of the covariance matrix, as well as the eigenvalues of the covariance matrix determine the axis lengths. If $e$ is close to zero, this implies that the degeneration between the two parameters is not very significant, irrespective of the value of $\theta$. On the other hand, if $\theta$ is a multiple of $\pi/2$ (lie on any of the two the axes), both parameters are uncorrelated, and consequently there is not any degeneracy between them. Consequently, both parameters are necessary to confirm whether there exists a degeneracy or not. In fact, we can quantify the level of degeneracy between parameters via the Pearson's correlation coefficient, which mathematically reflects both $e$ and $\theta$ effects. Formally,

\begin{equation}
r_{xy}=\frac{\sum\limits_{i=1}^n (x_i-\bar{x})(y_i-\bar{y})}{\sqrt{\sum\limits_{i=1}^n (x_i-\bar{x})^2 \sum\limits_{i=1}^n(y_i-\bar{y})^2}},
\label{eq:pearson}
\end{equation}
where $x_i$ and $y_i$ denote the value pairs of the parameters (age, metallicity, extinction) with means $\bar{x}$ and $\bar{y}$ respectively. The closer is $r_{xy}$ to $1$ (to $-1$), the larger is the correlation (anti-correlation), i.~e. the degeneracy between parameters; on the contrary, a value close to $0$ ($-0.1 \la r_{xy} \la 0.1$) suggests that the parameters are uncorrelated, and there would not exist any degeneracy.

Regarding the parameter ranges, we take bins in age of $0.5 \leq \mathrm{Age_\mathrm{L}} \leq 1.0$, $3.0 \leq \mathrm{Age_\mathrm{L}} \leq 4.5$, and $7.0 \leq \mathrm{Age_\mathrm{L}} \leq 10.0$~Gyr, whereas for metallicity we take $-0.8 \leq \mathrm{[Fe/H]_\mathrm{L}} \leq -0.6$, $-0.4 \leq \mathrm{[Fe/H]_\mathrm{L}} \leq -0.2$, and $-0.1 \leq \mathrm{[Fe/H]_\mathrm{L}} \leq 0.1$. These bins have been chosen to evaluate how the degeneracies vary along different stellar population parameters. Note that we do not establish any extinction bin, because we have previously checked that the degeneracies at different extinction bins present negligible differences in $e$ and $\theta$. We have also studied whether different redshifts can alter the degeneracy effects, as the redshift determines the observed spectral range of the ALHAMBRA SEDs. We find that at higher redshift some degeneracies tend to decrease, specially for young and low metallicity galaxies, but in general the degeneracies remains alike (less dispersion as well, see Sect.~\ref{sec:uncertainties}). Thus, the three bins in age and in metallicity define the nine intervals where the degeneracies of our parameters are explored. In each interval, we confront the age-metallicity, age-extinction and metallicity-extinction degeneracies.

\begin{figure*}
\resizebox{\hsize}{!}{\includegraphics{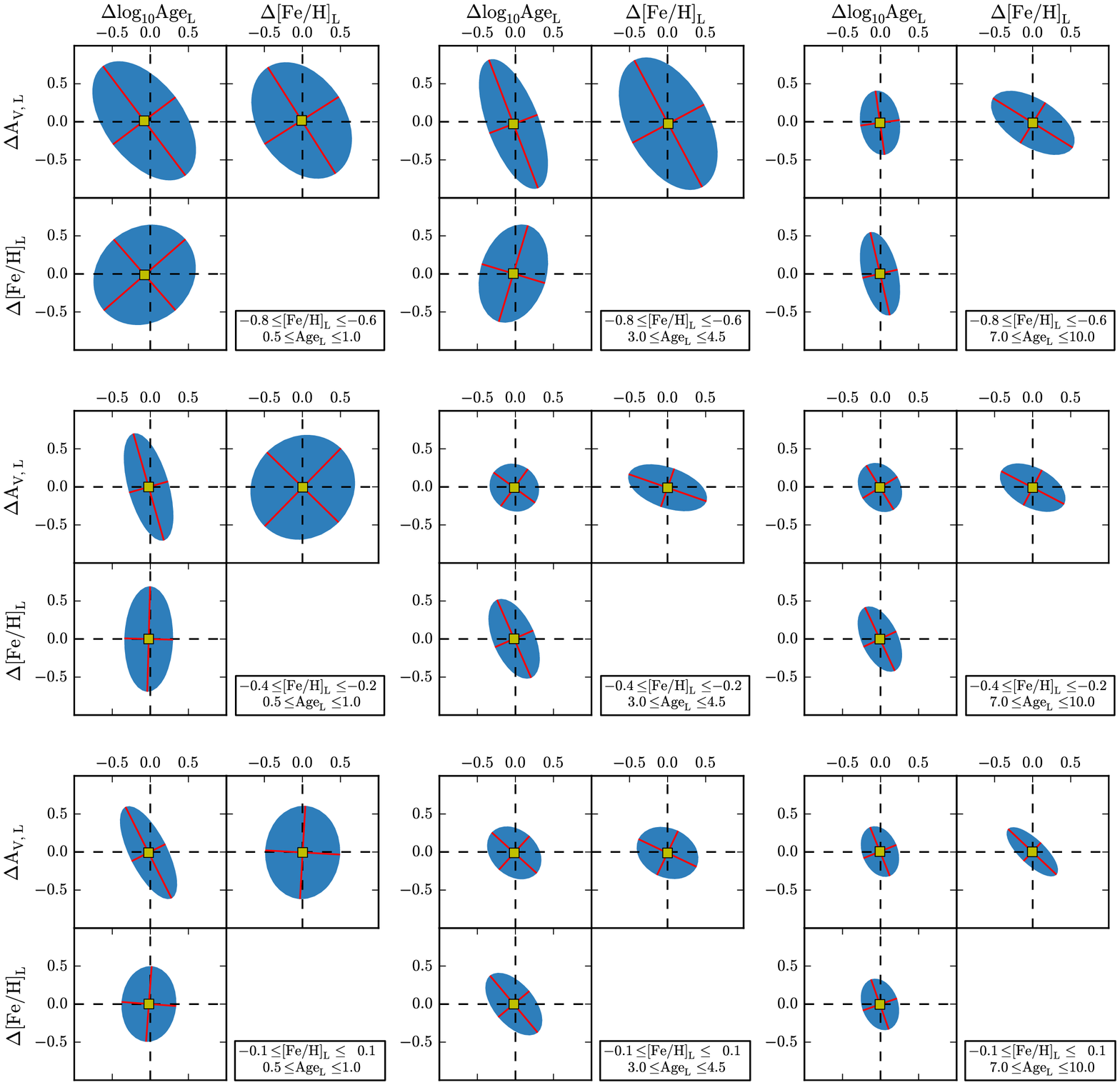}}
\caption{Covariance error ellipses, at the 95\% confidence level, of the stellar population parameters provided by the Monte Carlo approach in \textit{MUFFIT} with respect to their most probable values, for different age and metallicity bins (see inner panels). Yellow squares indicate the ellipse centres and the red lines illustrate the minor and major axis of each ellipse.}
\label{fig:degeneracies}
\end{figure*}

In Fig.~\ref{fig:degeneracies} we present the covariance error ellipses (in blue) that enclose, at the $95$\% confidence level, the distribution of the provided parameters during the Monte Carlo approach for all the ALHAMBRA RS galaxies in the nine age and metallicity ranges (see inner panels), and at redshift $z \leq 0.4$. The semi-minor and semi-major axes of each ellipse are shown in red, whereas their centres are represented with a yellow square. Table~\ref{tab:degeneracies} provides $\theta$, $e$, and $r_{xy}$ for the same age and metallicity regimes.

\begin{table*}
\caption{Summary of the confidence error ellipses in Fig.~\ref{fig:degeneracies} for different parameter bins and MIUSCAT (top) and BC03 (bottom) models. $\theta$ is the angle between the X--axis and the semi-major axis $a$ in degrees (counter clockwise), whereas $e$ is the ellipticity. To quantify the degeneracy between parameters, we provide the Pearson's correlation coefficient, $r_{xy}$, where values close to $0$ correspond to uncorrelated parameters.}
\label{tab:degeneracies}
\centering
\begin{tabular}{clrccrccrcc}
\hline
\hline
 &   & \multicolumn{3} {c} {$0.5 \leq \mathrm{Age}\ \mathrm{[Gyr]} \leq 1.0$} & \multicolumn{3} {c} {$3.0 \leq \mathrm{Age}\ \mathrm{[Gyr]} \leq 4.5$} & \multicolumn{3} {c} {$7.0 \leq \mathrm{Age}\ \mathrm{[Gyr]} \leq 10.0$} \\
&   & $r_{xy}$ & $\theta$ & $e$ & $r_{xy}$ & $\theta$ & $e$ & $r_{xy}$ & $\theta$ & $e$ \\
\hline
\parbox[t]{2mm}{\multirow{15}{*}{\rotatebox[origin=c]{90}{MIUSCAT}}} & \multirow{2}{*}{$\Delta \log_{10} \mathrm{Age}$ vs $\Delta$[Fe/H]}& & & & & & & & &  \\
&  & & & & & & & & &  \\
\cline{2-11}
& $ -0.8 \leq \mathrm{[Fe/H]} \leq -0.6$ & $0.11$ & $36$ & $0.11$ & $0.24$ & $70$ & $0.31$ & $-0.40$ & $103$ & $0.58$ \\
& $ -0.4 \leq \mathrm{[Fe/H]} \leq -0.2$ & $0.03$ & $88$ & $0.52$ & $-0.49$ & $113$ & $0.51$ & $-0.51$ & $116$ & $0.50$ \\
& $ -0.1 \leq \mathrm{[Fe/H]} \leq 0.1$ & $0.04$ & $85$ & $0.22$ & $-0.55$ & $129$ & $0.46$ & $-0.26$ & $108$ & $0.35$ \\
\cline{2-11}
& \multirow{2}{*}{$\Delta$[Fe/H] vs $\Delta A_V$}& & & & & & & & &  \\
&  & & & & & & & & &  \\
\cline{2-11}
& $ -0.8 \leq \mathrm{[Fe/H]} \leq -0.6$ & $-0.30$ & $119$ & $0.30$ & $-0.46$ & $114$ & $0.47$ & $-0.54$ & $148$ & $0.49$ \\
& $ -0.4 \leq \mathrm{[Fe/H]} \leq -0.2$ & $0.05$ & $58$ & $0.05$ & $-0.37$ & $161$ & $0.46$ & $-0.41$ & $150$ & $0.40$ \\
& $ -0.1 \leq \mathrm{[Fe/H]} \leq 0.1$ & $0.02$ & $87$ & $0.20$ & $-0.16$ & $152$ & $0.18$ & $-0.76$ & $136$ & $0.63$ \\
\cline{2-11}
& \multirow{2}{*}{$\Delta \log_{10} \mathrm{Age}$ vs $\Delta A_V$}& & & & & & & & &  \\
&  & & & & & & & & &  \\
\cline{2-11}
& $ -0.8 \leq \mathrm{[Fe/H]} \leq -0.6$ & $-0.50$ & $126$ & $0.44$ & $-0.69$ & $112$ & $0.67$ & $-0.13$ & $97$ & $0.38$ \\
& $ -0.4 \leq \mathrm{[Fe/H]} \leq -0.2$ & $-0.53$ & $106$ & $0.63$ & $-0.18$ & $136$ & $0.17$ & $-0.25$ & $121$ & $0.25$ \\
& $ -0.1 \leq \mathrm{[Fe/H]} \leq 0.1$ & $-0.70$ & $117$ & $0.63$ & $-0.37$ & $134$ & $0.32$ & $-0.28$ & $111$ & $0.34$ \\
\hline
\hline
\parbox[t]{2mm}{\multirow{15}{*}{\rotatebox[origin=c]{90}{BC03}}}& \multirow{2}{*}{$\Delta \log_{10} \mathrm{Age}$ vs $\Delta$[Fe/H]}& & & & & & & & &  \\
&  & & & & & & & & &  \\
\cline{2-11}
& $ -0.8 \leq \mathrm{[Fe/H]} \leq -0.6$ & $-0.07$ & $93$ & $0.42$ & $-0.24$ & $104$ & $0.38$ & $-0.19$ & $96$ & $0.53$ \\
& $ -0.4 \leq \mathrm{[Fe/H]} \leq -0.2$ & $0.01$ & $89$ & $0.39$ & $-0.46$ & $112$ & $0.49$ & $-0.32$ & $120$ & $0.31$ \\
& $ -0.1 \leq \mathrm{[Fe/H]} \leq 0.1$ & $-0.05$ & $91$ & $0.54$ & $-0.57$ & $118$ & $0.53$ & $-0.48$ & $136$ & $0.41$ \\
\cline{2-11}
& \multirow{2}{*}{$\Delta$[Fe/H] vs $\Delta A_V$}& & & & & & & & &  \\
&  & & & & & & & & &  \\
\cline{2-11}
& $ -0.8 \leq \mathrm{[Fe/H]} \leq -0.6$ & $-0.63$ & $145$ & $0.55$ & $-0.22$ & $118$ & $0.23$ & $-0.77$ & $150$ & $0.68$ \\
& $ -0.4 \leq \mathrm{[Fe/H]} \leq -0.2$ & $-0.58$ & $141$ & $0.49$ & $-0.16$ & $167$ & $0.31$ & $-0.34$ & $158$ & $0.40$ \\
& $ -0.1 \leq \mathrm{[Fe/H]} \leq 0.1$ & $-0.59$ & $144$ & $0.51$ & $-0.10$ & $174$ & $0.37$ & $-0.25$ & $104$ & $0.39$ \\
\cline{2-11}
& \multirow{2}{*}{$\Delta \log_{10} \mathrm{Age}$ vs $\Delta A_V$}& & & & & & & & &  \\
&  & & & & & & & & &  \\
\cline{2-11}
& $ -0.8 \leq \mathrm{[Fe/H]} \leq -0.6$ & $-0.45$ & $118$ & $0.43$ & $-0.62$ & $114$ & $0.60$ & $-0.16$ & $104$ & $0.28$ \\
& $ -0.4 \leq \mathrm{[Fe/H]} \leq -0.2$ & $-0.50$ & $117$ & $0.48$ & $-0.43$ & $126$ & $0.38$ & $-0.46$ & $144$ & $0.41$ \\
& $ -0.1 \leq \mathrm{[Fe/H]} \leq 0.1$ & $-0.46$ & $108$ & $0.53$ & $-0.40$ & $140$ & $0.35$ & $-0.44$ & $116$ & $0.44$ \\
\hline
\end{tabular}
\end{table*}

As intuitively expected, in all cases age and extinction are anti-correlated, in the sense that a reddening by extinction can mimic an older age, and viceversa. However, the behaviour of the age-metallicity and metallicity-extinction degeneracies is not so immediate. This is clearly a consequence of the role that extinction plays in the analysis as a third freedom degree, absorbing partially the weight of metallicity in the classical age-metallicity degeneracy problem. Whilst older galaxies exhibit, as expected, clear anti-correlated age-metallicity and metallicity-extinction degeneracies, such anti-correlations may turn very mild or even positive correlations depending on the range of age and metallicity. For instance, at the lower metallicity range, young and intermediate-age galaxies exhibit a positive degeneracy between age and metallicity, turning mild or negligible for young galaxies with intermediate and high metallicities. Interestingly, young metal rich galaxies are essentially only subject to the extinction-age degeneracy. 

Finally, we have checked that the general degeneracy trends presented in this section do not vary qualitatively when computed on the basis of the BC03 models (see Table.~\ref{tab:degeneracies}). For ages  lower than $\sim1$~Gyr, there is a clear degeneracy of both age and metallicity with extinction; and for ages older, there is also an age-metallicity degeneracy.


\section{Testing the perfomance of the code with ALHAMBRA galaxy data}\label{sec:testing}
Once the technical details of \textit{MUFFIT} have been described in detail, and the typical uncertainties and degeneracies among the derived parameters have been studied for the case of the ALHAMBRA survey, in this section we apply the code to different sub-samples of galaxies in ALHAMBRA. The ultimate goal of this section is not to provide a thorough study of the stellar populations of ALHAMBRA galaxies, but rather to test the reliability of the stellar populations, emission line equivalent widths (hereafter EW), stellar masses, extinctions and redshifts derived from our code in comparison with those published in previous work for either similar or identical galaxy samples. A forthcoming paper \citetext{D\'{\i}az-Garc\'{\i}a et al. 2015, in prep.} will present a complete analysis of the stellar populations of galaxies in ALHAMBRA making use of \textit{MUFFIT}.


\subsection{Stellar masses and photo-\textit{z} in the COSMOS survey}\label{sec:cosmos}

Since the Field--4 in the ALHAMBRA survey partly overlaps with the COSMOS field, we can construct a sub-sample of RS galaxies (see Sect.~\ref{sec:rs}) in common between both surveys.  After removing all the sources labelled as stars in COSMOS \citep[point-like sources,][]{Ilbert2009} and in ALHAMBRA, we end up with a sub-sample of $767$ common galaxies up to redshift $z \le 1.6$.

This galaxy sub-sample has indeed an important added value for our testing goals, as it has spectroscopic data in the zCOSMOS 10k-bright catalogue \citep{Lilly2007,Lilly2009}, allowing for a calibration of the derived photo-\textit{z} using the \textit{Le PHARE} code. In addition, using broad and medium bands and following a SED-fitting technique with BC03 models, \citet{Ilbert2010} estimated the stellar masses of the COSMOS galaxies. They assumed a fixed Chabrier IMF \citep[similar to the IMF of][]{Kroupa2001}, a star formation rate $\propto \mathrm{e}^{-t/\tau}$ (with $0.1 \leq \tau \leq 30$~Gyr), a unique solar metallicity, an age grid of $0.1$--$14.5$~Gyr, and the \citet{Calzetti2000} extinction law with $0.0 \leq E(B-V) \leq 0.5$ (large extinction values are only allowed for galaxies with high star formation). 

For the analysis of this sample we use \textit{MUFFIT}, that imposes a mixture of two SSPs rather than an exponential star formation rate. We select MIUSCAT models with a Kroupa IMF (which slightly differs from the Chabrier IMF), for a wider range in metallicity ($-1.31 \leq \mathrm{[Fe/H]} \leq 0.22$). For our analysis we use the \citet{Fitzpatrick1999} extinction law rather than the \citet{Calzetti2000} law, as the latter is overall more adequate for central star forming regions \citep{Calzetti1997}.

\begin{figure*}
\centering
\includegraphics[width=17cm]{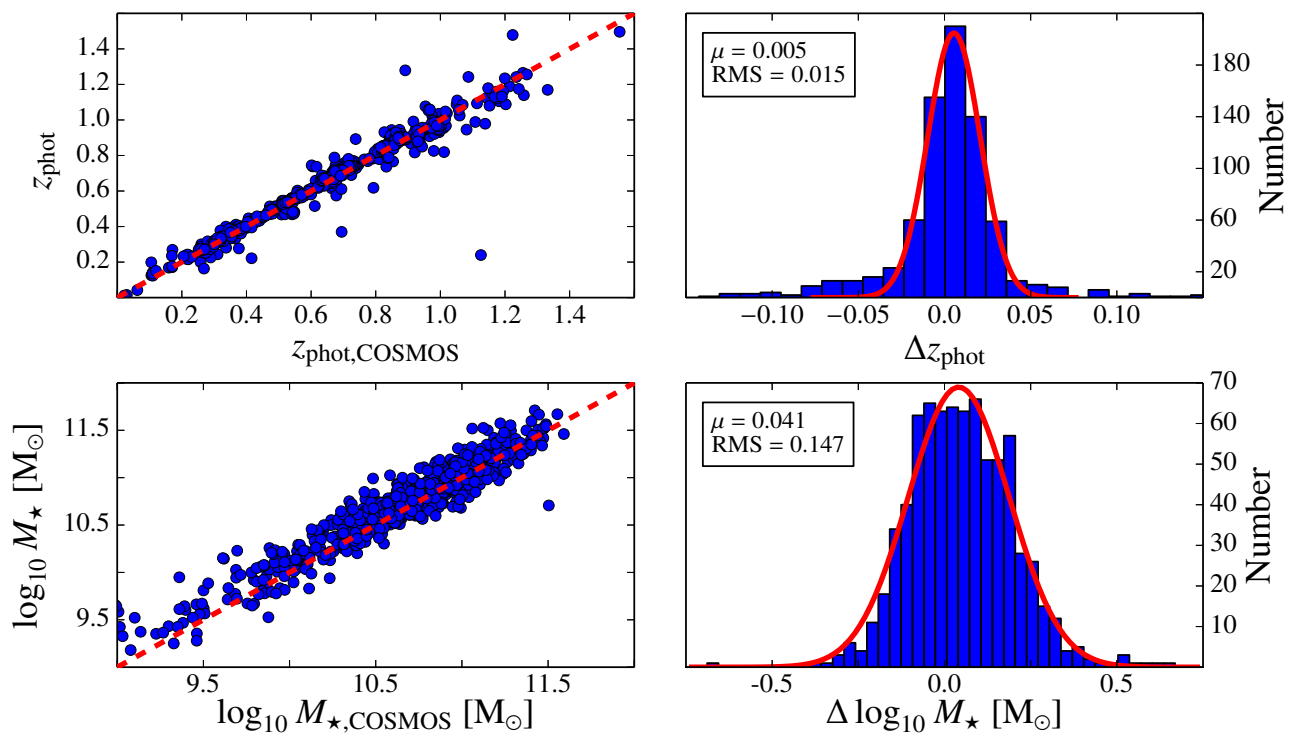}
\caption{Differences between the photo-\textit{z} (top panels) and stellar masses (bottom panels), computed with \textit{MUFFIT} (Y--axis) and the values provided in COSMOS catalogues \citep[X--axis;][]{Ilbert2009, Ilbert2010}, for a sub-sample of RS galaxies. On the right, we show the histograms of the differences, which are fitted to a Gaussian distribution.}
\label{fig:cosmos}
\end{figure*}

In Fig.~\ref{fig:cosmos} we present a one-to-one comparison of the photo-\textit{z} and stellar masses obtained with our code on the ALHAMBRA data and those presented in the above work on the COSMOS data \citep{Ilbert2009,Ilbert2010} for the sub-sample of $767$ RS galaxies in common. Having already constrained the reliability and accuracy of the ALHAMBRA photo-\textit{z} with respect to available spectroscopic redshifts (see Sect.~\ref{sec:photoz} and ~\ref{sec:photoz_uncer}), in the top panels of Fig.~\ref{fig:cosmos} we confront our outcomes with the COSMOS photo-\textit{z}, not only because of the qualitative similarity of both techniques, but also to check if there are systematics in our photo-\textit{z} measurements that might cause any kind of systematic in the retrieved ALHAMBRA stellar masses. From these plots we can see that both photo-\textit{z} estimations are in very good agreement, with one-to-one differences having an $\mathrm{RMS}$ of just $0.015$, and not finding statistically significant differences between both samples. The bottom panels of Fig.~\ref{fig:cosmos} are devoted to the stellar mass comparison. We find that the mean value of the one-to-one stellar mass differences is $0.04$~dex, with a dispersion of $0.15$~dex. 

As we explained in Sect.\ref{sec:photoz_uncer}, the typical uncertainties of our ALHAMBRA photo-\textit{z} may have an impact in the dispersion of the retrieved stellar masses of up to $0.026$ (see Fig.~\ref{fig:zphot_uncer}), well below the observed value. On the other hand, the offset between masses cannot be completely explained by the mild offset between redshifts ($\mu = 0.005$), as using Eq.~\ref{eq:mass} this difference implies a shift in mass $\la 0.015$~dex. Previous work have already reported the non-negligible effect of using different model sets on the absolute values of the derived stellar masses \citep[see e.~g.][]{Pozzetti2007,Ilbert2010}. To check this in our particular case, we have repeated the above analysis for the same sub-sample of galaxies with the BC03 models, instead of MIUSCAT. The new mean difference between the stellar masses of ALHAMBRA and COSMOS galaxies is now $-0.03$~dex, with a similar dispersion of $\mathrm{RMS}=0.15$~dex. Consequently, the mild systematic between stellar masses is probably due to the SSP model choice. Irrespective of the input set of SSP models, the dispersion between the stellar masses provided by COSMOS and the retrieved from ALHAMBRA data after running \textit{MUFFIT} remains $\mathrm{RMS} \sim 0.15 $~dex. This can be easily explained as the quadratic sum of the stellar mass uncertainties retrieved from both COSMOS and our proper technique in the common sample, being $\sim0.09$~dex in COSMOS and $\sim0.11$~dex in ALHAMBRA.

As a general conclusion, despite the differences in the analysis techniques presented in this paper for ALHAMBRA galaxies and those performed by previous work for the COSMOS data, we find a remarkable agreement between the photo-\textit{z} and the stellar masses derived for galaxies in common, showing the reliability and robustness of our code. We confirm that the choice of SSP models and extinction laws has an impact in the absolute values of the derived stellar masses. In particular, we find that the stellar masses derived using the MIUSCAT models lead to mean values $0.07$~dex larger that using BC03 models.


\subsection{Photometric EWs of emission lines}\label{sec:ew}

During the SED fitting process, the detection and subsequent removal of the bands affected by nebular emission lines may be crucial to determine reliable properties of the underlying stellar content of the galaxy under study. The way in which the affected bands are detected and removed from the analysis is already described in Sect.\ref{sec:elines}. Here we analyse to which extent the emission line residuals retrieved from our fittings, based on photometric data, are reliable and still keep meaningful information on the true EWs of the nebular lines derived from classical spectroscopy.

To build up a comparison galaxy sample, we first take all the ALHAMBRA galaxies that i) are in common with the data catalogues of the MPA/JHU \footnote{\url{www.mpa-garching.mpg.de/SDSS/}} (hereafter MPA/JHU catalogues) and ii) present nebular emission lines in their spectra \citep{Brinchmann2004,Tremonti2004}. This catalogue contains EWs and flux measurements of nebular lines for galaxies in the SDSS DR7 \citep{Abazajian2009}. Such measurements already account for their corresponding underlying stellar absorptions, as they were calculated after the subtraction of appropriate SSP models.

Since our stellar population code is focused towards the analysis of galaxies whose SEDs are dominated by their stellar content, for a fair comparison we systematically remove from the sample all the AGNs and QSOs (\textit{AGN}, \textit{AGN\_BROADLINE}, \textit{QSO}, and \textit{T2} types in SDSS), even if some of them could still be well interpreted by our code. In addition, galaxies with a signal-to-noise ratio lower than $5$ in the EW continuum are removed, as well as those galaxies in the redshift ranges $0.112<z<0.114$, $0.123<z<0.125$, and $0.146<z<0.148$, to avoid EW contaminations due to the sky line \ion{O}{I}\ $\lambda5577$, and all the galaxies larger than $4\arcsec$ (the SDSS fibers have $3\arcsec$ diameter) to minimize strong aperture effects in the photometry. Under the above constraints, there are  $92$ galaxies in common between ALHAMBRA and the MPA/JHU catalogues of SDSS. 

The detection and classification of emission lines in multi-filter surveys is clearly limited by the low spectral resolution of the data. For instance, at the ALHAMBRA resolution, and depending on the redshift, the emission line pairs H$\beta$--[\ion{O}{III}], H$\alpha$--[\ion{N}{II}], and even H$\alpha$--[\ion{S}{II}] can be unresolved because of their proximity in wavelength. To try to overcome this intrinsic limitation, rather than comparing the EWs of individual lines, we compare the total flux in excess along the observed spectral-range ($\lambda\lambda\ 3500$--$9700$~\AA\ for ALHAMBRA) with the total sum of the EWs measured in the MPA/JHU catalogues for the following strong nebular lines: \ion{H}{$\beta$}\ $\lambda4861$, [\ion{O}{III}]\ $\lambda4959$, [\ion{O}{III}]\ $\lambda5007$, \ion{H}{$\alpha$}\ $\lambda6563$, [\ion{N}{II}]\ $\lambda6548$, [\ion{N}{II}]\ $\lambda6584$, [\ion{S}{II}]\ $\lambda6717$, and [\ion{S}{II}]\ $\lambda6731$. We do not account for weaker lines as they might not be detected under the ALHAMBRA resolution. As explained in Section~\ref{sec:elines}, we set the detection limits of emission lines in ALHAMBRA to a flux excess of $\Delta m_\mathrm{EL} = 0.1$ and a signal-to-noise ratio with respect to the photometric error of the filter of $\sigma_\mathrm{EL} = 2.5$. Out of the $92$ galaxies in common, there are $44$ galaxies in ALHAMBRA for which our code detects an emission line in at least one filter, hence constituting the final galaxy sub-sample for the sake of EW comparisons.

Since our SED-fitting technique is based on a $\chi^2$ minimization technique between the filter fluxes of models and real galaxies over the full spectral range of the data, we must be aware of the fact that the best fitting solution may not match perfectly the local continuum around the emission lines, leading to random over/underestimations of the flux line. To minimize this effect for the sake of this study, given the filter that contains the emission line, $f_l$, we define a \textit{local continuum} for this band, $f_l^c$, as the mean value of the flux in the contiguous bands not affected by the emission line. For the model band that contains the modeled, corresponding stellar absorption, $t_l$, we similarly define a continuum for the model, $t_l^c$, in the same bands where $f_l^c$ is calculated. Following the formalism for spectroscopic EWs, the equation for photometric EWs (in \AA) is
\begin{equation}
EW_\mathrm{T, phot} = \sum\limits_{X}{\left( \frac{f_l}{f_l^c} - \frac{t_l}{t_l^c} \right) \Delta\lambda_l}\ ,
\label{eq:EW_phot}
\end{equation}
where $\Delta\lambda_l$ is the width of the band ($\sim 300$~\AA\ for the optical filters in ALHAMBRA) that contains the emission line, and the sum applies to all the bands affected by strong emission lines, $X$.

\begin{figure}
\resizebox{\hsize}{!}{\includegraphics{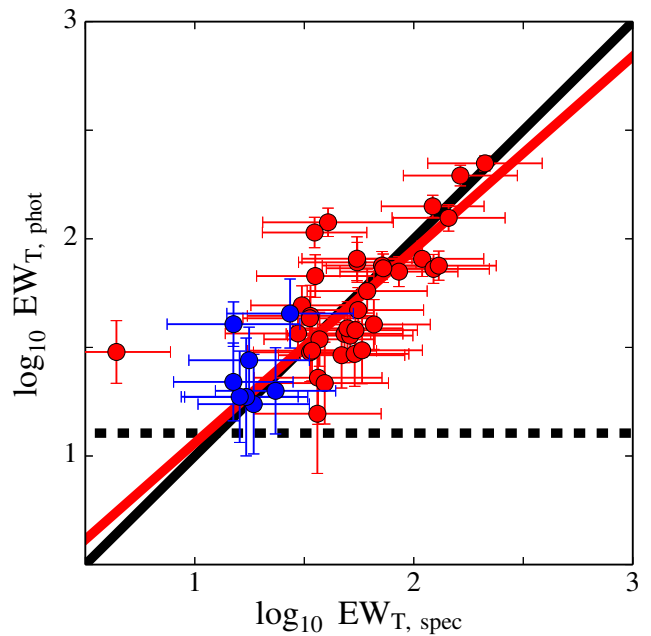}}
\caption{Comparison between the emission line EWs measured in galactic SDSS spectroscopic data taken from the MPA/JHU catalogues, $EW_\mathrm{T,\ spec}$, and the emission line EWs derived from the residuals of our fitting code on the same galaxies in ALHAMBRA, $EW_\mathrm{T,\ phot}$. Blue dots indicate galaxies with only one strong emission line in their spectra, i.~e. with a reasonable specific or individual emission line measurement in ALHAMBRA, whereas red dots illustrate galaxies with more than one emission line. The black line is the one-to-one relationship, whereas the red line is a linear regression to all the points. The dashed black line is the minimum $EW_\mathrm{T,\ phot}$ that we can detect imposing $\Delta m_\mathrm{EL}=0.1$ for a unique emission line in one ALHAMBRA filter. See more details in the text.}
\label{fig:EW}
\end{figure}

In Fig.~\ref{fig:EW} we present the photometric $EW_\mathrm{T,\ phot}$ derived with our spectral-fitting techniques on ALHAMBRA data versus the spectroscopic $EW_\mathrm{T,\ spec}$ computed from MPA/JHU catalogues, for the $44$ galaxies in common that fulfilled the above selection criteria. Blue dots indicate galaxies with only one strong emission line in their spectra, i.~e. corresponding to a reasonable specific or individual emission line measurement in ALHAMBRA, whereas red dots illustrate galaxies with more than one emission line. The black line is the one-to-one relationship. The dashed black line represents the minimum $EW_\mathrm{T,\ phot}$ that we can detect in the ALHAMBRA data imposing $\Delta m_\mathrm{EL}=0.1$, for a unique emission line along the SED and a top hat filter with FWHM$\sim300$~\AA. The red line is the linear regression to all the data points, considering both photometric and spectroscopic uncertainties. Overall, we obtain a good agreement between photometric and spectroscopic EWs, where the bias between both measurements is $\mu = 0.018$~dex and $\mathrm{RMS=0.234}$~dex, demonstrating the feasibility and reliability of our code to determine the EWs of emission lines above a certain strength ($\ga 13$~\AA). In fact, our method can provide more robust determinations under certain conditions. This is the case, for instance, of the galaxy that deviates on the left side of the panel. The SDSS spectrum of this galaxy, at redshift $z=0.299$, exhibits both H$\alpha$ and H$\beta$ in emission. H$\beta$ emission is too weak to be detected in ALHAMBRA with our criteria, whilst H$\alpha$, at $\lambda8526$, falls in a very noisy region of the spectrum plenty of sky emission line residuals and telluric bands, which are hardly corrected in the SDSS data. This is not the case for the ALHAMBRA data, for which the continuum is better determined and where the absolute flux excess, for this particular case, becomes more reliable. 

Finally we pay attention to the $48$ galaxies for which our code does not detect any emission line in the ALHAMBRA data with the detection limits set in this paper. We find that $35$ galaxies ($\sim75\%$) present $\log_{10}EW_\mathrm{T,\ spec} \leq 1.11$ ($\sim13$~\AA), that corresponds to our detection limit $\Delta m_\mathrm{EL} = 0.1$, i.~e. that indeed remain imperceptible under our detection constrain. About the remaining $13$ galaxies, $10$ of them have $\log_{10}EW_\mathrm{T,\ spec} > 1.11$, but distributed along different lines, being all the lines individually under the detection limit. Finally, for $3$ galaxies we do not detect properly the emission lines for two reasons: first, since one of the emission lines (in this particular case H$\alpha$) is right in between two filters and the flux is split into both of them, not fulfilling the detection criteria in any of the filters, and secondly, for a wrong determination of the $z_\mathrm{phot}$, that prevents the code from looking for emission lines in the right filters, besides the fact that a wrong redshift determination affects the quality of the derived continuum yielding a line residual under $\Delta m_\mathrm{EL} = 0.1$.

To conclude, we demonstrate that despite \textit{MUFFIT} being mainly optimized for the analysis of the stellar populations of galaxies dominated by their stellar content, it is still reliable to detect and characterise the strength of strong emission lines under certain conditions that depend on the type of multi-filter data (e.g.~filter width, signal-to-noise ratios per filter) we are working with. In future versions of the code we expect to improve the algorithms of detection of emission lines with additional techniques and criteria (e.~g, assuming intrinsic relations among lines), but this is out of the scope of the present paper.


\subsection{The stellar populations of M32}\label{sec:m32}

As a first step to test the reliability of the stellar populations derived with our code, we analyse the stellar content of M32, as this is one of the best known galaxies in terms of its resolved and unresolved stellar populations. The spectrum of M32 used for this study is taken from the compilation of \citet{Santos2002}, and it has been convolved with the ALHAMBRA filter set for the sake of this test, as if it had been observed in ALHAMBRA. Since the spectral range of this spectrum, $\lambda\lambda\ 3500$--$10\,000$~\AA, is shorter than the filter coverage of ALHAMBRA, both the bluest optical filter and the three NIR filters have been rejected from the whole fitting procedure described in Sect.~\ref{sec:bigcode}. In addition, to be able to explore the parameter space that is compatible with the best solution due to uncertainties in the photometry, and given that we have created a fake ALHAMBRA spectrum from a higher resolution spectrum, we add a synthetic error of $\sigma_{\rm AB} = 0.025$ in each filter, which is the expected error in the photometric calibration of ALHAMBRA (equivalent to a signal-to-noise ratio $\sim40$).

\begin{figure}
\resizebox{\hsize}{!}{\includegraphics{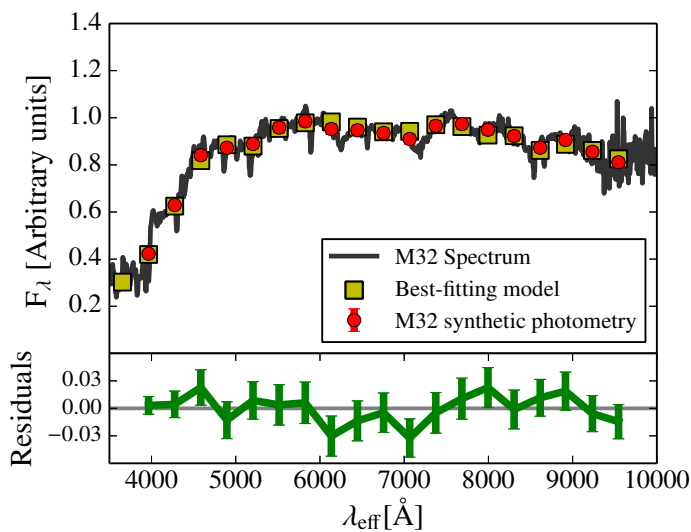}}
\caption{Spectral fitting of M32 as seen by ALHAMBRA using the MIUSCAT SSP SEDs as template models with the analysis explained above. The synthetic photometry of M32 is plotted in red, whereas the best fitting of a mixture of two SSPs to the spectrum of M32 is plotted in yellow. The bottom pannel shows the residuals of the best-fitting.}
\label{fig:m32}
\end{figure}

In Fig.~\ref{fig:m32} we present the complete spectrum of M32, in black, the M32 spectrum at the ALHAMBRA resolution, in red, and the best fitting derived from our code with a mixture of two MIUSCAT SSPs and a Kroupa IMF to the M32 ALHAMBRA spectrum, in yellow. As consequence of the spectral range ($\lambda\lambda\ 3\,500$--$10\,000$~\AA), the bluest and NIR ALHAMBRA filters have been rejected in the analysis. The obtained residuals are shown in the lower panel. 

It is clear from Fig.~\ref{fig:m32} that the best fitting derived from our code reproduces well the observed spectrum at both low and high frequencies. The best fitting solution to a single SSP, as derived from our code in the first step, corresponds to a MIUSCAT model of $3.7 \pm 1.3$~Gyr and around solar metallicity ([Fe/H]$=0.02 \pm 0.14$~dex). When the code is run completely for the mixture of two SSPs, we obtain a luminosity-weighted age of $6.8 \pm 2.2$~Gyr, a slightly sub-solar metallicity ([Fe/H]=$-0.08 \pm 0.14$~dex), and extinction $A_V = 0.28 \pm 0.08$. Looking at the individual results for the two SSPs, we derive that the spectrum of M32 is well reproduced by an intermediate age population of $2.1 \pm 0.5$~Gyr and an older population of $11.5 \pm 3.4$~Gyr. We find that the weight on the stellar mass of the young population is $\sim20\%$. Previous work, e.~g. \citet{Coelho2009} and \citet{Monachesi2012}, obtain similar results, in the sense that M32 is not composed of a unique SSP of intermediate age, but its stars were formed in at least two episodes of star formation, one ancient and the other one at intermediate ages. 

Note that, even though we are using $19$ filters (instead of $23$ in a typical ALHAMBRA photo-spectrum), we still get a good agreement between the retrieved parameters and those derived by previous work making use of detailed spectroscopic studies, showing the great power of this kind of multi-filter surveys for stellar population studies.


\subsection{Ages and metallicities of early-types in the local Universe}\label{sec:local}

Disentangling the stellar populations of early-type galaxies and their assembling histories is a key question for our understanding of galaxy evolution. However it is not our intention to address this point in this section. The aim of this section is only to explore the stellar content of a sub-sample of early-type galaxies in the nearby Universe from the ALHAMBRA survey, making use of \textit{MUFFIT}, and compare our results with previous findings in the literature. Once again, this is an additional check to assess the reliability of the stellar populations derived from our techniques. In a forthcoming paper \citetext{D\'iaz-Garc\'ia et al. 2015, in prep.} we will carry out a more complete and systematic analysis of all the galaxies in ALHAMBRA, allowing to face these and other related questions. 

Our reference work is the paper by \citet[][in the following G05]{Gallazzi2005} and, in particular, the ages and metallicities derived from spectroscopic analysis techniques for a sample of early-type galaxies located at $z < 0.22$ in SDSS. Their spectra were drawn from the SDSS DR4 \citep[$3\arcsec$ diameter fibres,][]{Adelman-McCarthy2006} spanning the full range of galaxy types (from actively star-forming to early-type galaxies), covering the range $\lambda\lambda\ 3\,800$--$9\,200$~\AA\ with a resolution of $R\sim1\,800$, and Petrosian magnitudes in the \textit{r}-band range $14.5<r<17.77$. To construct our sub-sample of early-types galaxies in ALHAMBRA we make use of the morphological catalogue provided by \citet{Povic2013}, built using the code \emph{galSVM} \citep[designed to deal with low-resolution images at low and high redshifts, ][]{Huertas2008} that, following a Bayesian approach, classifies the galaxies morphologically. This catalogue contains more than $\sim1\,500$ early-type galaxies with redshifts down to $z \la 0.5$ and a contamination lower than $10$\%, up to magnitude $F613W \leq 22$. To guarantee a fair comparison study, we only select from the above catalogue those early-type galaxies in ALHAMBRA with $z \le 0.22$, i.~e., in the same redshift interval as in G05. With this constraint in redshift, we end up with a reliable sub-sample of $\sim400$ early-type galaxies (mean signal-to-noise ratios S/N$>14$ in all cases), in which a significant part ($\sim65$\%) also reside in the RS.

\begin{figure*}
\centering
\includegraphics[width=17cm]{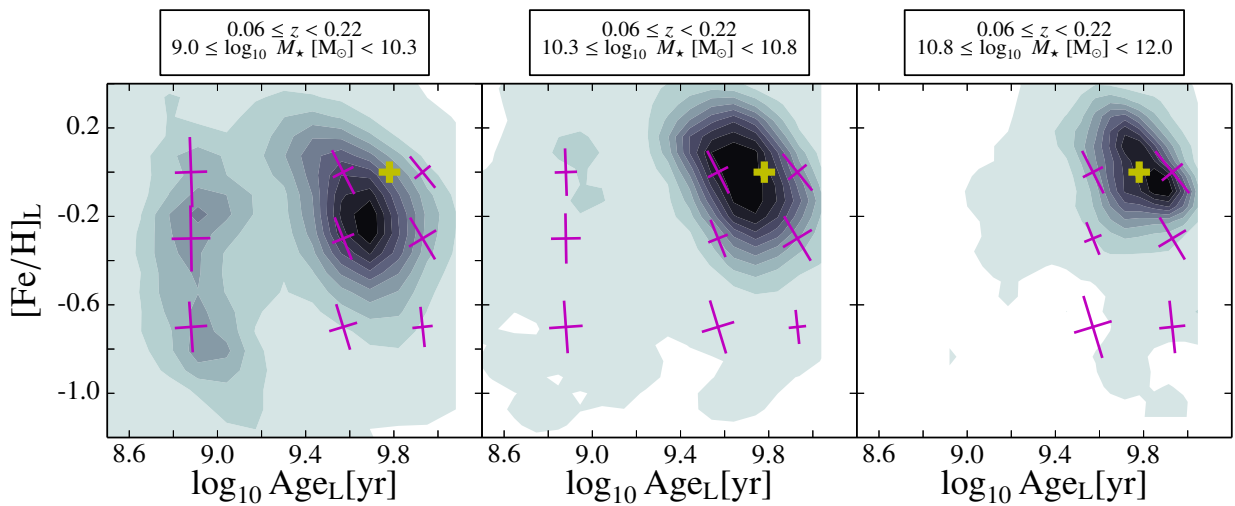}
\caption{Luminosity-weighted ages and metallicities derived from \textit{MUFFIT}, and the BC03 models, for a sub-sample of ALHAMBRA early-type galaxies at different stellar mass bins and up to $z \le 0.22$. Purple crosses illustrate the semi-axes of the degeneracy ellipses for $1-\sigma$ confidence level for the BC03 SSP models, as computed in Sect.~\ref{sec:degeneracies}, in the same age and metallicites ranges. To guide the eye, yellow crosses illustrate an age of $6$~Gyr with solar metallicity.}
\label{fig:mirjana}
\end{figure*}

Following Figure 12 in G05, in Fig.\ref{fig:mirjana} we present the density contours of our results on luminosity-weighted ages, Age$_{\rm L}$, and metallicities, [Fe/H]$_{\rm L}$, derived for our subsample of early-type galaxies in ALHAMBRA up to $z = 0.22$ using BC03 models and the photo-\textit{z} constraints provided by the Gold catalogue. As in G05, the galaxy sample is split in stellar mass bins as indicated in the top labels of Fig.\ref{fig:mirjana}. For each galaxy, rather than using the weighted values retrieved from the simulations, the whole set of results from the Monte Carlo simulations (see Sect.~\ref{sec:mc}) are included in the plot. Darker colours correspond to the ages and metallicity regions with higher population densities. Purple crosses illustrate the semi-axes of the degeneracy ellipses for $1$-$\sigma$ confidence level for the BC03 SSP models, as computed in Sect.~\ref{sec:degeneracies}, in the same age and metallicity ranges.

From a purely comparative point of view, it is highly remarkable the similarity of the age and metallicity results derived from our code and the ones presented in Figure 12 of G05 on the basis of totally different spectroscopic analysis techniques. This is an excellent prove that the reliability of the stellar population techniques presented in this paper for analysing multi-filter galaxy data is comparable to the ones employed during the last decades for the analysis of spectroscopic data, like diagnostic diagrams based on line-strength indices, at least when they were applied to large volumes of data.

From Fig.\ref{fig:mirjana} we infer that low mass early-types, $\log_{10}M_\star \la 10.3\ \mathrm{dex}$, show a bimodal distribution in their stellar populations. There is a population of younger early-types with slightly lower metallicities that does not seem to exist at higher masses. On the other hand, there is a main population of older and, on average, more metal-rich galaxies at the same stellar mass bin. The set of younger early-types of mild metallicities may be contaminated by lenticular galaxies \citep{Poggianti2001}, and that are also found in G05. Using the rest-frame colours (see Sect.~\ref{sec:kcorr}), we check that almost the totality of these "young" galaxies corresponds to galaxies that reside in the blue-cloud, and probably they are mainly composed of star-forming bulge-dominated galaxies. It is worth noticing that, despite this analysis is based on photometric data, the use of \textit{MUFFIT} and ALHAMBRA allows us to be sensitive to this population and to characterise quite well their stellar populations. Not only obtaining a great agreement with spectroscopic studies, but also opening the possibility of extending the mass limit up to lower stellar masses than typical spectroscopic surveys. We reinforce this result by repeating the analysis using the MIUSCAT SSP models instead of the BC03, getting the same result. Since the detailed analysis of the stellar populations is out of the scope of this paper, this point will be addressed in a forthcoming work \citetext{D\'iaz-Garc\'ia et al. 2015; in prep.}.

For intermediate stellar masses ($10.3 \la \log_{10} M_\star [\mathrm{M_\odot}] \la 10.8$), the "young" population tends to disappear, and consequently, the number of youngish and less metal-rich galaxies decreases, being negligible for the higher stellar masses ($10.8 \la \log_{10} M_\star [\mathrm{M_\odot}] \la 12.0$), for which there is a clear predominance of old and metal-rich galaxies. For the lowest stellar masses, the spread in age and metallicity is apparently larger than for the most massive cases. Overall, our results suggest that massive galaxies are in average more metal-rich than less massive ones \citep{Tremonti2004,Gallazzi2005,Gallazzi2006}, hence the abundance of metals in a galaxy is related or linked with its stellar mass, showing a larger spread in the low mass end (also found in G05), except for the young metal-poor population. We also observe that the mean ages of massive early-types tends to be slightly older than their less massive counterparts, and, consequently, were formed at earlier epochs (higher redshifts) than the low mass galaxies, in agreement with the "downsizing" scenario \citep{Cowie1996,Jimenez2007}. The increase in the mean ages and metallicities for massive early-types was also found in G05 (equivalent results with very similar age-metallicity relations using SDSS spectroscopy). Despite these stellar population differences are quite mild for the early-type galaxies (at least in comparison with late-type galaxies, see G05), it is worth remarking that \textit{MUFFIT}, running on the ALHAMBRA data, is still sensitive to the subtle changes in age and metallicity. On average, for the whole galaxy population in Fig.\ref{fig:mirjana}, we obtain that the increase in age from the low massive galaxies ($9.0 \la \log_{10} M_\star [\mathrm{M_\odot}] \la 10.3$) up to the most massive ones ($10.8 \la \log_{10} M_\star [\mathrm{M_\odot}] \la 12.0$) is $\sim3$~Gyr, being their mean ages $3$ and $6$~Gyr respectively. Similarly, the mean metallicity progressively increases from $-0.35$ to about solar metallicity.

We perform several tests to assess whether the bi-modality in the populations of the less massive galaxies, $9.0 \le \log_{10} M_\star < 10.3$~dex, is driven by age-metallicity degeneracies. First, we observe the degeneracies on the age-metallicity parameters for the whole sample of RS galaxies in ALHAMBRA at the parameter ranges of both distributions (Sect.~\ref{sec:degeneracies}, see Fig.~\ref{fig:degeneracies} and Table~\ref{tab:degeneracies}). For both sub-populations ($\mathrm{Age}_\mathrm{L}\sim 0.8$~Gyr with $\mathrm{[Fe/H]}_\mathrm{L}\sim -0.7$~dex; $\mathrm{Age}_\mathrm{L}\sim 4$~Gyr with solar metallicity), the degeneracy contours do not present bi-modalities and they are well constrained by a unique ellipse (illustrated by purple crosses in Fig.~\ref{fig:mirjana}). In addition, we confirm that all the Monte Carlo realizations of each individual galaxy in the left panel of Fig.~\ref{fig:mirjana} clearly belongs to only one of the two galaxy sub-populations, evidencing that degeneracies are not responsible for the younger and less metal rich population at the low mass regime. Moreover, we add noise ($\sigma_\mathrm{AB} = 0.05$--$0.20$, corresponding to a signal-to-noise ratio S/N$\sim 20$--$5$) to the high-mass galaxy sample and  analyse them again with \textit{MUFFIT} to see if there is any hint of bi-modality driven by degeneracies at low signal-to-noise regimes. Even in the worst case (S/N$\sim5$), less than $3$\% of the galaxies end up being at the young and metal poor sub-population region, not exhibiting any bimodal pattern in the distribution. We conclude that there exists a true sub-population of "young" early-type galaxies in the stellar mass regime $9.0 \le \log_{10} M_\star < 10.3$~dex, not being a consequence of parameter degeneracies and the use of probability distribution functions.


\subsection{Comparison with spectroscopic stellar-population studies}\label{sec:sdss}

A definitive step forward in the above analysis rests on the one-to-one comparison of spectroscopic galaxy ages and metallicities with the ones derived from \textit{MUFFIT}. Interestingly, there is a sub-sample of galaxies in the MPA/JHU catalogues for which individual spectroscopic estimations of ages and metallicities are provided (obtained following the methodology explained in G05), and also imaged in the ALHAMBRA fields. G05 performed an age and metallicity diagnostic method based on a simultaneous fitting to $5$ absorption line strength indices, most of them in the Lick system \citep{Gorgas1993,Worthey1994b}, constituted by age-sensitive indices like D$4000$ \citep{Balogh1999}, H$\beta$ and H$\delta_\mathrm{A}$+H$\gamma_\mathrm{A}$, and by metal-sensitive indices like $[\mathrm{Mg}_2\mathrm{Fe}]$ \citep{Bruzual2003} and $[\mathrm{Mg}\mathrm{Fe}]$' \citep{Thomas2003}, the later weakly dependent on non-solar [$\alpha$/Fe] abundances. Each set of $5$ spectral features is compared, through a $\chi^2$-test, with the values provided by a set of models randomly generated from BC03, with different bursts of star formation and different fractions relative to the total stellar mass in a velocity dispersion range, to finally construct the PDF of the parameters being the weight of each model $\propto \exp(-\chi^2/2)$ (see details in G05). By crossmatching the ALHAMBRA galaxy catalogue with the above work, we find $80$ RS galaxies (not spectroscopically classified as either \textit{BROADLINE} or \textit{AGN}) in common between both studies, with a mean signal-to-noise ratio per pixel in SDSS larger than $9$ (under this constraint, the signal-to-noise ratio of the common ALHAMBRA galaxies is larger than $18$ in all cases). We establish this as a minimum threshold to obtain meaningful stellar population results from spectroscopic diagnostics based on line-strength indices. Despite G05 stated that  metallicity is well constrained for spectra whose signal-to-noise ratio is larger than $20$, this more permissive restriction in the SDSS signal-to-noise ratio increases the number of common galaxies, allowing us to explore the age compatibility of both methods for a larger number of RS galaxies, where the age accuracy is not so affected by the signal-to-noise ratio of the SDSS spectra (these details are extensively tested in G05).

To keep the model consistency with G05, we feed \textit{MUFFIT} with the SSP models of BC03 to explore, via ALHAMBRA data, the stellar content of these $80$ RS galaxies in common. In addition, to explore the impact of different SSP models in the retrieved parameters, we repeat the same analysis with the MIUSCAT SSP models instead (see at the end of this section).

\begin{figure*}
\centering
\includegraphics[width=19cm]{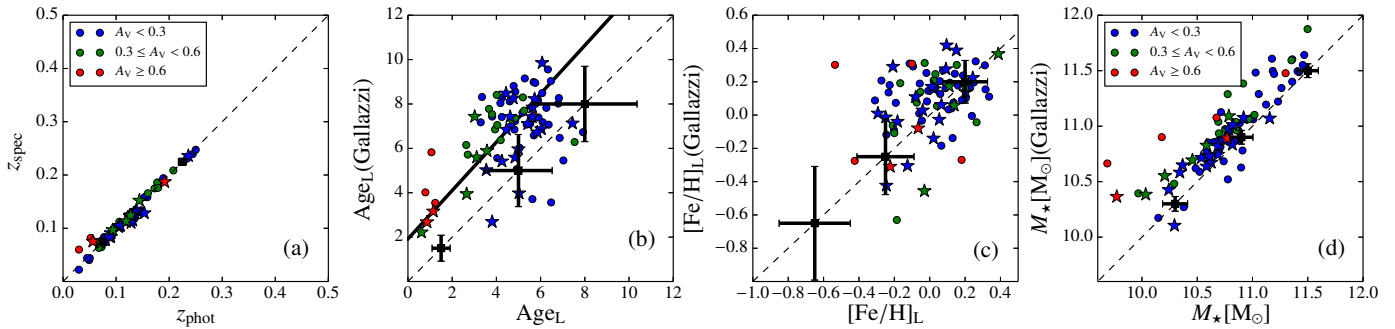}
\caption{Comparison of redshifts, ages, metallicities, and stellar masses between the spectroscopic study of \citet{Gallazzi2005}, Y--axis, and the stellar populations retrieved from ALHAMBRA with BC03 models, X--axis. In red, we present the galaxies for which our techniques stablish that may have large extinctions; in green, intermediate extinctions; and in blue, low extinctions. The dashed black line is the one-to-one relationship. The solid black line show the fitting between spectroscopic and photometric ages, accounting for the uncertainties in both measurements. The black crosses represent the average uncertainties of both techniques at different ranges. The star-shape markers are the galaxies with a radial size below $2\arcsec$.}
\label{fig:gallazzi}
\end{figure*}

Fig.~\ref{fig:gallazzi} presents a one-to-one comparison of the spectroscopic redshifts, luminosity-weighted ages, luminosity-weighted metallicities, and stellar masses given in the work by G05 for the sub-sample of $80$ SDSS galaxies in common with ALHAMBRA, and the photometric values determined from \textit{MUFFIT} using the BC03 SSP models for the same galaxies and the ALHAMBRA data. In all panels, the dashed black line represents the one-to-one relationship, and the error bars at different parameter ranges indicate the typical $1\sigma$ uncertainties in the parameters from both methods. Different colors indicate different extinction ranges ($A_V < 0.3$; $0.3 \le A_V < 0.6$; $A_V > 0.6$), as inferred from our code. In addition, to discuss aperture effects (SDSS spectra were taken in a $3\arcsec$ aperture, while ALHAMBRA photo-spectra are not restricted to a fixed aperture, which is determined by the synthetic band \emph{F814W}), star-shape markers are assigned to galaxies with apertures below $4\arcsec$ in ALHAMBRA, which are, a priori, less affected by any potential aperture bias.

Except for two galaxies, that have been removed from the plots since their ALHAMBRA photometry has been confirmed to be affected by nearby and bright stars, the spectroscopic redshifts in Fig.~\ref{fig:gallazzi}$a$ show an excellent agreement with our photo-\textit{z}, with an RMS of $\sim 0.008$. As we expect from Sect.~\ref{sec:photoz}, we rule out that any difference in the stellar populations between the two sources can be due to uncertainties in the photo-\textit{z} (see Sect.~\ref{sec:photoz_uncer}).

Concerning the luminosity-weighted age comparison, in Fig.~\ref{fig:gallazzi}$b$ we find a good qualitative agreement between both methods given the uncertainties of both methods (see black crosses in Fig.~\ref{fig:gallazzi}), in the sense that lower (higher) spectroscopic ages correspond, respectively, with lower (higher) photometric ages from \textit{MUFFIT}. Interestingly, according to \textit{MUFFIT}, young galaxies also tend to be more reddened by dust than old galaxies. From our results in this test-sample, the mean age (luminosity weighted) of the dusty galaxies ($A_V > 0.6$) is $1.8$~Gyr, increasing to $3.9$~Gyr for the intermediate extinction range, and rising up to $5.4$~Gyr for galaxies for which we retrieve low dust contents ($A_V \le 0.3$). This is an expected result, as it is well known that younger galaxies may still have remnants of gas and dust from recent star formation events, whereas older/quiescent galaxies use to have less dust content. We notice that this trend could be in principle explained by the age-extinction degeneracy (Section~\ref{sec:degeneracies}). However, if this were the case we would not find any qualitative relation with the spectroscopic ages, as the line indices, by construction, are not significantly affected by extinction. In addition, we emphasize that the galaxies in ALHAMBRA for which \textit{MUFFIT} retrieves young stellar populations, for ages down to $2.6$~Gyr, are also classified as \textit{STARFORMING} in SDSS. This also supports the idea that the retrieved extinctions with \textit{MUFFIT} are very robust and they are not dominated by the degeneracy with age, as star-forming galaxies may present young populations with significant amounts of dust. In fact, previous similar work \citep[e.~g.][]{Fontana2006,Pozzetti2007,Ilbert2010} assumed in their codes that models with large extinctions are only allowed for star-forming galaxies. The key point out of this result is that \textit{MUFFIT} is able to retrieve intrinsically the extinction of the stellar populations without assuming any prior on the models or the galaxy type. Despite the qualitative agreement between \textit{MUFFIT} and spectroscopic ages, with an RMS of $\sim 1.6$~Gyr, we also notice an offset between the ages of the two samples, with the spectroscopic ages being $\sim2$~Gyr older than the ones derived from our code for the ALHAMBRA data. We discuss on possible reasons for this offset at the end of this section.

In Fig.~\ref{fig:gallazzi}$c$ we can see that the metallicities present a qualitative good agreement (despite the metallicity is more affected by uncertainties, see black crosses in Fig.~\ref{fig:gallazzi}), with an RMS of $\sim 0.22$~dex, although there is also a very small shift in the sense that our retrieved metallicities in ALHAMBRA tend to be smaller ($\Delta[\mathrm{Fe/H}] \sim-0.08$~dex) than the spectroscopic ones derived for SDSS galaxies in G05. It is noticeable that the metallicities of ALHAMBRA galaxies within an aperture of $4\arcsec$ (star-shape markers) present a better agreement with the spectroscopic measurements ($\Delta[\mathrm{Fe/H}] \la -0.05$~dex) than the galaxies with larger apertures (dot markers; $\Delta[\mathrm{Fe/H}] \sim -0.15$~dex). As it is shown in G05, aperture effects and typical metallicity gradients can lead to up to $\sim0.15$--$0.20$~dex differences in metallicity for galaxies with $\ga 10^{10} \mathrm{M_\odot}$. This reinforces the consistency and good agreement between both metallicity predictions (from SDSS and ALHAMBRA), since \textit{MUFFIT} retrieves on average lower metallicities with respect to SDSS, and not larger, with a similar difference to those measured by G05 ($\sim 0.17$~dex) owing to possible aperture effects. Interestingly, as it is pointed out by G05, a signal-to-noise ratio larger than $20$ is required for a reliable constraint of the metallicity in SDSS spectra. Unfortunately, the adopted signal-to-noise ratio limit substantially restricts our sample, excluding low luminosity galaxies with potential sub-solar metallicities. Moreover, in G05 it is also mentioned that low-metallicity galaxies are more affected by uncertainties coming from their weak absorption features of $[\mathrm{Mg}_2\mathrm{Fe}]$ and $[\mathrm{Mg}\mathrm{Fe}]$'. Indeed, in our sub-sample, the lower the metallicity the larger the dispersion in the spectroscopic metallicity, as illustrated by the error bars. Instead, the metallicities provided by \textit{MUFFIT} using ALHAMBRA data at the same regime are slightly better constrained, as, in the end, it is the overall stellar continuum what mainly determines the retrieved stellar populations.

Fig.~\ref{fig:gallazzi}$d$ exhibits a good agreement between the stellar masses of the two methods, with an RMS of $\sim 0.19$~dex, despite there is an offset of $\sim 0.18$~dex. This offset in the stellar mass can be mainly explained by the systematic differences of $\sim 2$~Gyr between the ages of both methods, as this implies a shift in the mass-luminosity relation in the sense that older galaxies, at the same apparent magnitude, are also more massive galaxies.

Before concluding, we aim at investigating on the potential origin of the $\sim2$~Gyr offset in age derived in Fig.~\ref{fig:gallazzi}$b$. There are several potential reasons that could explain this offset: 
i) The age-extinction degeneracy; ii) The way to compute luminosity-weighted ages; iii) Aperture effects; and 
iv) Intrinsic systematic differences between both analysis techniques.

i) Unlike SED fitting techniques, absorption line-strength indices are basically not sensitive to extinction, as they are defined in short wavelength ranges. If the ages derived from \textit{MUFFIT} were severely affected by the age-extinction degeneracy, we would expect galaxies with very low extinction values, $A_V < 0.05$ to present a better agreement in the age comparison of Fig.~\ref{fig:gallazzi}$b$. However this is not the case. By exploring the ages and metallicities of the galaxies with low extinction values ($A_V < 0.05$, according to \textit{MUFFIT} and the ALHAMBRA data), checking that both metallicities, spectroscopic and photometric, remain in agreement without great differences, we obtain that the age difference is still $\sim 2$~Gyr. This discards the potential impact of the age-metallicity degeneracy, as well as the influence of using different extinction laws.

ii) In G05, luminosity-weighted ages are computed according to the total \textit{r}-band flux, whereas in \textit{MUFFIT} this is done using the whole flux in all the ALHAMBRA bands. To explore whether this normalization difference could drive the age offset, we have recomputed the \textit{MUFFIT} luminosity-weighted ages using the ALHAMBRA \textit{F644W} band, which has the most similar effective wavelength to the SDSS \textit{r}-band. The results are essentially the same, hence not explaining the observed age offset.

iii) It is well known that early-type galaxies may show radial variations of their stellar population properties, showing gradients in metallicity and/or age. We already discussed above how the combination of different photometric apertures and the existence of metallicity gradients has an impact on the metallicity comparison of Fig.~\ref{fig:gallazzi}$c$). It is worth noting however that, in general, age gradients use to be shallower than metallicity gradients \citep{Wu2005,Sanchezblazquez2007,LaBarbera2012,Eigenthaler2013}, so aperture effects are expected to be smaller as well. Nevertheless, shallow gradients in all the parameters can also be found \citetext{e.~g. Gonz\'alez-Delgado et al.~2015, in prep.}. To assess this effect, we focus on galaxies whose photometric apertures in ALHAMBRA are down to $4\arcsec$ (star-shape markers in Fig.~\ref{fig:gallazzi}), not  far from the SDSS fiber aperture. We observe that the age offset is significant even for these galaxies, hence aperture effects are rejected to explain the age offset too. 

\begin{figure}
\resizebox{\hsize}{!}{\includegraphics{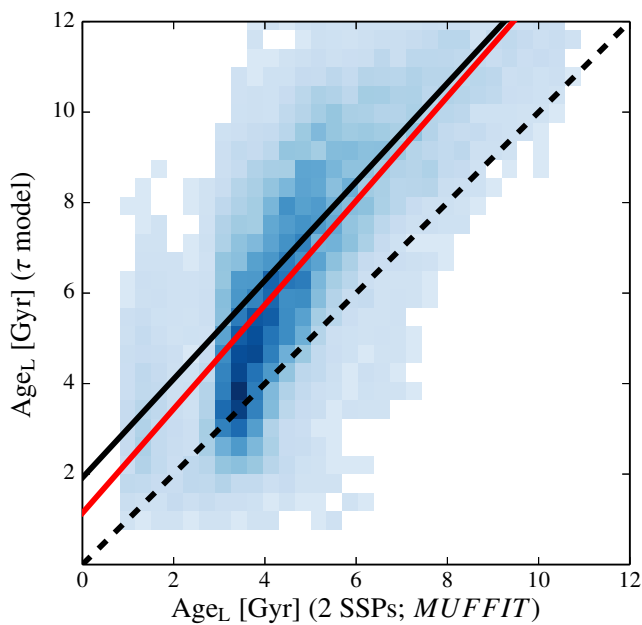}}
\caption{Comparison between the luminosity-weighted ages of a subset of RS $\tau$-models from SSAG and the ones derived by \textit{MUFFIT} for the same models employing a mixture of 2 SSPs. The dashed black line represents the one-to-one relationship, while the red line is the simple linear regression of the data points. The solid black line illustrates the fitting between the ages provided by G05 and the ones provided by \textit{MUFFIT} using ALHAMBRA data (see Fig.~\ref{fig:gallazzi}b).}
\label{fig:tau}
\end{figure}

iv) After the negative results of the three previous tests, the existence of intrinsic systematic differences between the two methods seems to be the most plausible reason for the different absolute values of the derived ages. The discrepancies between the analysis of spectral features versus colours, together with the assumptions of different SFHs (exponentially declining tau models in G05, versus a mixture of young$+$old SSPs in this work), may be responsible for the age offset. To shed light in this last item, we aim at constraining a purely mathematical problem: the potential differences between the luminosity ages derived from parametric $\tau$-models and the ones derived from a non-parametric mixture of two SSPs. To do this, we make use of the set of $\tau$-models \textit{Synthetic Spectral Atlas of Galaxies} \citep[SSAG, ][priv.~comm.]{Gladis2014} very similar to the $\tau$-models employed in G05. SSAG models are based in the recipes described in \citet{Chen2012} and have been constructed following a exponentially declining SFH and BC03 models, that may randomly suffer an instantaneous and random burst during different periods of time. The SSAG also includes intrinsic extinctions, following the dust model of \citet{Charlot2000}, and different velocity dispersions. To create a sub-sample of RS galaxies, we select all the SSAG galaxy models whose colours satisfy $U-V \ga 2.0$ (AB-system). After convolving SSAG models with the ALHAMBRA filter set, we run \textit{MUFFIT} using as input the same BC03 models, in concordance with SSAG, and compare derived luminosity-weighted ages with the input SSAG ones. The result of the comparison is exhibited in Fig.~\ref{fig:tau}. There appears a systematic offset between ages of $\Delta \mathrm{Age_L} \sim 1.8\pm1.7$~Gyr (red line in Fig.~\ref{fig:tau}), which fully explains (qualitatively and quantitatively) the $\sim 2\pm1.6$~Gyr offset (solid black line in Fig.~\ref{fig:tau}) found in the previous comparison between the spectroscopic ages of G05 and the ones retrieved using \textit{MUFFIT} and ALHAMBRA data, as due to the mathematical differences in the diagnostic input models (mixture of SSP models vs exponentially declining SFH models).

\begin{figure*}
\centering
\includegraphics[width=19cm]{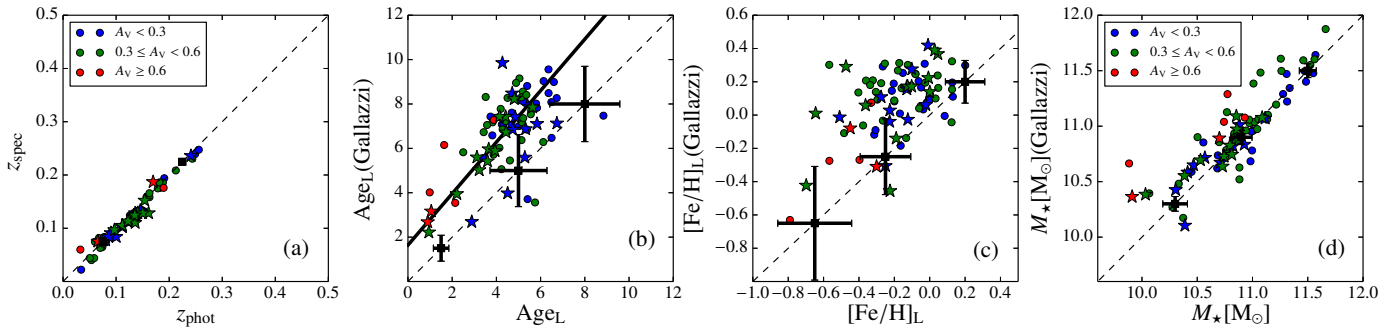}
\caption{As Fig.~\ref{fig:gallazzi} but using the MIUSCAT SSP models instead of the BC03 ones.}
\label{fig:gallazzi_mm}
\end{figure*}

Finally, we have also investigated the impact of using a different set of SSP models for the stellar population comparison. In Fig.~\ref{fig:gallazzi_mm} we present the same comparison of Fig.~\ref{fig:gallazzi}, but in this case having used the MIUSCAT SSP models, instead of BC03, to analyse the ALHAMBRA data with \textit{MUFFIT}. Except for a slightly larger difference in metallicity, in the sense that MIUSCAT models tend to predict lower metallicities, the rest of parameters compare similarly to Fig.~\ref{fig:gallazzi}. This is important to assess the impact of different SSP models in the absolute values of the derived parameters.


\section{Summary and conclusions}\label{sec:conclusions}

The arrival of present and future large-scale multi-filter surveys (e.g. COMBO-17, COSMOS, ALHAMBRA, SHARDS, J-PLUS, and J-PAS) promises the availability of formidable datasets for many purposes in Cosmology and Astrophysics. These photometric surveys, based on the mapping of different regions of the sky with a set of contiguous intermediate/narrow band filters, provide low resolution photo-spectra for each region of the sky (hence performing like a low resolution IFU with PSF-limited spatial resolution), with the survey depth as the only selection criterion and without the typical spectroscopic uncertainties in the flux calibration. This opens an unprecedented way to progress in our understanding about galaxy evolution through the study of millions of homogeneous galaxy SEDs, both spatially-resolved in the more nearby Universe and integrated.

This paper is devoted to presenting \textit{MUFFIT} (\textit{MUlti-Filter FITting in photometric surveys}) a generic code specifically designed for analysing the stellar content of galaxies with available multi-filter data (dealing with the technical peculiarities and the big amount of high-quality photometric data available in multi-filter surveys), as well as to show its functionalities, set the accuracy and typical uncertainties in the retrieved stellar population parameters, and ultimately test it with real data. In this sense, we make use of the ALHAMBRA database as a test-bench for \textit{MUFFIT}, not with the aim of performing a thorough stellar population analysis of the galaxies in ALHAMBRA (which constitutes the matter of the next papers of this series), but to compare the stellar population results derived from \textit{MUFFIT} with similar studies in the literature, allowing us to assess on its reliability and on the feasibility of this kind of techniques to accurately explore the stellar content of galaxies.

In the following items the main conclusions of this work are summarized:

\begin{itemize}
\item Using as input a set of SSP models that explores different stellar population parameters, \textit{MUFFIT} builds photometric predictions of bands at different redshifts and extinctions. For the present work, the stellar population parameters that are considered are just age and metallicity, although in a general case even the IMF slope and the $\alpha$-enhancement can also be retrieved if SSP models properly account for them. In addition, the survey photometry is corrected of MW dust effects, as the colour terms introduced by MW dust may play an important role not only in the stellar masses derived using SED-fitting techniques, but also in the retrieved stellar population parameters.

\item \textit{MUFFIT} compares the multi-filter fluxes of a given galaxy with the photometric predictions of a reasonable mixture of two SSPs, respectively one younger and one older than the mean age provided by a single SSP fitting. The mixture of two SSPs determined by the last prior, specific for each individual galaxy, is a relevant improvement with respect to the fitting of one SSP only, since it better represents a population mainly composed by an old component that suffer a later burst or star-formation episode. The stellar population parameters (in this work the age and metallicity weighted by both luminosity and mass, extinction, redshift, and stellar mass) provided by \textit{MUFFIT} are constrained by the use of an error-weighted $\chi^2$-test. During the fitting process, \textit{MUFFIT} removes those bands that are affected by emission lines, improving the quality of the fitting and restricting the plausible redshift space, as in a general case the redshift of the galaxy is treated as another free parameter to be determined. \textit{MUFFIT} is not limited to providing the parameters of the best fitting model, but also explores the parameter space using the proper photometric uncertainties in each band by a Monte Carlo method, reinforcing the parameter predictions as it provides their statistical uncertainties too. In addition, \textit{MUFFIT} also computes and provides the $k$-corrections of each galaxy from the same mixture of models at rest-frame. 

\item Specifically for the ALHAMBRA data, we study the intrinsic uncertainties in redshift, extinction, age, metallicity and stellar mass that appear when diagnosed by \textit{MUFFIT}. Using the typical distribution of errors for the RS galaxies in ALHAMBRA, we construct mock galaxies with an average $S/N$ per filter of $20$, obtaining typical uncertainties (RMS) of $\sigma_z \sim 0.01$, $\sigma_{A_V} \sim 0.11$, $\sigma_\mathrm{Age} \sim 0.10$~dex, $\sigma_\mathrm{[Fe/H]} \sim 0.16$~dex, and $\sigma_{M_\star} \sim 0.08$~dex. In no case there are systematic errors that are statistically significant. 

\item Despite \textit{MUFFIT} is not a generic photo-$z$ code, using as input the redshift PDFs provided by external photo-$z$ codes, \textit{MUFFIT} returns fine-tuned redshift values whose accuracy is improved by $\sim 10-20\%$. We also obtain that the photo-$z$ accuracy reached in ALHAMBRA, $\sigma_\mathrm{NMAD} \la 0.009$, presents a negligible impact on the main stellar population parameters retrieved using \textit{MUFFIT}, being more crucial the typical uncertainties in the photometry. 

\item We have studied with \textit{MUFFIT} the age-metallicity-extinction degeneracy of the ALHAMBRA data at different parameter ranges, and having fixed the IMF slope. The age-extinction anti correlation is present in all ranges of age and metallicity. However, the well known age-metallicity anti correlation may turn into a positive correlation for young and/or metal poor populations due to the role of the extinction in reddening the spectral energy distributions of galaxies.

\item The stellar mass predictions provided by \textit{MUFFIT} for a common sample of RS galaxies in ALHAMBRA are in wonderful agreement with the stellar masses computed for the same galaxies in COSMOS. The dispersion of the comparison, with an RMS of $\Delta \log_{10} M_\star \sim 0.15$~dex, can be fully explained by the intrinsic uncertainties of both methods.

\item \textit{MUFFIT} offers a reliable way to explore emission lines in multi-filter surveys. Using a set of emission line galaxies in common between SDSS and ALHAMBRA, we demonstrate that the residuals provided by \textit{MUFFIT} for the filters affected by emission lines in ALHAMBRA are correlated with the strengths of the main emission lines.

\item The age-metallicity loci provided by \textit{MUFFIT} for a sample of $z \le 0.22$ early-type galaxies in ALHAMBRA at different stellar mass bins are in very good agreement with the ones determined from SDSS data on the basis of spectroscopic diagnostics. When we analyse the stellar content of these galaxies in ALHAMBRA using their photometric data and \textit{MUFFIT}, our results point out that the more massive early-types ($\ga 10^{11} M_\star [\mathrm{M_\sun}]$) were formed in an earlier epoch than their low-mass counterparts ($\la 10^{10} M_\star [\mathrm{M_\sun}]$) with a larger content in metals, being these differences $\Delta \mathrm{Age} \sim 3 $~Gyr and $\Delta [\mathrm{Fe/H}] \sim 0.3$~dex. This result agrees with the "downsizing" scenario as well.

\item For a subsample of galaxies in common between ALHAMBRA and SDSS, a one-to-one comparison between the redshifts, ages, metallicities, and stellar masses derived spectroscopically for the SDSS data \citep[provided by][]{Gallazzi2005} and those determined from \textit{MUFFIT} and ALHAMBRA reveal good qualitative agreements in all the parameters given the uncertainties of both methods, with typical RMS for the distribution of differences between both diagnostics of $\sigma_z^{\ SDSS} \sim 0.008$, $\sigma_\mathrm{Age}^{\ SDSS} \sim 1.6$~Gyr, $\sigma_\mathrm{[Fe/H]}^{\ SDSS} \sim 0.2$~dex, and $\sigma_{M_\star}^{\ SDSS} \sim 0.19$~dex; hence reinforcing the strengths of multi-filter galaxy data and optimized analysis techniques, like \textit{MUFFIT}, to conduct reliable stellar population studies. Despite the qualitative agreement between ages, in the sense that young (old) spectroscopic ages in SDSS are also found to be young (old) photometric ages in ALHAMBRA using \textit{MUFFIT}, there exists a systematic difference of $\sim 2$~Gyr between the two samples that is explained by the differences of using mixtures of SSPs instead of $\tau$-models. Despite the good agreement between metallicities, it is noticeable that the metallicities of ALHAMBRA galaxies within an aperture of $\le 4\arcsec$ present a better agreement with the spectroscopic measurements ($\Delta [\mathrm{Fe/H}] \la 0.05$~dex) than the galaxies with larger apertures ($\Delta [\mathrm{Fe/H}] \la 0.15$~dex), pointing towards the possibility that aperture differences between SDSS and ALHAMBRA and the existence of metallicity gradients drive the observed differences (in agreement with G05). There is also a good agreement between stellar masses, with a minor shift of $\sim 0.18$~dex that can be explained by the observed offset in age.
\end{itemize}

To conclude, we demonstrate that \textit{MUFFIT} is a reliable stellar population code for multi-filter galaxy data, which is suited and optimized to analyse the stellar content of galaxies in ALHAMBRA-like surveys. This opens a new way to explore and address stellar population studies of galaxies with multiple photometric bands or colours, as long as the effective spectral resolution is at least the one of ALHAMBRA, allowing us to accurately extract the stellar content of thousands of galaxies at higher redshifts, benefited by the large-number statistics in comparison with typical spectroscopic datasets at same redshift. With the arrival of the new generation large-scale multi-filter surveys, like J-PLUS and J-PAS, codes like \textit{MUFFIT} will contribute greatly to  shed light in our understanding about the formation and evolution of galaxies.

%
\begin{acknowledgements}

We thank referee for his/her valuable suggestions and comments, which have reinforced part of our results. LADG acknowledges support from the "Caja Rural de Teruel" to develop this research. AJC is a Ram\'on y Cajal Fellow of the Spanish Ministry of Science and Innovation. This work has been supported by the “Programa Nacional de Astronom\'ia y Astrof\'isica” of the Spanish Ministry of Economy and Competitiveness (MINECO) under grant AYA2012-30789, as well as from FEDER funds and the Government of Arag\'on, through the Research Group E103. LADG also thanks the Mullard Space Science Laboratory (MSSL) and Royal Astronomical Society (RAS) for bringing the opportunity to support and develop part of this research in collaboration with IF. MINECO grants AYA2010-15081, AYA2010-15169, AYA2010-22111-C03-01, AYA2010-22111-C03-02, AYA2011-29517-C03-01, AYA2013-40611-P, AYA2013-42227-P, AYA2013-43188-P, AYA2013-48623-C2-1, AYA2013-48623-C2-2 and AYA2014-58861-C3-1 are also acknowledged, together with Generalitat Valenciana projects Prometeo 2009/064 and PROMETEOII/2014/060, and Junta de Andaluc\'ia grants TIC114, JA2828, and P10-FQM-6444. MP acknowledges financial support from JAE-Doc program of the Spanish National Research Council (CSIC), co-funded by the European Social Fund. We thank N. Gruel for his initial advices in synthetic photometry. The Max Planck Institute for Astrophysics and the Johns Hopkins University are acknowledged for making publicly available their catalogues (MPA/JHU catalogues) with physical properties for galaxies in SDSS employed in this work. Through this research, we make use of the \texttt{Matplotlib} package \citep{Hunter2007}, a 2D graphics package used for \texttt{Python} that is designed for interactive scripting and quality image generation.

\end{acknowledgements}


\bibliographystyle{aa}
\bibliography{sp_alh_v3.5}

\begin{thebibliography}{128}
\expandafter\ifx\csname natexlab\endcsname\relax\def\natexlab#1{#1}\fi

\bibitem[{{Abazajian} {et~al.}(2009){Abazajian}, {Adelman-McCarthy},
  {Ag{\"u}eros}, {Allam}, {Allende Prieto}, {An}, {Anderson}, {Anderson},
  {Annis}, {Bahcall}, \& et~al.}]{Abazajian2009}
{Abazajian}, K.~N., {Adelman-McCarthy}, J.~K., {Ag{\"u}eros}, M.~A., {et~al.}
  2009, \apjs, 182, 543

\bibitem[{{Adelman-McCarthy} {et~al.}(2006){Adelman-McCarthy}, {Ag{\"u}eros},
  {Allam}, {Anderson}, {Anderson}, {Annis}, {Bahcall}, {Baldry}, {Barentine},
  {Berlind}, {Bernardi}, {Blanton}, {Boroski}, {Brewington}, {Brinchmann},
  {Brinkmann}, {Brunner}, {Budav{\'a}ri}, {Carey}, {Carr}, {Castander},
  {Connolly}, {Csabai}, {Czarapata}, {Dalcanton}, {Doi}, {Dong}, {Eisenstein},
  {Evans}, {Fan}, {Finkbeiner}, {Friedman}, {Frieman}, {Fukugita}, {Gillespie},
  {Glazebrook}, {Gray}, {Grebel}, {Gunn}, {Gurbani}, {de Haas}, {Hall},
  {Harris}, {Harvanek}, {Hawley}, {Hayes}, {Hendry}, {Hennessy}, {Hindsley},
  {Hirata}, {Hogan}, {Hogg}, {Holmgren}, {Holtzman}, {Ichikawa}, {Ivezi{\'c}},
  {Jester}, {Johnston}, {Jorgensen}, {Juri{\'c}}, {Kent}, {Kleinman}, {Knapp},
  {Kniazev}, {Kron}, {Krzesinski}, {Kuropatkin}, {Lamb}, {Lampeitl}, {Lee},
  {Leger}, {Lin}, {Long}, {Loveday}, {Lupton}, {Margon},
  {Mart{\'{\i}}nez-Delgado}, {Mandelbaum}, {Matsubara}, {McGehee}, {McKay},
  {Meiksin}, {Munn}, {Nakajima}, {Nash}, {Neilsen}, {Newberg}, {Newman},
  {Nichol}, {Nicinski}, {Nieto-Santisteban}, {Nitta}, {O'Mullane}, {Okamura},
  {Owen}, {Padmanabhan}, {Pauls}, {Peoples}, {Pier}, {Pope}, {Pourbaix},
  {Quinn}, {Richards}, {Richmond}, {Rockosi}, {Schlegel}, {Schneider},
  {Schroeder}, {Scranton}, {Seljak}, {Sheldon}, {Shimasaku}, {Smith}, {Smol{\v
  c}i{\'c}}, {Snedden}, {Stoughton}, {Strauss}, {SubbaRao}, {Szalay},
  {Szapudi}, {Szkody}, {Tegmark}, {Thakar}, {Tucker}, {Uomoto}, {Vanden Berk},
  {Vandenberg}, {Vogeley}, {Voges}, {Vogt}, {Walkowicz}, {Weinberg}, {West},
  {White}, {Xu}, {Yanny}, {Yocum}, {York}, {Zehavi}, {Zibetti}, \&
  {Zucker}}]{Adelman-McCarthy2006}
{Adelman-McCarthy}, J.~K., {Ag{\"u}eros}, M.~A., {Allam}, S.~S., {et~al.} 2006,
  \apjs, 162, 38

\bibitem[{{Aparicio Villegas} {et~al.}(2010){Aparicio Villegas}, {Alfaro},
  {Cabrera-Ca{\~n}o}, {Moles}, {Ben{\'{\i}}tez}, {Perea}, {del Olmo},
  {Fern{\'a}ndez-Soto}, {Crist{\'o}bal-Hornillos}, {Husillos}, {Aguerri},
  {Broadhurst}, {Castander}, {Cepa}, {Cervi{\~n}o}, {Gonz{\'a}lez Delgado},
  {Infante}, {M{\'a}rquez}, {Masegosa}, {Mart{\'{\i}}nez}, {Prada}, {Quintana},
  \& {S{\'a}nchez}}]{Aparicio2010}
{Aparicio Villegas}, T., {Alfaro}, E.~J., {Cabrera-Ca{\~n}o}, J., {et~al.}
  2010, \aj, 139, 1242

\bibitem[{{Arnouts} {et~al.}(2013){Arnouts}, {Le Floc'h}, {Chevallard},
  {Johnson}, {Ilbert}, {Treyer}, {Aussel}, {Capak}, {Sanders}, {Scoville},
  {McCracken}, {Milliard}, {Pozzetti}, \& {Salvato}}]{Arnouts2013}
{Arnouts}, S., {Le Floc'h}, E., {Chevallard}, J., {et~al.} 2013, \aap, 558, A67

\bibitem[{{Arnouts} {et~al.}(2002){Arnouts}, {Moscardini}, {Vanzella},
  {Colombi}, {Cristiani}, {Fontana}, {Giallongo}, {Matarrese}, \&
  {Saracco}}]{Arnouts2002}
{Arnouts}, S., {Moscardini}, L., {Vanzella}, E., {et~al.} 2002, \mnras, 329,
  355

\bibitem[{{Baldry} {et~al.}(2004){Baldry}, {Glazebrook}, {Brinkmann},
  {Ivezi{\'c}}, {Lupton}, {Nichol}, \& {Szalay}}]{Baldry2004}
{Baldry}, I.~K., {Glazebrook}, K., {Brinkmann}, J., {et~al.} 2004, \apj, 600,
  681

\bibitem[{{Balogh} {et~al.}(1999){Balogh}, {Morris}, {Yee}, {Carlberg}, \&
  {Ellingson}}]{Balogh1999}
{Balogh}, M.~L., {Morris}, S.~L., {Yee}, H.~K.~C., {Carlberg}, R.~G., \&
  {Ellingson}, E. 1999, \apj, 527, 54

\bibitem[{{Baum}(1959)}]{Baum1959}
{Baum}, W.~A. 1959, \pasp, 71, 106

\bibitem[{{Bell} {et~al.}(2004){Bell}, {Wolf}, {Meisenheimer}, {Rix}, {Borch},
  {Dye}, {Kleinheinrich}, {Wisotzki}, \& {McIntosh}}]{Bell2004}
{Bell}, E.~F., {Wolf}, C., {Meisenheimer}, K., {et~al.} 2004, \apj, 608, 752

\bibitem[{{Benitez} {et~al.}(2014){Benitez}, {Dupke}, {Moles}, {Sodre},
  {Cenarro}, {Marin-Franch}, {Taylor}, {Cristobal}, {Fernandez-Soto}, {Mendes
  de Oliveira}, {Cepa-Nogue}, {Abramo}, {Alcaniz}, {Overzier},
  {Hernandez-Monteagudo}, {Alfaro}, {Kanaan}, {Carvano}, {Reis}, {Martinez
  Gonzalez}, {Ascaso}, {Ballesteros}, {Xavier}, {Varela}, {Ederoclite},
  {Vazquez Ramio}, {Broadhurst}, {Cypriano}, {Angulo}, {Diego}, {Zandivarez},
  {Diaz}, {Melchior}, {Umetsu}, {Spinelli}, {Zitrin}, {Coe}, {Yepes}, {Vielva},
  {Sahni}, {Marcos-Caballero}, {Shu Kitaura}, {Maroto}, {Masip}, {Tsujikawa},
  {Carneiro}, {Gonzalez Nuevo}, {Carvalho}, {Reboucas}, {Carvalho}, {Abdalla},
  {Bernui}, {Pigozzo}, {Ferreira}, {Chandrachani Devi}, {Bengaly}, {Campista},
  {Amorim}, {Asari}, {Bongiovanni}, {Bonoli}, {Bruzual}, {Cardiel}, {Cava},
  {Cid Fernandes}, {Coelho}, {Cortesi}, {Delgado}, {Diaz-Garcia}, {Espinosa},
  {Galliano}, {Gonzalez-Serrano}, {Falcon-Barroso}, {Fritz}, {Fernandes},
  {Gorgas}, {Hoyos}, {Jimenez-Teja}, {Lopez-Aguerri}, {Lopez-San Juan},
  {Mateus}, {Molino}, {Novais}, {OMill}, {Oteo}, {Perez-Gonzalez}, {Poggianti},
  {Proctor}, {Ricciardelli}, {Sanchez-Blazquez}, {Storchi-Bergmann}, {Telles},
  {Schoennell}, {Trujillo}, {Vazdekis}, {Viironen}, {Daflon},
  {Aparicio-Villegas}, {Rocha}, {Ribeiro}, {Borges}, {Martins}, {Marcolino},
  {Martinez-Delgado}, {Perez-Torres}, {Siffert}, {Calvao}, {Sako}, {Kessler},
  {Alvarez-Candal}, {De Pra}, {Roig}, {Lazzaro}, {Gorosabel}, {Lopes de
  Oliveira}, {Lima-Neto}, {Irwin}, {Liu}, {Alvarez}, {Balmes}, {Chueca},
  {Costa-Duarte}, {da Costa}, {Dantas}, {Diaz}, {Fabregat}, {Ferrari},
  {Gavela}, {Gracia}, {Gruel}, {Gutierrez}, {Guzman}, {Hernandez-Fernandez},
  {Herranz}, {Hurtado-Gil}, {Jablonsky}, {Laporte}, {Le Tiran}, {Licandro},
  {Lima}, {Martin}, {Martinez}, {Montero}, {Penteado}, {Pereira}, {Peris},
  {Quilis}, {Sanchez-Portal}, {Soja}, {Solano}, {Torra}, \&
  {Valdivielso}}]{Benitez2014}
{Benitez}, N., {Dupke}, R., {Moles}, M., {et~al.} 2014, ArXiv e-prints

\bibitem[{{Ben{\'{\i}}tez} {et~al.}(2009){Ben{\'{\i}}tez}, {Moles}, {Aguerri},
  {Alfaro}, {Broadhurst}, {Cabrera-Ca{\~n}o}, {Castander}, {Cepa},
  {Cervi{\~n}o}, {Crist{\'o}bal-Hornillos}, {Fern{\'a}ndez-Soto}, {Gonz{\'a}lez
  Delgado}, {Infante}, {M{\'a}rquez}, {Mart{\'{\i}}nez}, {Masegosa}, {Del
  Olmo}, {Perea}, {Prada}, {Quintana}, \& {S{\'a}nchez}}]{Benitez2009}
{Ben{\'{\i}}tez}, N., {Moles}, M., {Aguerri}, J.~A.~L., {et~al.} 2009, \apjl,
  692, L5

\bibitem[{{Bernardi} {et~al.}(2006){Bernardi}, {Nichol}, {Sheth}, {Miller}, \&
  {Brinkmann}}]{Bernardi2006}
{Bernardi}, M., {Nichol}, R.~C., {Sheth}, R.~K., {Miller}, C.~J., \&
  {Brinkmann}, J. 2006, \aj, 131, 1288

\bibitem[{{Bessell}(2005)}]{Bessel2005}
{Bessell}, M.~S. 2005, \araa, 43, 293

\bibitem[{{Bolzonella} {et~al.}(2000){Bolzonella}, {Miralles}, \&
  {Pell{\'o}}}]{Bolzonella2000}
{Bolzonella}, M., {Miralles}, J.-M., \& {Pell{\'o}}, R. 2000, \aap, 363, 476

\bibitem[{{Brammer} {et~al.}(2008){Brammer}, {van Dokkum}, \&
  {Coppi}}]{Brammer2008}
{Brammer}, G.~B., {van Dokkum}, P.~G., \& {Coppi}, P. 2008, \apj, 686, 1503

\bibitem[{{Brinchmann} {et~al.}(2004){Brinchmann}, {Charlot}, {White},
  {Tremonti}, {Kauffmann}, {Heckman}, \& {Brinkmann}}]{Brinchmann2004}
{Brinchmann}, J., {Charlot}, S., {White}, S.~D.~M., {et~al.} 2004, \mnras, 351,
  1151

\bibitem[{{Bruzual} \& {Charlot}(2003)}]{Bruzual2003}
{Bruzual}, G. \& {Charlot}, S. 2003, \mnras, 344, 1000

\bibitem[{{Burstein} {et~al.}(1984){Burstein}, {Faber}, {Gaskell}, \&
  {Krumm}}]{Burstein1984}
{Burstein}, D., {Faber}, S.~M., {Gaskell}, C.~M., \& {Krumm}, N. 1984, \apj,
  287, 586

\bibitem[{{Calzetti}(1997)}]{Calzetti1997}
{Calzetti}, D. 1997, in American Institute of Physics Conference Series, Vol.
  408, American Institute of Physics Conference Series, ed. W.~H. {Waller},
  403--412

\bibitem[{{Calzetti} {et~al.}(2000){Calzetti}, {Armus}, {Bohlin}, {Kinney},
  {Koornneef}, \& {Storchi-Bergmann}}]{Calzetti2000}
{Calzetti}, D., {Armus}, L., {Bohlin}, R.~C., {et~al.} 2000, \apj, 533, 682

\bibitem[{{Cardelli} {et~al.}(1989){Cardelli}, {Clayton}, \&
  {Mathis}}]{Cardelli1989}
{Cardelli}, J.~A., {Clayton}, G.~C., \& {Mathis}, J.~S. 1989, \apj, 345, 245

\bibitem[{{Cassata} {et~al.}(2007){Cassata}, {Guzzo}, {Franceschini},
  {Scoville}, {Capak}, {Ellis}, {Koekemoer}, {McCracken}, {Mobasher},
  {Renzini}, {Ricciardelli}, {Scodeggio}, {Taniguchi}, \&
  {Thompson}}]{Cassata2007}
{Cassata}, P., {Guzzo}, L., {Franceschini}, A., {et~al.} 2007, \apjs, 172, 270

\bibitem[{{Cenarro} {et~al.}(2001a){Cenarro}, {Cardiel}, {Gorgas}, {Peletier},
  {Vazdekis}, \& {Prada}}]{Cenarro2001a}
{Cenarro}, A.~J., {Cardiel}, N., {Gorgas}, J., {et~al.} 2001a, \mnras, 326, 959

\bibitem[{{Cenarro} {et~al.}(2001b){Cenarro}, {Gorgas}, {Cardiel}, {Pedraz},
  {Peletier}, \& {Vazdekis}}]{Cenarro2001b}
{Cenarro}, A.~J., {Gorgas}, J., {Cardiel}, N., {et~al.} 2001b, \mnras, 326, 981

\bibitem[{{Cenarro} {et~al.}(2002){Cenarro}, {Gorgas}, {Cardiel}, {Vazdekis},
  \& {Peletier}}]{Cenarro2002}
{Cenarro}, A.~J., {Gorgas}, J., {Cardiel}, N., {Vazdekis}, A., \& {Peletier},
  R.~F. 2002, \mnras, 329, 863

\bibitem[{{Cenarro} {et~al.}(2007){Cenarro}, {Peletier},
  {S{\'a}nchez-Bl{\'a}zquez}, {Selam}, {Toloba}, {Cardiel},
  {Falc{\'o}n-Barroso}, {Gorgas}, {Jim{\'e}nez-Vicente}, \&
  {Vazdekis}}]{Cenarro2007}
{Cenarro}, A.~J., {Peletier}, R.~F., {S{\'a}nchez-Bl{\'a}zquez}, P., {et~al.}
  2007, \mnras, 374, 664

\bibitem[{{Cervantes} \& {Vazdekis}(2009)}]{Cervantes2009}
{Cervantes}, J.~L. \& {Vazdekis}, A. 2009, \mnras, 392, 691

\bibitem[{{Chabrier}(2003)}]{Chabrier2003}
{Chabrier}, G. 2003, \pasp, 115, 763

\bibitem[{{Charlot} \& {Fall}(2000)}]{Charlot2000}
{Charlot}, S. \& {Fall}, S.~M. 2000, \apj, 539, 718

\bibitem[{{Chen} {et~al.}(2012){Chen}, {Kauffmann}, {Tremonti}, {White},
  {Heckman}, {Kova{\v c}}, {Bundy}, {Chisholm}, {Maraston}, {Schneider},
  {Bolton}, {Weaver}, \& {Brinkmann}}]{Chen2012}
{Chen}, Y.-M., {Kauffmann}, G., {Tremonti}, C.~A., {et~al.} 2012, \mnras, 421,
  314

\bibitem[{{Cid Fernandes} {et~al.}(2005){Cid Fernandes}, {Mateus}, {Sodr{\'e}},
  {Stasi{\'n}ska}, \& {Gomes}}]{Cid2005}
{Cid Fernandes}, R., {Mateus}, A., {Sodr{\'e}}, L., {Stasi{\'n}ska}, G., \&
  {Gomes}, J.~M. 2005, \mnras, 358, 363

\bibitem[{{Coelho} {et~al.}(2009){Coelho}, {Mendes de Oliveira}, \& {Cid
  Fernandes}}]{Coelho2009}
{Coelho}, P., {Mendes de Oliveira}, C., \& {Cid Fernandes}, R. 2009, \mnras,
  396, 624

\bibitem[{{Conroy} \& {van Dokkum}(2012)}]{Conroy2012}
{Conroy}, C. \& {van Dokkum}, P. 2012, \apj, 747, 69

\bibitem[{{Cooper} {et~al.}(2011){Cooper}, {Aird}, {Coil}, {Davis}, {Faber},
  {Juneau}, {Lotz}, {Nandra}, {Newman}, {Willmer}, \& {Yan}}]{Cooper2011}
{Cooper}, M.~C., {Aird}, J.~A., {Coil}, A.~L., {et~al.} 2011, \apjs, 193, 14

\bibitem[{{Cowie} {et~al.}(1996){Cowie}, {Songaila}, {Hu}, \&
  {Cohen}}]{Cowie1996}
{Cowie}, L.~L., {Songaila}, A., {Hu}, E.~M., \& {Cohen}, J.~G. 1996, \aj, 112,
  839

\bibitem[{{Crist{\'o}bal-Hornillos} {et~al.}(2009){Crist{\'o}bal-Hornillos},
  {Aguerri}, {Moles}, {Perea}, {Castander}, {Broadhurst}, {Alfaro},
  {Ben{\'{\i}}tez}, {Cabrera-Ca{\~n}o}, {Cepa}, {Cervi{\~n}o},
  {Fern{\'a}ndez-Soto}, {Gonz{\'a}lez Delgado}, {Husillos}, {Infante},
  {M{\'a}rquez}, {Mart{\'{\i}}nez}, {Masegosa}, {del Olmo}, {Prada},
  {Quintana}, \& {S{\'a}nchez}}]{Cristobal2009}
{Crist{\'o}bal-Hornillos}, D., {Aguerri}, J.~A.~L., {Moles}, M., {et~al.} 2009,
  \apj, 696, 1554

\bibitem[{{Davis} {et~al.}(2007){Davis}, {Guhathakurta}, {Konidaris}, {Newman},
  {Ashby}, {Biggs}, {Barmby}, {Bundy}, {Chapman}, {Coil}, {Conselice},
  {Cooper}, {Croton}, {Eisenhardt}, {Ellis}, {Faber}, {Fang}, {Fazio},
  {Georgakakis}, {Gerke}, {Goss}, {Gwyn}, {Harker}, {Hopkins}, {Huang},
  {Ivison}, {Kassin}, {Kirby}, {Koekemoer}, {Koo}, {Laird}, {Le Floc'h}, {Lin},
  {Lotz}, {Marshall}, {Martin}, {Metevier}, {Moustakas}, {Nandra}, {Noeske},
  {Papovich}, {Phillips}, {Rich}, {Rieke}, {Rigopoulou}, {Salim},
  {Schiminovich}, {Simard}, {Smail}, {Small}, {Weiner}, {Willmer}, {Willner},
  {Wilson}, {Wright}, \& {Yan}}]{Davis2007}
{Davis}, M., {Guhathakurta}, P., {Konidaris}, N.~P., {et~al.} 2007, \apjl, 660,
  L1

\bibitem[{{Eigenthaler} \& {Zeilinger}(2013)}]{Eigenthaler2013}
{Eigenthaler}, P. \& {Zeilinger}, W.~W. 2013, \aap, 553, A99

\bibitem[{{Faber}(1973)}]{Faber1973}
{Faber}, S.~M. 1973, \apj, 179, 731

\bibitem[{{Faber} {et~al.}(1985){Faber}, {Friel}, {Burstein}, \&
  {Gaskell}}]{Faber1985}
{Faber}, S.~M., {Friel}, E.~D., {Burstein}, D., \& {Gaskell}, C.~M. 1985,
  \apjs, 57, 711

\bibitem[{{Faber} {et~al.}(2007){Faber}, {Willmer}, {Wolf}, {Koo}, {Weiner},
  {Newman}, {Im}, {Coil}, {Conroy}, {Cooper}, {Davis}, {Finkbeiner}, {Gerke},
  {Gebhardt}, {Groth}, {Guhathakurta}, {Harker}, {Kaiser}, {Kassin},
  {Kleinheinrich}, {Konidaris}, {Kron}, {Lin}, {Luppino}, {Madgwick},
  {Meisenheimer}, {Noeske}, {Phillips}, {Sarajedini}, {Schiavon}, {Simard},
  {Szalay}, {Vogt}, \& {Yan}}]{Faber2007}
{Faber}, S.~M., {Willmer}, C.~N.~A., {Wolf}, C., {et~al.} 2007, \apj, 665, 265

\bibitem[{{Falc{\'o}n-Barroso} {et~al.}(2011){Falc{\'o}n-Barroso},
  {S{\'a}nchez-Bl{\'a}zquez}, {Vazdekis}, {Ricciardelli}, {Cardiel}, {Cenarro},
  {Gorgas}, \& {Peletier}}]{FalconBarroso2011}
{Falc{\'o}n-Barroso}, J., {S{\'a}nchez-Bl{\'a}zquez}, P., {Vazdekis}, A.,
  {et~al.} 2011, \aap, 532, A95

\bibitem[{{Ferreras} {et~al.}(2013){Ferreras}, {La Barbera}, {de la Rosa},
  {Vazdekis}, {de Carvalho}, {Falc{\'o}n-Barroso}, \&
  {Ricciardelli}}]{Ferreras2013}
{Ferreras}, I., {La Barbera}, F., {de la Rosa}, I.~G., {et~al.} 2013, \mnras,
  429, L15

\bibitem[{{Ferreras} \& {Silk}(2000)}]{Ferreras2000}
{Ferreras}, I. \& {Silk}, J. 2000, \apjl, 541, L37

\bibitem[{{Fitzpatrick}(1999)}]{Fitzpatrick1999}
{Fitzpatrick}, E.~L. 1999, \pasp, 111, 63

\bibitem[{{Fontana} {et~al.}(2006){Fontana}, {Salimbeni}, {Grazian},
  {Giallongo}, {Pentericci}, {Nonino}, {Fontanot}, {Menci}, {Monaco},
  {Cristiani}, {Vanzella}, {de Santis}, \& {Gallozzi}}]{Fontana2006}
{Fontana}, A., {Salimbeni}, S., {Grazian}, A., {et~al.} 2006, \aap, 459, 745

\bibitem[{{Fritz} {et~al.}(2014){Fritz}, {Scodeggio}, {Ilbert}, {Bolzonella},
  {Davidzon}, {Coupon}, {Garilli}, {Guzzo}, {Zamorani}, {Abbas}, {Adami},
  {Arnouts}, {Bel}, {Bottini}, {Branchini}, {Cappi}, {Cucciati}, {De Lucia},
  {de la Torre}, {Franzetti}, {Fumana}, {Granett}, {Iovino}, {Krywult}, {Le
  Brun}, {Le F{\`e}vre}, {Maccagni}, {Ma{\l}ek}, {Marulli}, {McCracken},
  {Paioro}, {Polletta}, {Pollo}, {Schlagenhaufer}, {Tasca}, {Tojeiro},
  {Vergani}, {Zanichelli}, {Burden}, {Di Porto}, {Marchetti}, {Marinoni},
  {Mellier}, {Moscardini}, {Nichol}, {Peacock}, {Percival}, {Phleps}, \&
  {Wolk}}]{Fritz2014}
{Fritz}, A., {Scodeggio}, M., {Ilbert}, O., {et~al.} 2014, \aap, 563, A92

\bibitem[{{Gallazzi} {et~al.}(2006){Gallazzi}, {Charlot}, {Brinchmann}, \&
  {White}}]{Gallazzi2006}
{Gallazzi}, A., {Charlot}, S., {Brinchmann}, J., \& {White}, S.~D.~M. 2006,
  \mnras, 370, 1106

\bibitem[{{Gallazzi} {et~al.}(2005){Gallazzi}, {Charlot}, {Brinchmann},
  {White}, \& {Tremonti}}]{Gallazzi2005}
{Gallazzi}, A., {Charlot}, S., {Brinchmann}, J., {White}, S.~D.~M., \&
  {Tremonti}, C.~A. 2005, \mnras, 362, 41

\bibitem[{{Gawiser} {et~al.}(2006){Gawiser}, {van Dokkum}, {Herrera}, {Maza},
  {Castander}, {Infante}, {Lira}, {Quadri}, {Toner}, {Treister}, {Urry},
  {Altmann}, {Assef}, {Christlein}, {Coppi}, {Dur{\'a}n}, {Franx}, {Galaz},
  {Huerta}, {Liu}, {L{\'o}pez}, {M{\'e}ndez}, {Moore}, {Rubio}, {Ruiz}, {Toft},
  \& {Yi}}]{Gawiser2006}
{Gawiser}, E., {van Dokkum}, P.~G., {Herrera}, D., {et~al.} 2006, \apjs, 162, 1

\bibitem[{{Gladis} {et~al.}(2014){Gladis}, {Juan}, {Mateu}, {Bruzual},
  {Cabrera-Ziri}, \& {Mej{\'{\i}}a-Narv{\'a}ez}}]{Gladis2014}
{Gladis}, M.~C., {Juan}, M.~P., {Mateu}, C., {et~al.} 2014, ArXiv e-prints

\bibitem[{{Gonz{\'a}lez Delgado} {et~al.}(2005){Gonz{\'a}lez Delgado},
  {Cervi{\~n}o}, {Martins}, {Leitherer}, \& {Hauschildt}}]{GonzalezDelgado2005}
{Gonz{\'a}lez Delgado}, R.~M., {Cervi{\~n}o}, M., {Martins}, L.~P.,
  {Leitherer}, C., \& {Hauschildt}, P.~H. 2005, \mnras, 357, 945

\bibitem[{{Gorgas} {et~al.}(1993){Gorgas}, {Faber}, {Burstein}, {Gonzalez},
  {Courteau}, \& {Prosser}}]{Gorgas1993}
{Gorgas}, J., {Faber}, S.~M., {Burstein}, D., {et~al.} 1993, \apjs, 86, 153

\bibitem[{{Gorgas} {et~al.}(2007){Gorgas}, {Jablonka}, \&
  {Goudfrooij}}]{Gorgas2007}
{Gorgas}, J., {Jablonka}, P., \& {Goudfrooij}, P. 2007, \aap, 474, 1081

\bibitem[{{Heap} \& {Lindler}(2007)}]{Heap2007}
{Heap}, S.~R. \& {Lindler}, D.~J. 2007, in Astronomical Society of the Pacific
  Conference Series, Vol. 374, From Stars to Galaxies: Building the Pieces to
  Build Up the Universe, ed. A.~{Vallenari}, R.~{Tantalo}, L.~{Portinari}, \&
  A.~{Moretti}, 409

\bibitem[{{Hogg}(1999)}]{Hogg2000}
{Hogg}, D.~W. 1999, ArXiv Astrophysics e-prints

\bibitem[{{Huertas-Company} {et~al.}(2008){Huertas-Company}, {Rouan}, {Tasca},
  {Soucail}, \& {Le F{\`e}vre}}]{Huertas2008}
{Huertas-Company}, M., {Rouan}, D., {Tasca}, L., {Soucail}, G., \& {Le
  F{\`e}vre}, O. 2008, \aap, 478, 971

\bibitem[{Hunter(2007)}]{Hunter2007}
Hunter, J.~D. 2007, Computing In Science \& Engineering, 9, 90

\bibitem[{{Ilbert} {et~al.}(2006){Ilbert}, {Arnouts}, {McCracken},
  {Bolzonella}, {Bertin}, {Le F{\`e}vre}, {Mellier}, {Zamorani}, {Pell{\`o}},
  {Iovino}, {Tresse}, {Le Brun}, {Bottini}, {Garilli}, {Maccagni}, {Picat},
  {Scaramella}, {Scodeggio}, {Vettolani}, {Zanichelli}, {Adami}, {Bardelli},
  {Cappi}, {Charlot}, {Ciliegi}, {Contini}, {Cucciati}, {Foucaud}, {Franzetti},
  {Gavignaud}, {Guzzo}, {Marano}, {Marinoni}, {Mazure}, {Meneux}, {Merighi},
  {Paltani}, {Pollo}, {Pozzetti}, {Radovich}, {Zucca}, {Bondi}, {Bongiorno},
  {Busarello}, {de La Torre}, {Gregorini}, {Lamareille}, {Mathez}, {Merluzzi},
  {Ripepi}, {Rizzo}, \& {Vergani}}]{Ilbert2006}
{Ilbert}, O., {Arnouts}, S., {McCracken}, H.~J., {et~al.} 2006, \aap, 457, 841

\bibitem[{{Ilbert} {et~al.}(2009){Ilbert}, {Capak}, {Salvato}, {Aussel},
  {McCracken}, {Sanders}, {Scoville}, {Kartaltepe}, {Arnouts}, {Le Floc'h},
  {Mobasher}, {Taniguchi}, {Lamareille}, {Leauthaud}, {Sasaki}, {Thompson},
  {Zamojski}, {Zamorani}, {Bardelli}, {Bolzonella}, {Bongiorno}, {Brusa},
  {Caputi}, {Carollo}, {Contini}, {Cook}, {Coppa}, {Cucciati}, {de la Torre},
  {de Ravel}, {Franzetti}, {Garilli}, {Hasinger}, {Iovino}, {Kampczyk},
  {Kneib}, {Knobel}, {Kovac}, {Le Borgne}, {Le Brun}, {F{\`e}vre}, {Lilly},
  {Looper}, {Maier}, {Mainieri}, {Mellier}, {Mignoli}, {Murayama}, {Pell{\`o}},
  {Peng}, {P{\'e}rez-Montero}, {Renzini}, {Ricciardelli}, {Schiminovich},
  {Scodeggio}, {Shioya}, {Silverman}, {Surace}, {Tanaka}, {Tasca}, {Tresse},
  {Vergani}, \& {Zucca}}]{Ilbert2009}
{Ilbert}, O., {Capak}, P., {Salvato}, M., {et~al.} 2009, \apj, 690, 1236

\bibitem[{{Ilbert} {et~al.}(2010){Ilbert}, {Salvato}, {Le Floc'h}, {Aussel},
  {Capak}, {McCracken}, {Mobasher}, {Kartaltepe}, {Scoville}, {Sanders},
  {Arnouts}, {Bundy}, {Cassata}, {Kneib}, {Koekemoer}, {Le F{\`e}vre}, {Lilly},
  {Surace}, {Taniguchi}, {Tasca}, {Thompson}, {Tresse}, {Zamojski}, {Zamorani},
  \& {Zucca}}]{Ilbert2010}
{Ilbert}, O., {Salvato}, M., {Le Floc'h}, E., {et~al.} 2010, \apj, 709, 644

\bibitem[{{Jimenez} {et~al.}(2007){Jimenez}, {Bernardi}, {Haiman}, {Panter}, \&
  {Heavens}}]{Jimenez2007}
{Jimenez}, R., {Bernardi}, M., {Haiman}, Z., {Panter}, B., \& {Heavens}, A.~F.
  2007, \apj, 669, 947

\bibitem[{{Johnson} \& {Morgan}(1953)}]{Johnson1953}
{Johnson}, H.~L. \& {Morgan}, W.~W. 1953, \apj, 117, 313

\bibitem[{{J{\o}rgensen}(1999)}]{Jorgensen1999}
{J{\o}rgensen}, I. 1999, \mnras, 306, 607

\bibitem[{{Kaviraj} {et~al.}(2007){Kaviraj}, {Schawinski}, {Devriendt},
  {Ferreras}, {Khochfar}, {Yoon}, {Yi}, {Deharveng}, {Boselli}, {Barlow},
  {Conrow}, {Forster}, {Friedman}, {Martin}, {Morrissey}, {Neff},
  {Schiminovich}, {Seibert}, {Small}, {Wyder}, {Bianchi}, {Donas}, {Heckman},
  {Lee}, {Madore}, {Milliard}, {Rich}, \& {Szalay}}]{Kaviraj2007}
{Kaviraj}, S., {Schawinski}, K., {Devriendt}, J.~E.~G., {et~al.} 2007, \apjs,
  173, 619

\bibitem[{{Koleva} {et~al.}(2009){Koleva}, {Prugniel}, {Bouchard}, \&
  {Wu}}]{Koleva2009}
{Koleva}, M., {Prugniel}, P., {Bouchard}, A., \& {Wu}, Y. 2009, \aap, 501, 1269

\bibitem[{{Koleva} {et~al.}(2008){Koleva}, {Prugniel}, {Ocvirk}, {Le Borgne},
  \& {Soubiran}}]{Koleva2008}
{Koleva}, M., {Prugniel}, P., {Ocvirk}, P., {Le Borgne}, D., \& {Soubiran}, C.
  2008, \mnras, 385, 1998

\bibitem[{{Kriek} {et~al.}(2009){Kriek}, {van Dokkum}, {Labb{\'e}}, {Franx},
  {Illingworth}, {Marchesini}, \& {Quadri}}]{Kriek2009}
{Kriek}, M., {van Dokkum}, P.~G., {Labb{\'e}}, I., {et~al.} 2009, \apj, 700,
  221

\bibitem[{{Kroupa}(2001)}]{Kroupa2001}
{Kroupa}, P. 2001, \mnras, 322, 231

\bibitem[{{Kuntschner} {et~al.}(2001){Kuntschner}, {Lucey}, {Smith}, {Hudson},
  \& {Davies}}]{Kuntschner2001}
{Kuntschner}, H., {Lucey}, J.~R., {Smith}, R.~J., {Hudson}, M.~J., \& {Davies},
  R.~L. 2001, \mnras, 323, 615

\bibitem[{{La Barbera} {et~al.}(2012){La Barbera}, {Ferreras}, {de Carvalho},
  {Bruzual}, {Charlot}, {Pasquali}, \& {Merlin}}]{LaBarbera2012}
{La Barbera}, F., {Ferreras}, I., {de Carvalho}, R.~R., {et~al.} 2012, \mnras,
  426, 2300

\bibitem[{{Le Borgne} {et~al.}(2003){Le Borgne}, {Bruzual}, {Pell{\'o}},
  {Lan{\c c}on}, {Rocca-Volmerange}, {Sanahuja}, {Schaerer}, {Soubiran}, \&
  {V{\'{\i}}lchez-G{\'o}mez}}]{LeBorgne2003}
{Le Borgne}, J.-F., {Bruzual}, G., {Pell{\'o}}, R., {et~al.} 2003, \aap, 402,
  433

\bibitem[{{Lilly} {et~al.}(2009){Lilly}, {Le Brun}, {Maier}, {Mainieri},
  {Mignoli}, {Scodeggio}, {Zamorani}, {Carollo}, {Contini}, {Kneib}, {Le
  F{\`e}vre}, {Renzini}, {Bardelli}, {Bolzonella}, {Bongiorno}, {Caputi},
  {Coppa}, {Cucciati}, {de la Torre}, {de Ravel}, {Franzetti}, {Garilli},
  {Iovino}, {Kampczyk}, {Kovac}, {Knobel}, {Lamareille}, {Le Borgne}, {Pello},
  {Peng}, {P{\'e}rez-Montero}, {Ricciardelli}, {Silverman}, {Tanaka}, {Tasca},
  {Tresse}, {Vergani}, {Zucca}, {Ilbert}, {Salvato}, {Oesch}, {Abbas},
  {Bottini}, {Capak}, {Cappi}, {Cassata}, {Cimatti}, {Elvis}, {Fumana},
  {Guzzo}, {Hasinger}, {Koekemoer}, {Leauthaud}, {Maccagni}, {Marinoni},
  {McCracken}, {Memeo}, {Meneux}, {Porciani}, {Pozzetti}, {Sanders},
  {Scaramella}, {Scarlata}, {Scoville}, {Shopbell}, \& {Taniguchi}}]{Lilly2009}
{Lilly}, S.~J., {Le Brun}, V., {Maier}, C., {et~al.} 2009, \apjs, 184, 218

\bibitem[{{Lilly} {et~al.}(2007){Lilly}, {Le F{\`e}vre}, {Renzini}, {Zamorani},
  {Scodeggio}, {Contini}, {Carollo}, {Hasinger}, {Kneib}, {Iovino}, {Le Brun},
  {Maier}, {Mainieri}, {Mignoli}, {Silverman}, {Tasca}, {Bolzonella},
  {Bongiorno}, {Bottini}, {Capak}, {Caputi}, {Cimatti}, {Cucciati}, {Daddi},
  {Feldmann}, {Franzetti}, {Garilli}, {Guzzo}, {Ilbert}, {Kampczyk}, {Kovac},
  {Lamareille}, {Leauthaud}, {Borgne}, {McCracken}, {Marinoni}, {Pello},
  {Ricciardelli}, {Scarlata}, {Vergani}, {Sanders}, {Schinnerer}, {Scoville},
  {Taniguchi}, {Arnouts}, {Aussel}, {Bardelli}, {Brusa}, {Cappi}, {Ciliegi},
  {Finoguenov}, {Foucaud}, {Franceschini}, {Halliday}, {Impey}, {Knobel},
  {Koekemoer}, {Kurk}, {Maccagni}, {Maddox}, {Marano}, {Marconi}, {Meneux},
  {Mobasher}, {Moreau}, {Peacock}, {Porciani}, {Pozzetti}, {Scaramella},
  {Schiminovich}, {Shopbell}, {Smail}, {Thompson}, {Tresse}, {Vettolani},
  {Zanichelli}, \& {Zucca}}]{Lilly2007}
{Lilly}, S.~J., {Le F{\`e}vre}, O., {Renzini}, A., {et~al.} 2007, \apjs, 172,
  70

\bibitem[{{Liu} {et~al.}(2013){Liu}, {Lu}, {Chen}, {Du}, \& {Zhao}}]{Liu2013}
{Liu}, G., {Lu}, Y., {Chen}, X., {Du}, W., \& {Zhao}, Y. 2013, ArXiv e-prints

\bibitem[{{Lonoce} {et~al.}(2014){Lonoce}, {Longhetti}, {Saracco}, {Gargiulo},
  \& {Tamburri}}]{Lonoce2014}
{Lonoce}, I., {Longhetti}, M., {Saracco}, P., {Gargiulo}, A., \& {Tamburri}, S.
  2014, \mnras, 444, 2048

\bibitem[{{Maraston} {et~al.}(2009){Maraston}, {Str{\"o}mb{\"a}ck}, {Thomas},
  {Wake}, \& {Nichol}}]{Maraston2009}
{Maraston}, C., {Str{\"o}mb{\"a}ck}, G., {Thomas}, D., {Wake}, D.~A., \&
  {Nichol}, R.~C. 2009, \mnras, 394, L107

\bibitem[{{M{\'a}rmol-Queralt{\'o}} {et~al.}(2008){M{\'a}rmol-Queralt{\'o}},
  {Cardiel}, {Cenarro}, {Vazdekis}, {Gorgas}, {Pedraz}, {Peletier}, \&
  {S{\'a}nchez-Bl{\'a}zquez}}]{Marmol-Queralto2008}
{M{\'a}rmol-Queralt{\'o}}, E., {Cardiel}, N., {Cenarro}, A.~J., {et~al.} 2008,
  \aap, 489, 885

\bibitem[{{Martins} {et~al.}(2005){Martins}, {Gonz{\'a}lez Delgado},
  {Leitherer}, {Cervi{\~n}o}, \& {Hauschildt}}]{Martins2005}
{Martins}, L.~P., {Gonz{\'a}lez Delgado}, R.~M., {Leitherer}, C.,
  {Cervi{\~n}o}, M., \& {Hauschildt}, P. 2005, \mnras, 358, 49

\bibitem[{{Massa}(1987)}]{Massa1987}
{Massa}, D. 1987, \aj, 94, 1675

\bibitem[{{Mathis} {et~al.}(2006){Mathis}, {Charlot}, \&
  {Brinchmann}}]{Mathis2006}
{Mathis}, H., {Charlot}, S., \& {Brinchmann}, J. 2006, \mnras, 365, 385

\bibitem[{{Mathis}(1990)}]{Mathis1990}
{Mathis}, J.~S. 1990, \araa, 28, 37

\bibitem[{{McClure} \& {van den Bergh}(1968)}]{Mcclure1968}
{McClure}, R.~D. \& {van den Bergh}, S. 1968, \aj, 73, 313

\bibitem[{{Moles} {et~al.}(2008){Moles}, {Ben{\'{\i}}tez}, {Aguerri}, {Alfaro},
  {Broadhurst}, {Cabrera-Ca{\~n}o}, {Castander}, {Cepa}, {Cervi{\~n}o},
  {Crist{\'o}bal-Hornillos}, {Fern{\'a}ndez-Soto}, {Gonz{\'a}lez Delgado},
  {Infante}, {M{\'a}rquez}, {Mart{\'{\i}}nez}, {Masegosa}, {del Olmo}, {Perea},
  {Prada}, {Quintana}, \& {S{\'a}nchez}}]{Moles2008}
{Moles}, M., {Ben{\'{\i}}tez}, N., {Aguerri}, J.~A.~L., {et~al.} 2008, \aj,
  136, 1325

\bibitem[{{Molino} {et~al.}(2014){Molino}, {Ben{\'{\i}}tez}, {Moles},
  {Fern{\'a}ndez-Soto}, {Crist{\'o}bal-Hornillos}, {Ascaso},
  {Jim{\'e}nez-Teja}, {Schoenell}, {Arnalte-Mur}, {Povi{\'c}}, {Coe},
  {L{\'o}pez-Sanjuan}, {D{\'{\i}}az-Garc{\'{\i}}a}, {Varela}, {Stefanon},
  {Cenarro}, {Matute}, {Masegosa}, {M{\'a}rquez}, {Perea}, {Del Olmo},
  {Husillos}, {Alfaro}, {Aparicio-Villegas}, {Cervi{\~n}o}, {Huertas-Company},
  {Aguerri}, {Broadhurst}, {Cabrera-Ca{\~n}o}, {Cepa}, {Gonz{\'a}lez},
  {Infante}, {Mart{\'{\i}}nez}, {Prada}, \& {Quintana}}]{Molino2014}
{Molino}, A., {Ben{\'{\i}}tez}, N., {Moles}, M., {et~al.} 2014, \mnras, 441,
  2891

\bibitem[{{Monachesi} {et~al.}(2012){Monachesi}, {Trager}, {Lauer}, {Hidalgo},
  {Freedman}, {Dressler}, {Grillmair}, \& {Mighell}}]{Monachesi2012}
{Monachesi}, A., {Trager}, S.~C., {Lauer}, T.~R., {et~al.} 2012, \apj, 745, 97

\bibitem[{{Ocvirk} {et~al.}(2006){Ocvirk}, {Pichon}, {Lan{\c c}on}, \&
  {Thi{\'e}baut}}]{Ocvirk2006}
{Ocvirk}, P., {Pichon}, C., {Lan{\c c}on}, A., \& {Thi{\'e}baut}, E. 2006,
  \mnras, 365, 74

\bibitem[{{O'Donnell}(1994)}]{Odonnell1994}
{O'Donnell}, J.~E. 1994, \apj, 422, 158

\bibitem[{{Oke} \& {Gunn}(1983)}]{Oke1983}
{Oke}, J.~B. \& {Gunn}, J.~E. 1983, \apj, 266, 713

\bibitem[{{P{\'e}rez-Gonz{\'a}lez} {et~al.}(2013){P{\'e}rez-Gonz{\'a}lez},
  {Cava}, {Barro}, {Villar}, {Cardiel}, {Ferreras},
  {Rodr{\'{\i}}guez-Espinosa}, {Alonso-Herrero}, {Balcells}, {Cenarro}, {Cepa},
  {Charlot}, {Cimatti}, {Conselice}, {Daddi}, {Donley}, {Elbaz}, {Espino},
  {Gallego}, {Gobat}, {Gonz{\'a}lez-Mart{\'{\i}}n}, {Guzm{\'a}n},
  {Hern{\'a}n-Caballero}, {Mu{\~n}oz-Tu{\~n}{\'o}n}, {Renzini},
  {Rodr{\'{\i}}guez-Zaur{\'{\i}}n}, {Tresse}, {Trujillo}, \&
  {Zamorano}}]{PerezGonzalez2013}
{P{\'e}rez-Gonz{\'a}lez}, P.~G., {Cava}, A., {Barro}, G., {et~al.} 2013, \apj,
  762, 46

\bibitem[{{Pickles} \& {Depagne}(2010)}]{Pickles2010}
{Pickles}, A. \& {Depagne}, {\'E}. 2010, \pasp, 122, 1437

\bibitem[{{Poggianti} {et~al.}(2001){Poggianti}, {Bridges}, {Carter},
  {Mobasher}, {Doi}, {Iye}, {Kashikawa}, {Komiyama}, {Okamura}, {Sekiguchi},
  {Shimasaku}, {Yagi}, \& {Yasuda}}]{Poggianti2001}
{Poggianti}, B.~M., {Bridges}, T.~J., {Carter}, D., {et~al.} 2001, \apj, 563,
  118

\bibitem[{{Postman} {et~al.}(2012){Postman}, {Coe}, {Ben{\'{\i}}tez},
  {Bradley}, {Broadhurst}, {Donahue}, {Ford}, {Graur}, {Graves}, {Jouvel},
  {Koekemoer}, {Lemze}, {Medezinski}, {Molino}, {Moustakas}, {Ogaz}, {Riess},
  {Rodney}, {Rosati}, {Umetsu}, {Zheng}, {Zitrin}, {Bartelmann}, {Bouwens},
  {Czakon}, {Golwala}, {Host}, {Infante}, {Jha}, {Jimenez-Teja}, {Kelson},
  {Lahav}, {Lazkoz}, {Maoz}, {McCully}, {Melchior}, {Meneghetti}, {Merten},
  {Moustakas}, {Nonino}, {Patel}, {Reg{\"o}s}, {Sayers}, {Seitz}, \& {Van der
  Wel}}]{Postman2012}
{Postman}, M., {Coe}, D., {Ben{\'{\i}}tez}, N., {et~al.} 2012, \apjs, 199, 25

\bibitem[{{Povi{\'c}} {et~al.}(2013){Povi{\'c}}, {Huertas-Company}, {Aguerri},
  {M{\'a}rquez}, {Masegosa}, {Husillos}, {Molino}, {Crist{\'o}bal-Hornillos},
  {Perea}, {Ben{\'{\i}}tez}, {Olmo}, {Fern{\'a}ndez-Soto}, {Jim{\'e}nez-Teja},
  {Moles}, {Alfaro}, {Aparicio-Villegas}, {Ascaso}, {Broadhurst},
  {Cabrera-Ca{\~n}o}, {Castander}, {Cepa}, {Fernandez Lorenzo}, {Cervi{\~n}o},
  {Delgado}, {Infante}, {L{\'o}pez-Sanjuan}, {Mart{\'{\i}}nez}, {Matute},
  {Oteo}, {P{\'e}rez-Garc{\'{\i}}a}, {Prada}, \& {Quintana}}]{Povic2013}
{Povi{\'c}}, M., {Huertas-Company}, M., {Aguerri}, J.~A.~L., {et~al.} 2013,
  \mnras, 435, 3444

\bibitem[{{Pozzetti} {et~al.}(2007){Pozzetti}, {Bolzonella}, {Lamareille},
  {Zamorani}, {Franzetti}, {Le F{\`e}vre}, {Iovino}, {Temporin}, {Ilbert},
  {Arnouts}, {Charlot}, {Brinchmann}, {Zucca}, {Tresse}, {Scodeggio}, {Guzzo},
  {Bottini}, {Garilli}, {Le Brun}, {Maccagni}, {Picat}, {Scaramella},
  {Vettolani}, {Zanichelli}, {Adami}, {Bardelli}, {Cappi}, {Ciliegi},
  {Contini}, {Foucaud}, {Gavignaud}, {McCracken}, {Marano}, {Marinoni},
  {Mazure}, {Meneux}, {Merighi}, {Paltani}, {Pell{\`o}}, {Pollo}, {Radovich},
  {Bondi}, {Bongiorno}, {Cucciati}, {de la Torre}, {Gregorini}, {Mellier},
  {Merluzzi}, {Vergani}, \& {Walcher}}]{Pozzetti2007}
{Pozzetti}, L., {Bolzonella}, M., {Lamareille}, F., {et~al.} 2007, \aap, 474,
  443

\bibitem[{{Prevot} {et~al.}(1984){Prevot}, {Lequeux}, {Prevot}, {Maurice}, \&
  {Rocca-Volmerange}}]{Prevot1984}
{Prevot}, M.~L., {Lequeux}, J., {Prevot}, L., {Maurice}, E., \&
  {Rocca-Volmerange}, B. 1984, \aap, 132, 389

\bibitem[{{Pritchet}(1977)}]{Pritchet1977}
{Pritchet}, C. 1977, \apjs, 35, 397

\bibitem[{{Prugniel} \& {Soubiran}(2001)}]{Prugniel2001}
{Prugniel}, P. \& {Soubiran}, C. 2001, \aap, 369, 1048

\bibitem[{{Ricciardelli} {et~al.}(2012){Ricciardelli}, {Vazdekis}, {Cenarro},
  \& {Falc{\'o}n-Barroso}}]{Ricciardelli2012}
{Ricciardelli}, E., {Vazdekis}, A., {Cenarro}, A.~J., \& {Falc{\'o}n-Barroso},
  J. 2012, \mnras, 424, 172

\bibitem[{{Rogers} {et~al.}(2010){Rogers}, {Ferreras}, {Peletier}, \&
  {Silk}}]{Rogers2010}
{Rogers}, B., {Ferreras}, I., {Peletier}, R., \& {Silk}, J. 2010, \mnras, 402,
  447

\bibitem[{{S{\'a}nchez-Bl{\'a}zquez} {et~al.}(2007){S{\'a}nchez-Bl{\'a}zquez},
  {Forbes}, {Strader}, {Brodie}, \& {Proctor}}]{Sanchezblazquez2007}
{S{\'a}nchez-Bl{\'a}zquez}, P., {Forbes}, D.~A., {Strader}, J., {Brodie}, J.,
  \& {Proctor}, R. 2007, \mnras, 377, 759

\bibitem[{{S{\'a}nchez-Bl{\'a}zquez} {et~al.}(2006a){S{\'a}nchez-Bl{\'a}zquez},
  {Gorgas}, {Cardiel}, \& {Gonz{\'a}lez}}]{Sanchezblazquez2006a}
{S{\'a}nchez-Bl{\'a}zquez}, P., {Gorgas}, J., {Cardiel}, N., \& {Gonz{\'a}lez},
  J.~J. 2006a, \aap, 457, 809

\bibitem[{{S{\'a}nchez-Bl{\'a}zquez} {et~al.}(2006b){S{\'a}nchez-Bl{\'a}zquez},
  {Peletier}, {Jim{\'e}nez-Vicente}, {Cardiel}, {Cenarro},
  {Falc{\'o}n-Barroso}, {Gorgas}, {Selam}, \&
  {Vazdekis}}]{Sanchezblazquez2006b}
{S{\'a}nchez-Bl{\'a}zquez}, P., {Peletier}, R.~F., {Jim{\'e}nez-Vicente}, J.,
  {et~al.} 2006b, \mnras, 371, 703

\bibitem[{{Santos} {et~al.}(2002){Santos}, {Alloin}, {Bica}, \&
  {Bonatto}}]{Santos2002}
{Santos}, F.~C.~Jr., J., {Alloin}, D., {Bica}, E., \& {Bonatto}, C.~J. 2002, in
  IAU Symposium, Vol. 207, Extragalactic Star Clusters, ed. D.~P. {Geisler},
  E.~K. {Grebel}, \& D.~{Minniti}, 727

\bibitem[{{Sawicki}(2012)}]{Sawicki2012}
{Sawicki}, M. 2012, \pasp, 124, 1208

\bibitem[{{Schlafly} \& {Finkbeiner}(2011)}]{Schlafly2011}
{Schlafly}, E.~F. \& {Finkbeiner}, D.~P. 2011, \apj, 737, 103

\bibitem[{{Schlegel} {et~al.}(1998){Schlegel}, {Finkbeiner}, \&
  {Davis}}]{Schlegel1998}
{Schlegel}, D.~J., {Finkbeiner}, D.~P., \& {Davis}, M. 1998, \apj, 500, 525

\bibitem[{{Scoville} {et~al.}(2007){Scoville}, {Aussel}, {Brusa}, {Capak},
  {Carollo}, {Elvis}, {Giavalisco}, {Guzzo}, {Hasinger}, {Impey}, {Kneib},
  {LeFevre}, {Lilly}, {Mobasher}, {Renzini}, {Rich}, {Sanders}, {Schinnerer},
  {Schminovich}, {Shopbell}, {Taniguchi}, \& {Tyson}}]{Scoville2007}
{Scoville}, N., {Aussel}, H., {Brusa}, M., {et~al.} 2007, \apjs, 172, 1

\bibitem[{{Stone}(1996)}]{Stone1996}
{Stone}, R.~P.~S. 1996, \apjs, 107, 423

\bibitem[{{Strateva} {et~al.}(2001){Strateva}, {Ivezi{\'c}}, {Knapp},
  {Narayanan}, {Strauss}, {Gunn}, {Lupton}, {Schlegel}, {Bahcall}, {Brinkmann},
  {Brunner}, {Budav{\'a}ri}, {Csabai}, {Castander}, {Doi}, {Fukugita}, {Gy{\H
  o}ry}, {Hamabe}, {Hennessy}, {Ichikawa}, {Kunszt}, {Lamb}, {McKay},
  {Okamura}, {Racusin}, {Sekiguchi}, {Schneider}, {Shimasaku}, \&
  {York}}]{Strateva2001}
{Strateva}, I., {Ivezi{\'c}}, {\v Z}., {Knapp}, G.~R., {et~al.} 2001, \aj, 122,
  1861

\bibitem[{{Thomas} {et~al.}(2003){Thomas}, {Maraston}, \&
  {Bender}}]{Thomas2003}
{Thomas}, D., {Maraston}, C., \& {Bender}, R. 2003, \mnras, 339, 897

\bibitem[{{Thomas} {et~al.}(2005){Thomas}, {Maraston}, {Bender}, \& {Mendes de
  Oliveira}}]{Thomas2005}
{Thomas}, D., {Maraston}, C., {Bender}, R., \& {Mendes de Oliveira}, C. 2005,
  \apj, 621, 673

\bibitem[{{Tifft}(1963)}]{Tifft1963}
{Tifft}, W.~G. 1963, \aj, 68, 302

\bibitem[{{Tojeiro} {et~al.}(2007){Tojeiro}, {Heavens}, {Jimenez}, \&
  {Panter}}]{Tojeiro2007}
{Tojeiro}, R., {Heavens}, A.~F., {Jimenez}, R., \& {Panter}, B. 2007, \mnras,
  381, 1252

\bibitem[{{Trager} {et~al.}(1998){Trager}, {Worthey}, {Faber}, {Burstein}, \&
  {Gonz{\'a}lez}}]{Trager1998}
{Trager}, S.~C., {Worthey}, G., {Faber}, S.~M., {Burstein}, D., \&
  {Gonz{\'a}lez}, J.~J. 1998, \apjs, 116, 1

\bibitem[{{Tremonti} {et~al.}(2004){Tremonti}, {Heckman}, {Kauffmann},
  {Brinchmann}, {Charlot}, {White}, {Seibert}, {Peng}, {Schlegel}, {Uomoto},
  {Fukugita}, \& {Brinkmann}}]{Tremonti2004}
{Tremonti}, C.~A., {Heckman}, T.~M., {Kauffmann}, G., {et~al.} 2004, \apj, 613,
  898

\bibitem[{{Valdes} {et~al.}(2004){Valdes}, {Gupta}, {Rose}, {Singh}, \&
  {Bell}}]{Valdes2004}
{Valdes}, F., {Gupta}, R., {Rose}, J.~A., {Singh}, H.~P., \& {Bell}, D.~J.
  2004, \apjs, 152, 251

\bibitem[{{Vazdekis} \& {Arimoto}(1999)}]{Vazdekis1999}
{Vazdekis}, A. \& {Arimoto}, N. 1999, \apj, 525, 144

\bibitem[{{Vazdekis} {et~al.}(2003){Vazdekis}, {Cenarro}, {Gorgas}, {Cardiel},
  \& {Peletier}}]{Vazdekis2003}
{Vazdekis}, A., {Cenarro}, A.~J., {Gorgas}, J., {Cardiel}, N., \& {Peletier},
  R.~F. 2003, \mnras, 340, 1317

\bibitem[{{Vazdekis} {et~al.}(2012){Vazdekis}, {Ricciardelli}, {Cenarro},
  {Rivero-Gonz{\'a}lez}, {D{\'{\i}}az-Garc{\'{\i}}a}, \&
  {Falc{\'o}n-Barroso}}]{Vazdekis2012}
{Vazdekis}, A., {Ricciardelli}, E., {Cenarro}, A.~J., {et~al.} 2012, \mnras,
  424, 157

\bibitem[{{Vazdekis} {et~al.}(2010){Vazdekis}, {S{\'a}nchez-Bl{\'a}zquez},
  {Falc{\'o}n-Barroso}, {Cenarro}, {Beasley}, {Cardiel}, {Gorgas}, \&
  {Peletier}}]{Vazdekis2010}
{Vazdekis}, A., {S{\'a}nchez-Bl{\'a}zquez}, P., {Falc{\'o}n-Barroso}, J.,
  {et~al.} 2010, \mnras, 404, 1639

\bibitem[{{Walcher} {et~al.}(2011){Walcher}, {Groves}, {Budav{\'a}ri}, \&
  {Dale}}]{Walcher2011}
{Walcher}, J., {Groves}, B., {Budav{\'a}ri}, T., \& {Dale}, D. 2011, \apss,
  331, 1

\bibitem[{{Williams} {et~al.}(2009){Williams}, {Quadri}, {Franx}, {van Dokkum},
  \& {Labb{\'e}}}]{Williams2009}
{Williams}, R.~J., {Quadri}, R.~F., {Franx}, M., {van Dokkum}, P., \&
  {Labb{\'e}}, I. 2009, \apj, 691, 1879

\bibitem[{{Wolf} {et~al.}(2003){Wolf}, {Meisenheimer}, {Rix}, {Borch}, {Dye},
  \& {Kleinheinrich}}]{Wolf2003}
{Wolf}, C., {Meisenheimer}, K., {Rix}, H.-W., {et~al.} 2003, \aap, 401, 73

\bibitem[{{Wood}(1966)}]{Wood1966}
{Wood}, D.~B. 1966, \apj, 145, 36

\bibitem[{{Worthey}(1994a)}]{Worthey1994a}
{Worthey}, G. 1994a, \apjs, 95, 107

\bibitem[{{Worthey} {et~al.}(1994b){Worthey}, {Faber}, {Gonzalez}, \&
  {Burstein}}]{Worthey1994b}
{Worthey}, G., {Faber}, S.~M., {Gonzalez}, J.~J., \& {Burstein}, D. 1994b,
  \apjs, 94, 687

\bibitem[{{Wu} {et~al.}(2005){Wu}, {Shao}, {Mo}, {Xia}, \& {Deng}}]{Wu2005}
{Wu}, H., {Shao}, Z., {Mo}, H.~J., {Xia}, X., \& {Deng}, Z. 2005, \apj, 622,
  244

\end{thebibliography}

%


\end{document}